\documentclass[aps]{revtex4}

\usepackage{graphicx}
\usepackage{dcolumn}
\usepackage{bm}

\begin{document}

\title{\bf Clock Synchronization and Navigation in the
Vicinity of the Earth}

\author{Thomas B. Bahder}
\email[]{bahder@arl.army.mil}
\affiliation{U. S. Army Research Laboratory \\
2800 Powder Mill Road \\
Adelphi, Maryland, USA  20783-1197}

\date{\today}

\begin{abstract}
Clock synchronization is the backbone of applications such as
high-accuracy satellite navigation, geolocation, space-based
interferometry, and cryptographic communication systems.   The
high accuracy of synchronization needed over satellite-to-ground
and satellite-to-satellite distances requires the use of general
relativistic concepts. The role of geometrical optics and antenna
phase center approximations are discussed in high accuracy work.
The clock synchronization problem is explored from a general
relativistic point of view, with emphasis on the local measurement
process and the use of the tetrad formalism as the correct model
of relativistic measurements.  The treatment makes use of J. L.
Synge's world function of space-time as a basic coordinate
independent geometric concept. A metric is used for space-time in
the vicinity of the Earth, where coordinate time is proper time on
the geoid. The problem of satellite clock syntonization is
analyzed by numerically integrating the geodesic equations of
motion for low-Earth orbit (LEO), geosynchronous orbit (GEO), and
highly elliptical orbit (HEO) satellites. Proper time minus
coordinate time is computed for satellites in these orbital
regimes.   The frequency shift as a function of time is computed
for a signal observed on the Earth's geoid from a LEO, GEO, and
HEO satellite. Finally, the problem of geolocation in curved
space-time is briefly explored using the world function formalism.
\end{abstract}


\maketitle

\newpage

\tableofcontents

\newpage

\section{Introduction}
In the past several decades, there has been a dramatic improvement
in two technology areas: atomic clocks and lasers for free-space
optical communications. Future spacecraft are envisioned as
communicating by free-space laser links.  The synergy of
technological development in atomic clocks and free-space lasers
will lead to unprecedented advancements in high-accuracy
space-time navigation~\cite{Bahder2001,Bahder2003}, digital
communications,
geolocation~\cite{Ho1993,Fang1995,Niezgoda1994,Ho1997,Pattison2000},
surveillance using space-based
interferometers~\cite{burke1991,Steyskal2001}, coherent
distributed-aperture sensing at high
frequencies~\cite{Sovers1998,Boverie1970,overman2000,Hodge1999,levine1983,levine1990,sedwick1999,colavita1996},
and cryptographic communication systems.

Accurate clock synchronization is the backbone of these systems.
Consider the dependence of geolocation accuracy on clock
synchronization.  Consider geosynchronous satellites that receive
a signal from an emitter of electromagnetic radiation located on
the surface of the Earth. Assuming that the signals travel on the
line-of-sight,  the geometry leads to a maximum signal time
difference of arrival between two satellites that is 19.64 ms, see
Appendix A. The information on the difference of ranges, $\Delta
l$, between the emitter and each satellite, is contained in the
maximum time delay that is equal to 19.64 ms. If the clocks in two
satellites are synchronized only to an accuracy of, say 10 ns, the
order of magnitude in the position error of the emitter, $\Delta
x$, is given by the fraction of the range difference:
\begin{equation} \label{rangeaccuracy}
\Delta x  \sim   \frac{10 {\rm ns}}{19.64 {\rm ms}} \, \Delta l
\sim 3000 {\rm m} = 1.5 {\rm nm}
\end{equation}
An improvement in the clock synchronization translates directly
into an improvement in position accuracy.  For example, an
improvement in clock synchronization by three orders of magnitude
will produce a position accuracy on the order of  3 m.

Applications such as a space-based interferometer, can have even
more stringent requirements on clock synchronization. For example,
a multi-satellite space-based interferometer (distributed aperture
system) that operates at a wavelength $\lambda$ will require
accurate determination of relative satellite positions (nodes of
the interferometer) to better than $\Delta x = \lambda$.  For
practical purposes, say we will need $\Delta x =\lambda/10$, which
translates to a clock synchronization requirement $(\lambda/10
c)$, where $c$ is the speed of light. As an example of the
stringent requirements on time synchronization, consider operating
at 22~GHz, which is a frequency of interest to radio
astronomers~\cite{VSOP}.  At this frequency, the position of the
satellite nodes must be known accurately to $\lambda/(10
c)=$1.4~mm and time synchronization to 4.5 ps. See
Table~\ref{tab:TimeingPrecision} for a range of values
corresponding to different frequencies. For operating at optical
wavelengths of 500 nm, the required position must be known to 50
nm and time must be synchronized to 0.16 fs. These numbers
challenge and exceed the realm of possible time synchronization
accuracy that is available today. However, the synergy between
accurate clocks and optical free-space communication is expected
to continue, so that hardware may soon support  such stringent
time synchronization requirements. New time synchronization
schemes will then be required.

The emerging field of quantum information and quantum
computation~\cite{Ekert2000,Nielson2000} has potential to produce
new ultra-precise clock synchronization protocols. In fact,
several clock synchronization schemes have been proposed based on
quantum mechanical
concepts~\cite{MandelOu1987,Jozsa2000,Chuang2000,Yurtsever2000,burt2001,Jozsa2000a,preskill2000,Giovannetti2001,Bahder2004,Valencia2004}.
However, most of these schemes have neglected real features of the
clock synchronization problem that are essential for real-life
applications: {\it the clocks to be synchronized are in relative
motion, and at varying gravitational potentials}.  The fact that
clocks are affected by their motion and by the gravitational
potential are basic concepts that have their origin in Einstein's
special and general relativity theory. The examples of the
required accuracy of clock synchronization cited above, together
with the magnitudes of relativistic effects on satellites (see
Section IV, subsection E), show that the required accuracies can
only be met by theories that take into account the effect of
gravitational potential differences and relative motion. If
synchronization of clocks is to be achieved using quantum
information concepts, then certain features, such as relative
motion of clocks and effects of gravitational potential, must be
incorporated in the quantum information approach to clock
synchronization.

This article addresses the theoretical problem of clock
synchronizing and syntonization (making two clocks run at the same
rate), or correlating time on clocks on satellite platforms that
are orbiting Earth, or that are near-Earth.  However, Lorentz
transformations between two systems of coordinates in relative
motion show that space and time are really interwoven.  The
problem of clock synchronization is really part of the more
general problem of navigation in
space-time~\cite{Bahder2001,Bahder2003}. For example, in the
Global Positioning System (GPS), a user's receiver determines
three spatial coordinates as well as time~\cite{Bahder2003}.
Therefore, in this article we will focus on the complete problem
of navigation in space-time from the point of view of curved
space-time, such as is invoked in general relativity. The features
that relativity deals with, motion of clocks and effect of
gravitational potential, are also features that must also be
included in any classical or quantum theory of space-time
navigation.  We use the fact that space-time is described by a
four dimensional metric, but for the most part we do not
explicitly use the field equations of general relativity.  In this
sense our discussion is not restricted to general relativity.

\begin{table*}
\caption{\label{tab:TimeingPrecision} Frequency $f$, wavelength
$\lambda$, and  $\lambda/(10 c)$ time synchronization is shown for
various operating regimes of an interferometer.}
\begin{ruledtabular}
\begin{tabular}{ccc}
$f$        &  $\lambda$    & $\lambda/(10 c)$ time synchronization
\\ \hline
1500 kHz   &    200 m      &    66 ns  \\
20  MHz    &   15 m       &      5 ns  \\
120 MHz   &  2.5 m         &    0.83 ns  \\
20 GHz    &  1.5 cm       &  5.0 ps   \\
60 GHz    &  5 mm        &    1.6 ps  \\
30 THz    & 10 $\mu$ m    &  3.3 fs  \\
500 THz   & 500 nm        &  0.16 fs \\
\end{tabular}
\end{ruledtabular}
\end{table*}

In this article, several considerations are stressed in the
space-time navigation problem.  First, space-time navigation is
based on real measurements made by real physical devices.  Real
measurements are (spatially and temporally)  local quantities that
are invariants under change of space-time coordinates (see Section
VI). Historically, due to a lack of accurate measurements over
large distances (such as spacecraft to ground) measurement theory
has not played a large role in relativity theory. On the other
hand, measurements  are at the core of quantum theory.    Perhaps
this point of intersection between relativity and quantum theory
will help clarify how to augment quantum information theory with
relativistic ideas.

\section{Hardware Time, Proper Time, and Coordinate Time}

A clock is a physical device consisting of an oscillator running
at some angular frequency $\omega$,  and a counter that counts the
cycles. The period of the oscillator, $T = 2 \pi/\omega$, is
calibrated to some standard oscillator. The counter simply counts
the cycles of the oscillator.  Since some epoch, or the event at
which the count started, we say that a quantity of time equal to
$N T$  has elapsed, if $N$ cycles have been counted.

In this article I distinguish between three types of time:
hardware time $\tau^\ast$, proper time $\tau$, and  coordinate
time $t$. Hardware time is associated with a real physical device
that keeps time, which I call a {\it hardware clock}.
Specifically, {\it hardware time} $\tau^\ast = N T$ is the time
kept by a hardware clock, and is given in terms of the number of
cycles $N$  counted by the device. Two hardware clocks will
differ in the elapsed time that they indicate between two events,
because no two devices are exactly the same. Furthermore, heating,
cooling, and vibration typically affects real devices, and
consequently the value of the elapsed hardware time registered on
a real hardware clock can vary for these reasons.

Proper time is an idealized time interval occurring in the theory
of relativity. We imagine that there exists an ideal clock
(oscillator plus counter) that is unaffected by temperature or
vibration. However, based on Einstein's theory of
relativity~\cite{Synge1960,LLClassicalFields},  the ideal clock is
affected by gravitational fields, acceleration and velocities.
According to Einstein's general theory of relativity,
gravitational fields, acceleration and velocities affect all
physical processes, and hence these effects are associated with
the geometry of space and time. A basic tenet of the theory is
that between any two events that are infinitesimally separated in
space-time by $dx^i$, $i=0,1,2,3$, there exists an invariant
quantity $ds$ called the space-time interval~\cite{MyConventions}
\begin{equation}\label{interval}
  ds^2 = -g_{ij}dx^i dx^j
\end{equation}
where $g_{ij}$ is the metric of the 4-dimensional space-time. In
general relativity, if the two events are time-like, then, $ds^2 >
0$, and the events are ``separated farther in time than in space".
We interpret $ds = c d\tau$, where $d\tau$ is the proper time
elapsed on an ideal clock that moves between these two events.

The definition of an ideal clock is one that keeps proper time
intervals. More specifically, if a clock moves from point $P_1$ to
point $P_2$ on a 4-dimensional world line given by coordinates
$x^i(u)$, $i=0,1,2,3$, for $u_1\le u \le u_2$, where $u$ is some
parameter such that $P_1=\{ x^i(u_1) \} $ and $P_2=\{ x^i(u_2)\}$,
then the proper time interval $\Delta \tau$ between these two
events is given by the path integral in the space-time:
\begin{equation}\label{propertime}
\Delta \tau = \int_{u_1}^{u_2} \sqrt{-g_{ij}  \, \frac{dx^i}{du}
\frac{dx^j}{du}} \,du =\int_{x^0_1}^{x^0_2} \sqrt{-g_{ij}  \,
\frac{dx^i}{dx^0} \frac{dx^j}{dx^0}} \,dx^0
\end{equation}
where the second integral has been parametrized by the coordinate
time length $x^0 =c t$, where $t$ has units of time. From
Eq.~(\ref{propertime}), it appears that the proper time interval
$\Delta \tau$  depends on the metric of the space-time, $g_{ij}$,
and on the 4-velocity of the clock, $dx^i / ds$.  But in fact,
$\Delta \tau$ is a geometric quantity that depends on the path in
space-time, and is independent of the coordinates used to compute
$\Delta \tau$. For the purpose of applications,
Eq.~(\ref{propertime}) provides a functional relation between the
proper time interval measured on an ideal clock, $\Delta \tau$,
and the path which the clock traversed in 4-dimensional
space-time.  Under certain conditions, Eq.~(\ref{propertime}) can
provide a relation between elapsed proper time $\Delta \tau$ and
coordinate time interval, $\Delta x^0=x^0_2 -x^0_1$,
\begin{equation}\label{coordinateProperTimeRel}
\Delta \tau = f_1(x^0_1,x^0_2)
\end{equation}
which depends  on the coordinate time $x^0_1$ and $x^0_2$ of
events $P_1$ and $P_2$, respectively.  However, in general $\Delta
\tau$ depends on the whole path $x^i(u)$, and not just on the end
points of the path.

A good, real clock, which we will call a hardware clock, is
believed to provide an approximate measure of elapsed proper time.
Therefore, for a given real hardware clock, there is some
functional relation between the elapsed proper time, $\Delta
\tau$, and hardware time, $\Delta \tau^\ast$:
\begin{equation}\label{HardwareProperTime}
\Delta \tau^\ast = f_2(\Delta \tau)
\end{equation}
where the function $f_2()$ has some stochastic aspects and is
hardware dependent.

A peculiarity of the theory of relativity is that the coordinate
time, $x^0$, as well as the other coordinates $x^\alpha$,
$\alpha=1,2,3$, are not directly measureable
quantities~\cite{Brumberg1991}. The coordinates are only
mathematical constructs, and cannot be measured directly. Events
occurring in space-time have coordinates associated with them. Two
events are labelled with coordinates $P_1=\{x^i_1\}$ and
$P_2=\{x^i_2\}$ even though these quantities are not directly
measureable. However, the theory allows us to relate the
difference of coordinate time, $x^0_2 - x^0_1$, between two
events, to the proper time interval, by the path integral in
Eq.~(\ref{propertime}).  Of course, as mentioned previously,
neither of these quantities are measureable directly, instead,
only hardware time intervals, $\Delta \tau^\ast$, are measured
directly from hardware clocks. Everything else is calculated.

The relations in
Eq.~(\ref{propertime})--(\ref{HardwareProperTime}) permit, under
some circumstances, the relation of measured hardware times
$\Delta \tau^\ast$ to coordinates times $\Delta x^0$, which enter
into the theory or relativity.

\subsection{Synchronization versus Syntonization}

The relation between proper time $\Delta \tau$ and coordinate time
is such that they may ``run at different rates".  This is clearly
the case when the metric of the space-time is not a constant over
the integration path. For example, a good clock may measure proper
time intervals accurately, however, because of
Eq.~(\ref{propertime}), this clock will run at a rate that differs
from coordinate time $x^0$, because $d \tau / d x^0 \ne {\rm
constant}$.  While coordinate time is a global coordinate
quantity, valid (almost) everywhere in the space-time, the proper
time interval depends on the world line of the clock.  For
example, for an ideal clock that is stationary (has constant
spatial coordinates $x^\alpha$, $\alpha=1,2,3$) with world line
$x^i=(x^0,x^1,x^2,x^3)$, for $x^0_o < x^0 < x^0_o + dx^0$, the
proper time interval is $d \tau = \sqrt{-g_{00}} dx^0$. Therefore,
the rate of proper time, $d \tau/ d x^0$,  depends on position
thorough $g_{00}$.

Consider now two ideal clocks at the same location and assume that
these two clocks are synchronized to read the same starting time
at some epoch, or starting event. Next move the clocks apart
(hence they travel on different world lines) and then bring them
together once again to a common location.  On comparing the times
on these clocks, we find that different amounts of proper time
have elapsed on each of them.  In other words, proper time ran at
different rates for each of the clocks. We say that the two clocks
were not syntonized (i.e., they did not run at the same rate)
since, when they were brought back together a different amount of
proper time elapsed on each clock. The rate of each clock can be
compared at any instant to the underlying coordinate time (which
is a globally defined quantity), by using Eq.~(\ref{propertime}),
and in this way the time on one clock can be compared to the time
on the other clock.  In Section IX, we will find that the most
serious problem for practical satellite applications is the
syntonization of clocks.

\section{Choice of a Physical Theory}
A theory must be chosen as the basis for navigation and clock
synchronization.  The theory must deal with two areas: the
space-time in which all events occur, and the propagation of
electromagnetic fields in this space-time.  While these two areas
are coupled in the sense that electromagnetic fields are sources
for the curvature of space-time (within general relativity), we
will take these two areas separately.  More specifically, we will
neglect the (very small) effect of the electromagnetic field
creating space-time curvature.  We will assume that all curvature
is caused by the presence of mass.  For applications in the
vicinity of the Earth, this is an excellent approximation.

The most common theory to choose is general relativity, as
developed by Einstein~\cite{Synge1960,LLClassicalFields}. As has
already been discussed, this theory includes the effects of motion
on clocks and also the effects of gravitational potential on
clocks. In fact, there are several variants of relativistic
theories of gravitation.  However, Einstein's general theory of
relativity has so far passed all physical
tests~\cite{Will,WillGRReview1998Update}, and we shall subscribe
to it in this report.  Einstein's general theory of relativity can
be divided into two parts: first, the theory assumes that there
exists a space-time metric $g_{ij}$, given by
Eq.~(\ref{interval}), which relates proper time $d\tau=ds/c$ to
the space-time geometric interval $ds$, and coordinate difference
$dx^i$ between two events. This idea is quite general, and is
really an embodiment of the principle of
equivalence~\cite{Will,WillGRReview1998Update}. The principle of
equivalence essentially states that gravitational mass (in
Newton's universal law of gravitation)  and inertial mass (in
Newton's second law of motion) are the same quantity. Roughly
speaking, the equivalence principle says that two small test
bodies of different mass will fall along the same geodesic path,
i.e., the same distance in the same time. This principle has been
extensively tested~\cite{Will,WillGRReview1998Update}, and is the
basis of essentially all metric-based theories of gravity, because
they are based on geometric ideas embodied in
Eq.~(\ref{interval}). Physicists today generally agree that a
theory of gravity should be a metric theory in a pseudo-Riemannian
curved space-time with metric of the form given by
Eq.~(\ref{interval}).

The second part of Einstein's general relativity theory consists
of the field equations~\cite{Synge1960,LLClassicalFields}
\begin{equation}  \label{EinsteinFieldEquations}
G^{ij}= - \kappa T^{ij}
\end{equation}
where $\kappa=8 \pi G/c^2$, where $G$ is Newton's gravitational
constant, and $c$ is the speed of light in vacuum. The field
Eqs.~(\ref{EinsteinFieldEquations}) relate the matter
distribution, given by the stress energy tensor $T^{ij}$, to the
effect that this matter has on space-time via the Einstein tensor,
\begin{equation}  \label{EinsteinTensor}
G_{ij}=R_{ij}-\frac{1}{2}g_{ij} R
\end{equation}
where the Ricci tensor, $R_{ij}=R^k_{ijk}$, is related to the
Riemann tensor
\begin{equation}
R^i_{jkm} = \Gamma^i_{jm,k} -\Gamma^i_{jk,m} + \Gamma^a_{jm}
\Gamma^i_{ak} -\Gamma^a_{jk} \Gamma^i_{am} \label{RiemannTensor}
\end{equation}
and the affine connection
\begin{equation}
 \Gamma^i_{jk} = \frac{1}{2} g^{il}\left( g_{jl,k} + g_{kl,j} - g_{jk,l} \right)
\label{connection}
\end{equation}
is related to the metric $g_{ij}$.  So the field
Eqs.~(\ref{EinsteinFieldEquations}) are a set of equations for the
components of the metric tensor field $g_{ij}$.  In
Eq.~(\ref{connection}), ordinary partial derivatives with respect
to the coordinates are indicated by commas.

The field equations of Einstein, given in
Eq.~(\ref{EinsteinFieldEquations}), have only been solved and
tested in a limited number of cases.  Consequently, the field
equations are on a less-firm footing than the equivalence
principle. Fortunately, most of the applications of navigation and
clock synchronization in a curved space-time rely only on the fact
that space-time is a metric theory, given by Eq.~(\ref{interval}).
Therefore, our conclusions below regarding navigation and clock
synchronization transcend general relativity. Specifically, our
conclusions are based on the assumed-correctness of the
equivalence principle.

\subsection{Electromagnetic Waves}
The electromagnetic field plays a central role in experiments and
applications.  In technology applications, all  information is
currently carried by travelling electromagnetic fields.
Inter-satellite links, and ground to satellite links are all done
using electromagnetic radiation fields.  Time transfer between
stations on the ground and satellites is done by electromagnetic
fields. Since electromagnetic fields paly such a key role, in this
subsection we briefly outline the equations governing the
propagation of electromagnetic waves, namely the Maxwell equations
in flat space-time, and their generalization to curved space-time.

Electromagnetic fields in a medium, such as air, a dielectric, or
a magnet, are described by the macroscopic Maxwell's equations,
and in SI units, in {\it flat space-time}, these equations take
the form
\begin{eqnarray}
{\rm div} \, {\bf D} & = & \rho  \label{MaxwellEqn1} \\
 {\rm div} \, {\bf B} & = & 0  \label{MaxwellEqn2} \\
{\rm curl} \, {\bf H} & = & {\bf J} + \frac{\partial {\bf
D}}{\partial \, t}  \label{MaxwellEqn3} \\
{\rm curl} \, {\bf E} & = &  -\frac{\partial {\bf B}}{\partial \,
t} \label{MaxwellEqn4}
\end{eqnarray}
where ${\bf E}$ and ${\bf B}$ are the electric field and magnetic
induction (or magnetic field), respectively, and ${\bf D}$ and
${\bf H}$ are the electric displacement field and magnetic
intensity, respectively.  In a medium, the fields ${\bf E}$ and
${\bf D}$, and the fields ${\bf B}$ and ${\bf H}$ are related by
constitutive relations.  In a vacuum, these fields are simply
related by:
\begin{eqnarray}   \label{vacuumConstitutiveRel}
{\bf D} & = & \epsilon_o \, {\bf E} \\
{\bf B} & = & \mu_o \, {\bf H}
\end{eqnarray}
where $\epsilon_o$  and $\mu_o$ are the permitivitty and
permeability of vacuum, respectively.  In an isotropic (but not
necessarily homogeneous) medium such as the Earth's atmosphere,
the constitutive equations are
\begin{eqnarray}
{\bf D} & = & \epsilon \, {\bf E}  \label{ConstitutiveRelE} \\
{\bf B} & = & \mu \, {\bf H}  \label{ConstitutiveRelB}
\end{eqnarray}
where $\epsilon$  and $\mu$ are the permitivitty and permeability
of the medium, respectively.

In a curved space-time, the physics of electromagnetic wave
propagation has been less
explored~\cite{LLClassicalFields,Mo1971,Volkov1971,Mashhoon1973,robertson1968,Manzano1997}.
However, it is known that the gravitational field scatters and
diffracts electromagnetic waves, and that the plane of
polarization of an electromagnetic wave is rotated as the wave
propagates through a gravitational field. In general, a
gravitational field affects electromagnetic wave propagation
similarly to a dispersive medium~\cite{LLClassicalFields}. For
weak fields, such as exist in the vicinity of the earth, these
gravitational effects are smaller than the dispersive effects due
to the atmosphere (at low altitude). The general equations that
govern electromagnetic wave phenomena in curved space-time, in the
presence of a dielectric are given by~\cite{robertson1968}
\begin{equation}   \label{Max1}
F_{ij,k} +F_{jk,i} +F_{ki,j} = 0
\end{equation}
and
\begin{equation}   \label{Max2}
H^{ik}_{~~;k}  = J^i
\end{equation}
where $F_{ij}$ and $H^{ik}$ are two antisymmetric tensor fields,
 the comma indicates partial differentiation with respect to
the coordinates and the semicolon indicates covariant
differentiation with respect to the coordinates. All Roman indices
take values $i=0,1,2,3$. The two fields are related by a
constitutive relation
\begin{equation}   \label{constitutiveRel}
\sqrt{-g} \, \sqrt{-\gamma} \, H^{ik}  = c^2 \left(
\frac{\epsilon}{\mu} \right)^{\frac{1}{2}} \, \gamma^{i a} \,
\gamma^{k b} \, F_{a b}
\end{equation}
The constitutive relation in Eq~(\ref{constitutiveRel}) assumes
that the medium is isotropic, but not necessarily homogeneous, so
the permittivity $\epsilon$ and permeability $\mu$ are both
functions of position.  In Eq~(\ref{constitutiveRel}), we have
used the definitions of the contravariant components of the metric
tensor $g^{ij}$ and the effective metric
\begin{equation}   \label{MaxDefs}
\gamma^{ij} = g^{ij} - (n^2-1) \, u^i \, u^j
\end{equation}
where $n$ is the scalar index of refraction given by
\begin{equation}\label{indexDef}
n^2 \, = \, \frac{\epsilon \, \mu}{\epsilon_o \, \mu_o}
\end{equation}
where $g={\rm det} \, g_{ij}$ and $\gamma={\rm det} \,
\gamma_{ij}$,  $u^i$ is the local 4-velocity of the medium in our
system of coordinates, and $J^i$ is the 4-current density.  In the
proper frame of reference of the medium (where the medium is at
rest), the field tensors take the simple forms:
\begin{equation}\label{Ftensor}
F_{ik}= \left(
  \begin{array}{cccc}
  0   & -E_x  &    -E_y   &   -E_z \\
  E_x &    0  & c B_z     & -c B_y \\
  E_y &  -c B_z &  0 & c B_x \\
  E_z & c B_y & -c B_x & 0
  \end{array}  \right)
\end{equation}
and
\begin{equation}\label{Htensor}
H_{ik}= \left(
  \begin{array}{cccc}
  0   & -c D_x  &    -c D_y   &   -c D_z \\
  c D_x &    0  &  H_z     & -H_y \\
  c D_y &  -H_z &  0 &  H_x \\
  c D_z &  H_y & -H_x & 0
  \end{array}  \right),
\end{equation}
the current density is $J^i=(\rho_o c, J^\alpha)$, where $\rho_o$
is the proper charge density and $J^\alpha$ is the current.  In
this proper frame of reference of the medium,
Eqs.~(\ref{Max1})--(\ref{Max2}) reduce to
Eqs~(\ref{MaxwellEqn1})--(\ref{MaxwellEqn4}), and the consitutive
Eq.~(\ref{constitutiveRel}) reduces to relations in
Eq.~(\ref{ConstitutiveRelE})--(\ref{ConstitutiveRelB}).

If the medium is vacuum, then $n=1$ and $\gamma^{ij} = g^{ij}$,
and the tensors $F_{ik}$ and $H_{ik}$ are not independent, instead
they differ only by trivial constants of the vacuum.

Equations~(\ref{Max1})--(\ref{Max2}) govern the propagation of
electromagnetic fields in the presence of a medium in a
gravitational field.  In general, in the vicinity of the Earth,
the dispersive effects of the medium are larger than those of the
gravitational field. Due to the complexity of these equations,
they have not been explored in detail. Only several treatments
have been attempted, see for example
Refs.~\cite{Volkov1971,Mashhoon1973,Manzano1997}. The
Eqs.~(\ref{Max1})--(\ref{Max2}) or
(\ref{MaxwellEqn1})--(\ref{MaxwellEqn4}), form the basis for
applications such as time transfer, clock synchronization and
communication. These equations are valid in a system of
coordinates (frame of reference) that is in arbitrary motion. In
particular, these equations describe propagation of
electromagnetic waves, and the reception and transmission
properties of antennas in the radio portion of the spectrum.

\subsection{The Geometrical Optics Approximation}
When discussing precise measurements, it is important to state
precisely the theory and approximations used.   Up to this point
in time, the geometric optics approximation has been universally
used in discussions of time transfer, usually without stating its
use, and without examining the limitation of this approximation.
Below, we state the geometric optics approximation and how it
enters into time transfer ideas.

The geometric optics approximation consists of the assumption that
the wavelength of the travelling electromagnetic wave, $\lambda$,
is much smaller than the linear dimension $l$ of all objects
(length scales) in the physical problem under
consideration~\cite{LLClassicalFields}:
\begin{equation} \label{geometricOpticsApprox}
\lambda << l
\end{equation}
In physical terms, the limit of short wavelength waves
(geometrical optics limit) corresponds to waves that travel along
straight lines, so that diffraction (e.g., bending of waves around
boundary edges) is absent.   As an example of geometric optics at
work, consider the visible shadows cast on the ground by objects
in the path of the light from the sun to ground. The light travels
at approximately straight lines. However, if one looks very
closely near the edge of the shadow, the boundary between light
and dark areas is not sharp, and this is where geometrical optics
shows its limitation--there is diffraction, or bending of the rays
around sharp edges of an opaque material.

In the literature, statements are often made in the context of
special relativity theory that in flat space-time ``light travels
along a straight line", or in curved space-time, that ``light
travels along a geodesic". Both of these statements are true only
within the context of the geometrical optics
limit~\cite{Anile1976,Bicak1975,Mashhoon1986,faraoni1993}. In
recent years, there has been limited work to explore the nature of
travelling electromagnetic waves in a curved space-time, going
beyond the geometrical optics
approximation~\cite{Volkov1971,Mashhoon1973,Manzano1997,Anile1976,Bicak1975,Mashhoon1986,faraoni1993}.
The gravitational field creates complex effects such as
diffraction of the electromagnetic wave and rotating its
polarization.

\subsection{Signal Detection and Use of Antenna Phase Center}

In precision measurements, it is important to have a clear concept
of the point from which radiation emanates and the point at which
the radiation is detected.  Usually, such a discussion makes
implicit use of the geometric optics approximation.

One place where the geometric optics approximation is not
well-satisfied is for real (radio frequency) antennas, because
antennas are efficient at receiving and transmitting radiation at
wavelengths that are comparable to the antenna size, so
Eq.~(\ref{geometricOpticsApprox}) is not well-satisfied. The
desire to continue to use the (very convenient) geometric optics
approximation forces us to invoke the concept of antenna phase
center in precise time transfer or navigation applications.  We
then imagine that there is a unique emission point, from which
radiation emanates, and a unique reception point, where the
radiation is absorbed.

The antenna phase center is defined as the apparent point from
which radiation emanates (or is absorbed).  In the far-field
radiation region, for one vector component of the electric field
of antenna radiation, and for one polarization and  one frequency
$\omega$, the electric field can be expressed as
\begin{equation}\label{EField}
{\bf E} =  {\bf u} \, E(\theta,\phi) e^{\psi(\theta,\phi)}
\frac{e^{i k r}}{r} e^{-i \omega t}
\end{equation}
where the vector ${\bf u}$ is a real polarization unit vector,
$E(\theta, \phi)$ is the (real) electric field amplitude,
$\psi(\theta,\phi)$ is the (real) phase, $r$ is the distance from
the antenna, and $i=\sqrt{-1}$.  If a point can be found such that
$\psi(\theta,\phi)$ is independent of $\theta$ and $\phi$, i.e.,
independent of the direction of propagation,  then this point is
the antenna phase center. For most practical antennas, no such
point exists~\cite{Schupler1994,Balanis1997}. The reception of an
antenna is related to its transmission properties, with the same
phase center, by reciprocity
relations~\cite{Balanis1997,CollinZuckerBook1969}.

In the receive mode,  the open-circuit voltage $V$ induced in a
receiving
antenna~\cite{Sinclair1950,CollinZuckerBook1969,Price1986,Balanis1997}
is given by the scalar product between the radiation field from a
given satellite, ${\bf E}$, and the vector effective length ${\bf
h}({\bf n})$:
\begin{equation}\label{AntennaVoltage}
V = {\rm Re} \{ \, {\bf h}({\bf n}) \cdot {\bf E} \, \}
\end{equation}
where Re takes the real part of a complex expression, ${\bf E}$ is
the electric radiation field at the receiving antenna, and ${\bf
h}({\bf n})$ is the receiving antenna vector effective length
(sometimes called the effective height), which is a complex vector
quantity that characterizes the electromagnetic wave phase
relationship of the antenna in receive and transmit mode. The
receive and transmit modes are related by reciprocity
relations~\cite{Balanis1997,CollinZuckerBook1969,Sinclair1950,Price1986}

Despite the limitations of the concept of antenna ``phase center",
this concept is routinely invoked in practice in real antenna
systems. The practical complication is that the antenna ``phase
center" is not a fixed point, but its position depends on
electromagnetic wave frequency, $\omega$, polarization vector
${\bf u}$, and direction of travel with respect to the antenna
(both in receive and transmit modes). In other words the phase
center position is not fixed, instead it varies with $\omega$ and
${\bf u}$ over some range of coordinate values $\Delta x$ that is
on the order of a wavelength of the radiation, $\Delta x \sim
\lambda$. In other words, the point at which an antenna receives a
signal is only precisely defined (in geometrical optics) to within
a distance $\Delta x \sim \lambda$.  Consequently, when one
antenna transmits radiation and another antenna receives
radiation, the effective distance between these two antennas can
vary with frequency and polarization of the radiation, and with
the relative orientation of the antennas.  When the relative
orientation of one satellite changes with respect to another
satellite due to their relative motion, their effective separation
changes due to change of relative orientation of their antennas,
in addition to a real change of distance between them. Clearly,
the concept of a single emission point and a single reception
point is fuzzy on the order of the scale of a wavelength at both
transmit and receive end points.

In conclusion, the apparent distance between antennas changes with
frequency, polarization, and orientation. In a case where precise
navigation is to be carried out using radio frequencies, these
effects must be taken into account, so that an accuracy of better
than one wavelength of the radiation can be achieved.

As an alternative to using radio frequencies, we can transition to
satellites using optical frequencies, which have considerably
shorter wavelengths.  The transition to optical free-space
communications in satellites, with smaller wavelengths on the
order of 10 -- 60 nm, has the advantage that we can invoke the
geometric optics approximation, and suffer $\Delta x \sim \lambda$
errors that are much smaller, because of the much smaller optical
wavelengths. For example,  for optical wavelengths in the range 10
nm to 60 nm, position errors are comparable to the wavelength, and
this should be compared with radio frequencies, of say, 1 MHz to
10 GHz, with wavelength range of 3~cm to 300~m.

Actually, at optical frequencies, Eq.~(\ref{AntennaVoltage}) is
not the physical mechanism that is responsible for detecting
electromagnetic fields.  Instead, detectors of electromagnetic
radiation work either as bolometers or quantum detectors.
Bolometric detectors are based on the pyroelectric effect, which
produces a change of dielectric polarization with increase of
temperature, due to absorption of electromagnetic
radiation~\cite{bahder1993}. The polarization change is detected
electrically.

The second common mode of detecting optical radiation is based on
a quantum mechanical effect.  For example, in a semiconductor
material a photon (quantum) of the electromagnetic field is
absorbed by an electron, and the electron makes a transition from
a valence band quantum state to a conduction band state.  The
electron in the conduction band is then detected electrically. In
either case, the minimum volume that is needed for detection is
roughly of the dimensions of a wavelength of the optical
radiation, so similar criteria apply (to that of radio frequency)
for accuracy of the point of absorption.

As discussed in the Introduction, we believe that future satellite
systems will have optical links.   Such satellites may have an
interferometer that operates in the radio frequency portion of the
spectrum, but the navigation (positioning) and time
synchronization will be done optically.  Consequently, the
geometric approximation will be useful because of the (relatively)
small optical wavelengths compared to radio frequency wavelengths.
Therefore, we will freely use theory that assumes that
electromagnetic waves travel on geodesics in curved space-time,
which is the geometric optics approximation.

\section{World Function of Space-Time}

The geometric optics approximation is valid for the applications
addressed in this report.  Within the geometric optics
approximation we can say that electromagnetic waves travel on
geodesics in space-time. This is a purely geometric statement: the
emission and reception events of a signal are connected by a null
geodesic. Geodesics in the space are determined by the metric of
the space-time.

A useful quantity to deal with measurements in space-time is the
world function $\Omega$. The world function is simply one-half the
square of the space-time interval (see Eq.~(\ref{interval})),
measured along the geodesic  connecting two
points~\cite{Synge1960,Bahder2001,Bahder2003,LLClassicalFields}.

The world function was initially introduced into tensor calculus
by Ruse~\cite{Ruse1931a,Ruse1931b}, Synge~\cite{Synge1931}, Yano
and Muto~\cite{YanoandMuto1936}, and Schouten~\cite{Schouten1954}.
It was further developed and extensively used by Synge in
applications to problems dealing with measurement theory in
general relativity~\cite{Synge1960}. The world function has
generally received little attention in the literature, so we
provide a detailed definition here. Consider two points, $P_1$ and
$P_2$, in a general space-time, connected by a unique geodesic
path $\Gamma$ given by $x^i(u)$, where $u_1 \le u \le u_2$. A
geodesic is defined by a class of special parameters $u^\prime$
that are related to one another by linear transformations
$u^\prime = a u + b$, where $a$ and $b$ are constants.  Here, $u$
is a particular parameter from the class of special parameters
that define the geodesic $\Gamma$, and $x^i(u)$ satisfy the
geodesic equations
\begin{equation}
\frac{d^2 x^i}{du^2}+ \Gamma^i_{jk} \frac{dx^j}{du}
\frac{dx^k}{du} =0 \label{GeodesicDiffEq}
\end{equation}
The world function between $P_1$ and $P_2$ is defined as the
integral along $\Gamma$:
\begin{equation}
\Omega(P_1,P_2) = \frac{1}{2} (u_2 - u_1) \int^{u_2}_{u_1} \,
g_{ij} \frac{dx^i}{du} \frac{dx^j}{du} \, du
\label{WorldFunctionDef}
\end{equation}
The value of the world function has a geometric meaning: it is
one-half the square of the space-time distance between points
$P_1$ and $P_2$. Its value depends only on the eight coordinates
of the points $P_1$ and $P_2$. The value of the world function in
Eq.\ (\ref{WorldFunctionDef}) is independent of the particular
special parameter $u$ in the sense that under a transformation
from one special parameter $u$ to another, $u^\prime$, given by
$u=a u^\prime + b$, with $x^i(u)=x^i(u(u^\prime))$, the world
function definition in Eq.\ (\ref{WorldFunctionDef}) has the same
form (with $u$ replaced by $u^\prime$).

The world function is a two-point invariant in the sense that it
is invariant under independent transformation of coordinates at
$P_1$ and at $P_2$.  Consequently, the world function
characterizes the geometry of the space-time. For a given
space-time, the world function between points $P_1$ and $P_2$ has
the same value independent of the coordinates that are used.  As a
simple example of the world function for Minkowski space-time,
consider
\begin{equation}
\Omega(x^i_1,x^j_2) = \frac{1}{2}\, \eta_{ij} \, \Delta x^i  \,
\Delta x^j \label{MinkowskiWorldFunction}
\end{equation}
where $\eta_{ij}$ is the Minkowski metric with only non-zero
diagonal components $(-1,+1,+1,+1)$, and $ \Delta x^i  = (x_2^i -
x_1^i)$, $i=0,1,2,3$, where $x_1^i$ and $x_2^i$ are the
coordinates of points $P_1$ and $P_2$, respectively.  Up to a
sign, the world function gives one-half the square of the
geometric measure (the interval) in space-time. Calculations of
the world function for specific space-times can be found in
Refs.~\cite{Synge1960,John1984,John1989,Buchdahl79,Bahder2001,Bahder2003}
and application to Fermi coordinates in Synge\cite{Synge1960} and
Gambi et al.~\cite{Gambi1991}.

The world function for the Schwarzschild metric, linearized in
small parameter $GM/c^2 r$,  is given by~\cite{Bahder2001}:
\begin{eqnarray}
\Omega(x^i_1,x^j_2) & = & \frac{1}{2}\eta_{ij} \Delta x^i \Delta
x^j + \frac{GM}{c^2} \, \left[  |{\bf x}_2 - {\bf x}_1| +
   \frac{c^2 \Delta t^2}{|{\bf x}_2 - {\bf x}_1|} \right] \;
 \log \left( \frac{\tan(\frac{\theta_1}{2})}{\tan(\frac{\theta_2}{2})} \right)  \nonumber \\
  &  &  + \frac{GM}{c^2} |{\bf x}_2 - {\bf x}_1| \left(  \cos \theta_1 - \cos \theta_2
 \right)
\label{SchwarzschildWorldFunction}
\end{eqnarray}
where $c \Delta t \equiv x^0_2 - x^0_1$, and  $\theta_1$ and
$\theta_2$ are defined by
\begin{equation}
\cos \theta_a = \frac{{\bf x}_a \cdot ( {\bf x}_2 -  {\bf x}_1 )
}{|{\bf x}_a| |{\bf x}_2 - {\bf x}_1|}, \;\;\;\; a=1,2
\label{cosineDef}
\end{equation}
The world function in Eq.~(\ref{SchwarzschildWorldFunction})
assumes a spherical Earth ($J_2 = 0$), see
Eq.~(\ref{EarthPotential}). For most applications dealing with
delays of electromagnetic wave propagation, this approximation is
sufficient. A more accurate calculation of the world function for
Schwarzschild metric is given by Buchdahl and
Warner~\cite{Buchdahl79}.  Including the small effect of the
oblateness of the Earth will require additional computations and
is left for future work. For a pedagogic discussion of applying
the world function to space-time navigation,  see
Ref.~\cite{Bahder2001}.

\subsection{Navigation in Curved Space-time}

A satellite must be localized with respect to some system of
coordinates. The satellite can receive signals from four
electromagnetic beacons and use the measurements to solve for its
position. Alternatively, the satellite can send out
electromagnetic signals that are received at four locations (e.g.,
on the Earth surface) , and these four quantities can be used to
compute the satellite position at emission time.

In either case, the equations for navigation in a curved
space-time can be formulated in a covariant, and invariant way
(independent of coordinates) using the world
function~\cite{Bahder2001}. For example, consider a satellite with
unknown coordinates $x^i_o=(t_o,{\bf x}_o)$, $i=0,1,2,3$, which we
want to locate with respect to four electromagnetic beacons having
coordinates $x^i_a=(t_a,{\bf x}_a)$, $a=1,2,3,4$. Assume that the
satellite simultaneously receives the four signals at the
space-time event with coordinates $x^i_o$. We must solve the four
equations given by
\begin{equation}
\Omega(x^i_o,x^j_a) = 0, \;\;\;\; a=1,2,3,4
\label{CovariantNavigation}
\end{equation}
for the unknown satellite coordinates,  $x^i_o=(t_o,{\bf x}_o)$,
in terms of the known emission event coordinates, $x^i_a=(t_a,{\bf
x}_a)$. These equations state that the emission and reception
events are connected by null geodesics.  In addition to
Eq.~(\ref{CovariantNavigation}), the appropriate causality
conditions $t_o > t_a$, for $a=1,2,3,4$, must be added. The set of
relations in Eq.\ (\ref{CovariantNavigation}) are manifestly
covariant and invariant due to the transformation properties of
the world function under independent space-time coordinate
transformations at point $P_1$ and at $P_2$.
Equations~(\ref{CovariantNavigation}) neglect atmospheric effects,
although these could be included in future work.

From the definition of the world function, the intrinsic
limitations of navigation in a curved space-time are evident: the
world function $\Omega(P_1,P_2)$ must be a single-valued function
of $P_1$ and $P_2$. In general, if two or more geodesics connect
the points $P_1$ and $P_2$, then $\Omega(P_1,P_2)$ will not be
single-valued and the set of equations in Eq.\
(\ref{CovariantNavigation}) may have multiple solutions or no
solutions.  Such conjugate points $P_1$ and $P_2$ are known to
occur in applications to planetary orbits and in
optics~\cite{Synge1960}. However, when the points $P_1$ and $P_2$
are close together in space and in time and the curvature of
space-time is small, we expect the world function to be single
valued and the solution of Eq.\ (\ref{CovariantNavigation}) to be
unique. Therefore, navigation in curved space-time is limited by
the possibility of determining a set of four unique null geodesics
connecting four emission events to one reception event. In the
vicinity of the Earth, there is no ambiguity, due to the weak
gravitational field. However, in the case of strong gravitational
fields, as may exist in the vicinity of a black hole, or when the
(satellite) radio beacons are at large distances from the observer
in a space-time of small curvature, navigation by radio beacons
may not be possible in principle. In such cases, one may have to
supplement radio navigation by inertial techniques; see for
example the discussion by Sedov~\cite{Sedov1976}.

\section{Physical Measurements}

In order to extract information from a quantum mechanical system,
a measurement has to be performed. In a quantum mechanical theory,
the measurement process plays a central role. The quantum
mechanical measurement process is imagined as an interaction that
occurs between a classical, macroscopic apparatus and the quantum
system~\cite{QMLandauLifshitz1977,Ekert2000,Nielson2000}. Much
effort has been expended on detailed investigations of the
measurement process.

In contrast, in the general theory of relativity, comparatively
little attention has been devoted to the role of the measuring
process. In part, this is due to the fact that really
high-accuracy measurements were not carried out due to the
associated technological difficulties.  However, the presence of
high accuracy clocks in space make high accuracy measurements a
reality.

\subsection{Observations and Measurements}
In a real laboratory experiment, the process of taking data
consists of recording information that is observed on physical
meters, dials and gauges. For example, the reading on a volt meter
can be observed at some point in time and space.  The reading on
the volt meter is an example of an {\it observation}.  The
observation occurs at some point in space and time (an event),
marked by coordinates in some 4-dimensional system of space-time
coordinates. The space-time coordinates of an observation may not
be known to the person that is making the observation.

Physically useful observations usually consist of coincidences of
two (or more) things occurring  at the same place at the same
time. As an example of an observation, consider walking on a side
walk, and noting the time on your watch when passing a crack in
the sidewalk. The time  noted is an observation  of a coincidence:
the position of my watch with the position of the crack in the
sidewalk.  This coincidence occurred at a space-time event, which
has some coordinates.  The observation is idealized as occurring
at a given point in space-time, and each situation must be
analyzed whether the ``region of observation" satisfies, for the
given application, the accuracy requirement of the observation as
occurring at a point.

\begin{figure*}
\includegraphics{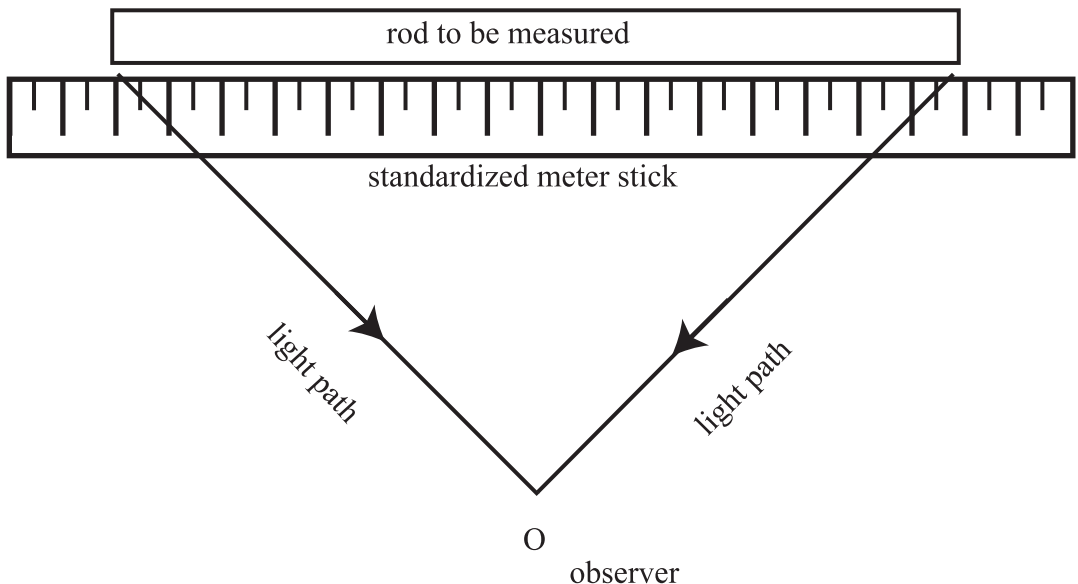}
\caption{\label{fig:measure} The standardized meter stick is shown
next to the rod to be measured.}
\end{figure*}

In contrast to an observation, a {\it measurement} is an
observation in which a comparison is made.  For example, consider
an a.c. electric current that is fed into a circuit where the
phase of the incoming current is compared  ({\it measured}) to the
phase of a reference a.c.  current. The measurement is the phase
difference between these two currents.  The measurement occurs at
a specific space-time event, with definite coordinates.  In the
real world, the event is not a point, but often, to a sufficient
degree of accuracy it can be modelled as occurring at one
space-time point.  Whether any given measurement can be regarded
as occurring at one point in space-time depends on the required
accuracy, and must be analyzed on a case by case basis.

As another example of a measurement process, consider measuring a
rod, by use of a standardized meter stick.  Light from the ends of
the rod comes to our eyes, along with light from the graduated
scale on the standardized meter stick.  The (simultaneous) event
of light entering our eye from the left and right sides of the rod
and meter stick constitute an observation, and because it is a
comparison, it is also a measurement. See Fig.~\ref{fig:measure}.
Measurements are a subset of observations.

Measurements are dimensionless ratios: the thing measured is
compared to a standard.  Furthermore, measurements are invariant
quantities.  In general relativity theory, the tetrad formalism
treats measurements as invariant quantities.

\subsection{Tetrad Formalism}

Consider two observers that are making spectral measurements on
light from the same star.  Assume that the two observers are in
relative motion, but that at the instant of measurement, they are
located at the same point in space. At this point, the observer
that is moving toward the star may measure predominantly blue
light emitted from the star.  On the other hand, the other
observer that is travelling away from the star may measure
predominantly red light.  So two measurements at the same place at
the same time lead to different results.  (There are other types
of measurements that may produce identical results for the two
observers.)  In the previous section, we stated that measurements
are invariant quantities. In what sense then are measurements
invariant?

In connection with measurements, there are two types of
transformations that must be considered.  First, the global
coordinates in the space, $x^i$,  can be transformed to new
coordinates, say using a transformation such as
\begin{equation}\label{transformMeasurement}
x^i \rightarrow y^i =f^i(x^k)
\end{equation}
where $f^i()$ are a set of transformation functions. Since
measurements are scalar quantities (see below),  they are always
invariant under the generalized coordinate transformations of the
type in Eq.~(\ref{transformMeasurement}).

\begin{figure}
\includegraphics{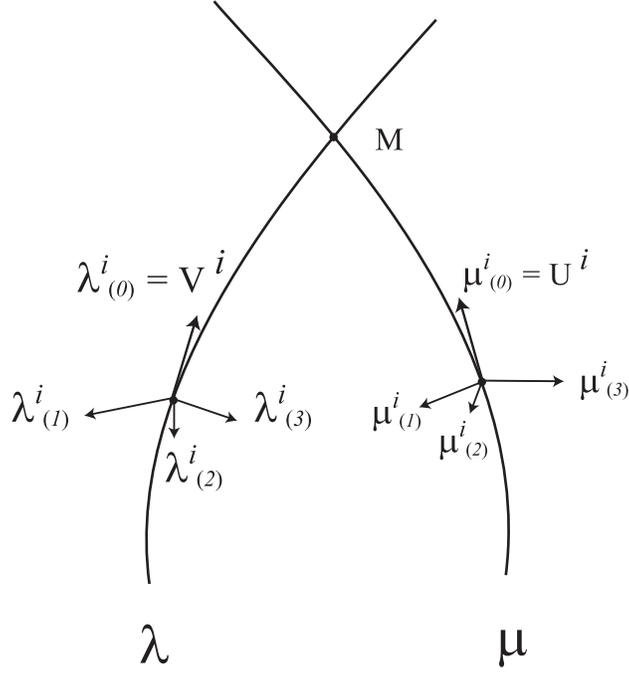}
\caption{\label{fig:ObserverIntersection} The  world lines of two
observers, $\lambda$ and $\mu$, are shown with their respective
tetrads, $\lambda^i_{(\alpha)}$ and $\mu^i_{(\alpha)}$.  At point
$M$, the observers are momentarily colocated and they make a
measurement of the same physical quantity. The results of their
measurements are related by the transformation between tetrad
basis vectors given in Eq.~(\ref{TetradTransform}).}
\end{figure}

The second type of transformation that must be considered in
connection with measurements is that two observers have different
world lines and consequently different tetrad basis vectors onto
which they project electromagnetic fields.  The projection onto
the tetrad is the measurement process.  Real measurements are
local quantities, and they can be compared when two observers are
colocated at the same space-time point, see
Figure~\ref{fig:ObserverIntersection}.
Transformations from the tetrad basis $\lambda^i_{(a)}$,
$a=0,1,2,3$,  of one observer to the tetrad basis of another
observer, $\mu^i_{(a)}$, can be contemplated:
\begin{equation}\label{TetradTransform}
\lambda^i_{(\alpha)} = H^{~~(\beta)}_{(\alpha)} \mu^i_{(\beta)}
\,\,\,\,{\rm for} \,\, \alpha,\beta=1,2,3
\end{equation}
where $H^{~~(\beta)}_{(\alpha)}$ is a 3$\times$3 rotation matrix
($\alpha,\beta=1,2,3$) that relates the spatial tetrad basis
vectors.  Note that the matrix $H^{(\beta)}_{(\alpha)}$ relates
three 4-vectors $\lambda^i_{(\alpha)}$ to three 4-vectors
$\mu^i_{(\alpha)}$. (For each observer, the 0th components of the
tetrad basis, $\lambda^i_{(0)}$ and $\mu^i_{(0)}$, are determined
by their respective 4-velocity (see below), so these vectors do
not enter the transformation.)

The key idea is that measurements are scalar quantities that are
projections on the local basis vectors carried by each observer.
Even though two observers coincide in time and space, their tetrad
basis vectors are different:  $\lambda^i_{(a)}$ for one observer
and $\mu^i_{(a)}$ for the other observer. Consequently, the two
observers obtain different values of the measurement. Measurements
are quantities that are projections on the observer's tetrad; they
are of the form
\begin{equation}\label{FijEq}
F_{ij} \, \lambda^i_{(a)} \lambda^j_{(a)}
\end{equation}
for one observer and of the form
\begin{equation}\label{FijEq2}
F_{ij} \, \mu^i_{(a)} \mu^j_{(a)}
\end{equation}
for the other observer. Each of these quantities is a scalar,
i.e., each is invariant under general coordinate transformations
given in Eq.~(\ref{transformMeasurement}). However, the measured
quantities depend on each observer's tetrad and so the
measurements are different.  At any point, the observer's tetrad
is determined by Fermi-Walker transport of the tetrad from some
initial point, along the world line of each observer.

The need for the tetrad formalism to relate experiment to theory,
as well as the problem of measurable quantities in general
relativity, is extensively discussed by Pirani~\cite{Pirani57},
Synge~\cite{Synge1960}, Soffel~\cite{Soffel89},
Brumberg~\cite{Brumberg91}, and more recently within the context
of metrology  by Guinot~\cite{Guinot97}.

The tetrad formalism was initially investigated for the case of
inertial observers that move on
geodesics~\cite{Pirani57,Fermi22,Walker,ManasseMisner,Li79,AshbyBertotti}.
Many observers are terrestrially based, or are based on
non-inertial platforms and the general theory for the case of
non-inertial observers has been investigated by
Synge~\cite{Synge1960}, who considered the case of non-rotating
observers moving along a time-like world line, and by
others~\cite{Li78,Ni78,Li79,Fukushima1988,Nelson87,Nelson90,Bahder98FermiCoord},
who considered accelerated, rotating observers. Perhaps the most
significant work for space-time navigation in the vicinity of the
Earth is contained in
Ref.~\cite{Bahder2001,Synge1960,AshbyBertotti,Fukushima1988,vyblyi1982,Marzlin1994,Nesterov1999}.

As an illustration of the relativistic measurement process,
consider an antenna that receives radio frequency electromagnetic
waves. The antenna converts an antisymmetric 4-dimensional tensor
of second rank, $F_{ij}$, into a scalar voltage reading on a
meter. The meter may have a digital readout of the measurement.
Consequently, the voltage is a scalar that does not transform
under Lorentz (or generalized coordinate transformations). The
voltage that is measured by a moving observer, $V(\tau)$, is a
function of the observer's proper time (since some starting
point), $\tau$, and depends on the observer's tetrad defined on
his world line.  The voltage measurement process can be modelled
as
\begin{eqnarray}
  V(\tau)  &  = & F_{ij} M^{ij} \label{VoltageMeasurementProcess0} \\
          & = & F_{ij} \lambda^i_{(a)} \lambda^j_{(b)} M^{(ab)} \label{VoltageMeasurementProcess1} \\
         &  =  & F_{(ab)} M^{(ab)}  \label{VoltageMeasurementProcess2}
\end{eqnarray}
where $F_{ij}$ is the electromagnetic field in the space-time, and
$M^{ij}$ is the measurement tensor that depends on the world line
of the observer.  The scalar product of the tensors in
Eq.~(\ref{VoltageMeasurementProcess0})  reduces the
electromagnetic field to a scalar quantity, $ V(\tau)$, which is
the measured voltage. This voltage $ V(\tau)$ is invariant under
coordinate transformations of the form in
Eq.~(\ref{transformMeasurement}). The projection of the
electromagnetic field tensor, $F_{ij}$, on the tetrad basis,
$\lambda^i_{(a)}$, yields a set of scalar numbers, $F_{(ab)}=
F_{ij} \lambda^i_{(a)}\lambda^j_{(b)}$. The quantity $M^{(ab)}$ is
an invariant matrix of numbers that characterizes the measuring
apparatus (the observer's antenna). A simple model for an antenna
is:
\begin{equation}\label{antennaMeasurentModel}
M^{(ab)} = \frac{1}{2} \left(
\begin{array}{cccc}
0 & -l_1 & -l_2 & -l_3 \\
l_1 & 0 & m_3 & -m_2  \\
l_2 & -m_3 & 0 & m_1 \\
l_3 & m_2 & -m_1 & 0 \\
\end{array}
\right)
\end{equation}
where the six quantities, $l_\alpha$ and $m_\alpha$, represent the
sensitivity of the antenna to electric and magnetic fields,
respectively.  The 3-vector $l_\alpha$ is simply the vector
effective length that characterizes the receiving antenna, see
Eq.~(\ref{AntennaVoltage}).

\subsubsection{Construction of the Tetrad: Fermi-Walker Transport}

\begin{figure*}
\includegraphics{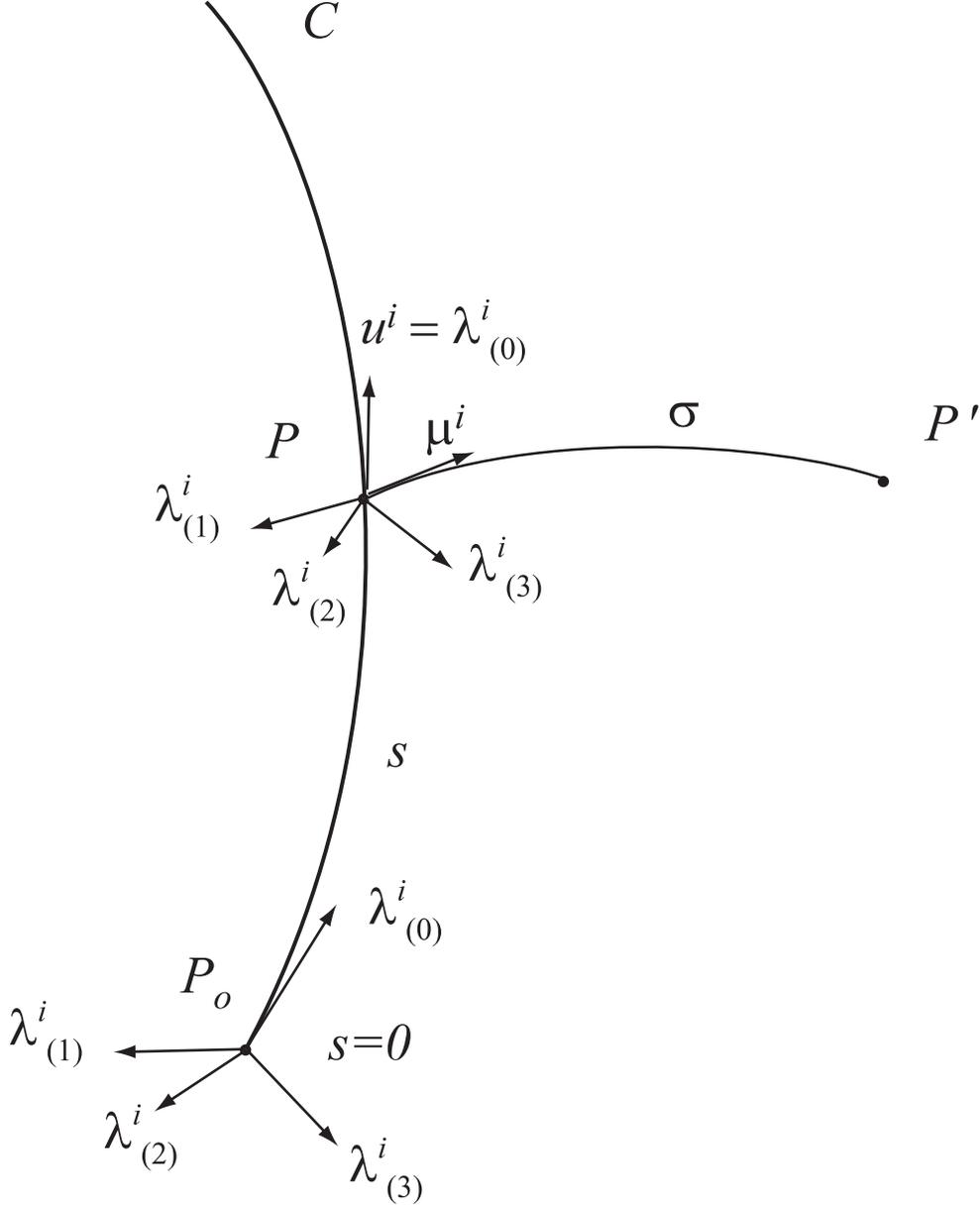}
\caption{\label{fig:WorldLine} The observer's world line $C$ is
shown with the initial tetrad basis vectors $\lambda^i_{(\alpha)}$
at $s=0$ at point $P_0$. The Fermi transported tetrad basis
vectors at proper time $s$ are shown at point $P$.}
\end{figure*}

The tetrad formalism for a given observer is constructed using the
world line of the observer.  Consider a time-like world line $C$
of an observer given by coordinates $x^i(u)$ with $u_1 \le u \le
u_2$, (see Fig.~\ref{fig:WorldLine}). Along this world line, the
observer has a four velocity $ u^i = d x^i / ds $ and acceleration
\begin{equation}
w^i =\frac{\delta u^i}{\delta s} = \frac{d u^i}{d s} + \Gamma^i_{j
k} u^j u^k  \label{fourAcceleration}
\end{equation}
where $\Gamma^i_{j k}$ is the affine connection and $\delta u^i /
\delta s$ indicates covariant differentiation along the world line
$x^i(s)$. The normalization of the 4-velocity, $u^i u_i = -1$,
provides the relation between the arc length, $s=c \tau$, where
$\tau$ is the proper time, and the parameter $u$.

As the observer moves on the time-like world line $C$, he carries
with him an ideal clock that keeps proper time $\tau$, and three
gyroscopes. At some initial coordinate time $x^0$, the observer is
at point $P_0$ at proper time $\tau=s/c=0$.  On his world line,
the observer carries with him three orthonormal tetrad basis
vectors, $\lambda_{(\alpha)}^{i}$, where $\alpha=1,2,3$, labels
the vectors and $i=0,1,2,3$ labels the components of these vectors
in some global system of coordinates. These vectors form the basis
for his measurements~\cite{Pirani57,Synge1960}, see
Fig.~\ref{fig:WorldLine}. The orientation of each basis vector is
held fixed with respect to each of the gyroscopes' axis of
rotation~\cite{MTW}. The fourth basis vector is taken to be the
observer 4-velocity, $\lambda_{(0)}^{i} = u^i$. The four unit
vectors $\lambda_{(a)}^{i}$, $a=0,1,2,3$, form the observer's
tetrad, which is an orthonormal set  of vectors at $P_0$
\begin{equation}
g_{ij} \, \lambda^i_{(a)} \, \lambda^j_{(b)} = \eta_{a b}
\label{initialTetrad}
\end{equation}
where the matrix $\eta_{(ab)}$ is the Minkowski metric, see
Appendix B.

At a later time $s=c \tau >0$, the observer is at a point $P$. The
observer's orthonormal set of basis vectors are related to his
tetrad basis at $P_0$ by Fermi-Walker transport. Fermi-Walker
transport preserves the lengths and relative angles of the
transported vectors. For an arbitrary vector with contravariant
components $f^i$, its components at $P$ are related to its
components at $P_0$ by the Fermi-Walker transport differential
equations~\cite{Synge1960}
\begin{equation}
\frac{\delta f^i}{\delta s} = W^{ij} f_j \label{FWtransport}
\end{equation}
where
\begin{equation}
W^{ij}= u^i w^j - w^i u^j \label{FWtensor}
\end{equation}
When we use Eq.\ (\ref{FWtransport}) to transport a vector $f^i$
that is orthogonal to the 4-velocity, $ u^i f_i = 0 $, the second
term in Eq.\ (\ref{FWtensor}) does not contribute. We refer to
transport of such space-like basis vectors as Fermi transport, and
$ W^{i j} \rightarrow \tilde{W}^{ij}= u^i w^j $.  The space-like
tetrad components satisfy $\lambda^i_{(\alpha)} u_i=0$, for
$\alpha=1,2,3$, and  at any point $P$ they  are found by
integrating the differential Eq.~(\ref{FWtransport}) over the
world line $x^i(s)$, using the initial conditions in
Eq.~(\ref{initialTetrad}) on the tetrad at point $P_o$, with
$W^{ij} \rightarrow \tilde{W}^{ij}$.

\subsubsection{Fermi Coordinates}

Associated with each Fermi-transported tetrad basis, there is a
set of Fermi coordinates, defined by the geometric construction
shown in Figure~\ref{fig:WorldLine}. Every event $P^\prime$ in
space-time has coordinates $x^{i \, \prime}$ in the global
 coordinate system. According to the observer moving on a
time-like world line, the same event has the Fermi coordinates
$X^{(a)}$, $a=0,1,2,3$. The first Fermi coordinate, $X^{(0)}=s$,
is taken to be the proper time (in units of length) associated
with the event $P^\prime$. The proper time for $P^\prime$ is
defined as the value of arc length $s$ such that a space-like
geodesic from point $P$ passes through event $P^\prime$, where the
tangent vector of this geodesic, $\mu^i$, is orthogonal to the
observer 4-velocity at $P$:
\begin{equation}
\mu^i \, u_i \vert_P = 0 \label{normal1}
\end{equation}
The orthogonality condition in Eq.\ (\ref{normal1}) is
\begin{equation}
g_{i j} \, \mu^i (s) \, \lambda^j_{(0)}(s) = 0
\label{normalCondition}
\end{equation}
and gives an implicit equation for $s$ for a given point
$P^\prime$. This orthogonality condition gives the first Fermi
coordinate of the point $P^\prime$
\begin{equation}
X^{(0)}=s \label{orthogCondition}
\end{equation}

The contravariant spatial Fermi coordinates, $X^{(\alpha)}$,
$\alpha=1,2,3$, are defined as~\cite{Synge1960}
\begin{equation}
X^{(\alpha)}= \sigma \mu^i  \lambda^{(\alpha)}_i = g_{i j} \,
\sigma(s)  \, \mu^i(s) \, \eta^{(\alpha \beta)} \,
\lambda^j_{(\beta)}(s) \label{FCcoordDef}
\end{equation}
where $\sigma$  is the  measure along the space-like geodesic
between $P$ and $P^\prime$,  $g_{ij}$ is the metric and  $\eta^{(i
j)}=\eta_{(i j)}=\eta_{ij}$ is the invariant Minkowski matrix, see
Appendix B.

\subsubsection{Metric in Fermi Coordinates}

All measurements made by a real observer  are done locally, at the
origin of Fermi coordinates. The measurements are projections on
the tetrad of the observer, and the tetrad is only defined on the
world line of the observer.  Never-the-less, Fermi coordinates can
be defined off the world line of the observer, and a corresponding
metric for Fermi coordinates can be defined. The space-time
interval in the Fermi coordinate system of the observer is
\begin{equation}
ds^2=-G_{(ij)} \, dX^{(i)} \, dX^{(j)}
\label{FermiCoordLineElement}
\end{equation}
where $G_{(ij)}$ are the metric tensor components when the
$X^{(i)}$ are used as coordinates~\cite{Synge1960}.

Despite the clear interpretation of measurement that the tetrad
formalism offers, the analysis of experiments has seldom been done
using the full tetrad formalism described above. In part, this is
due to limitations of data accuracy and theorist patience to carry
out the detailed computations.  There are, however, a few explicit
theoretical constructions of tetrads in the literature that
address the issues discussed
here~\cite{AshbyBertotti,Fukushima1988,Marzlin1994,Bahder98FermiCoord,Nesterov1999}.

\section{Reference Frames and Coordinate Systems}
The world function of space-time (see Section V) is a useful tool
for describing the problem of navigation in space-time in a simple
geometric way. The general equations are covariant, and invariant,
and there is no need to specify a system of space-time
coordinates. Navigation is carried out by making local
measurements, in the comoving frame of the apparatus that does the
measurement.

However, in actual applications, a specific implementation of a
system of space-time coordinates must be used. In most real-life
applications today, such as the GPS, a fully relativistic
4-dimensional space-time coordinate system is not used.  Instead,
a system of three-dimensional coordinates plus a set of clocks are
used. There are three common systems of three-dimensional
coordinates that play a role in the navigation problem:
Earth-centered inertial (ECI) coordinates, Earth-centered
Earth-fixed (ECEF) coordinates, and topocentric coordinates.  The
origin of ECI coordinates is at the Earth's center of mass, and
the orientation is determined with respect to distant objects--so
the coordinates do not rotate with respect to these distant
objects. However, the origin of ECI coordinates revolves around
the sun along with the Earth, so these coordinates are better
called quasi-inertial coordinates.   ECEF coordinates have the
same origin as ECI coordinates, but these coordinates rotate with
the Earth, so a point that is stationary on the Earth surface has
a constant value for its ECEF coordinates. Topocentric coordinates
have their origin on the Earth surface, with the $x$-axis pointing
South, the $y$-axis pointing East, and the $z$-axis pointing
radially away from Earth center (pointing up). Topocentric
coordinates are used for making radar and other observations on
the Earth surface. All three of these coordinate systems are
three-dimensional. In experimental observations, these three
dimensional coordinate systems are used together with clocks to
record the coordinates of space-time events. Great care must be
exercised when relativistic theories are used to analyze the data,
because the observations were not recorded with respect to a true
(relativistic) four-dimensional system of space-time coordinates.
As an example of potential errors, see the next section.

\subsection{Gravitational Warping of Coordinates}
The presence of a gravitational field leads to a warping of the
system of coordinates. More precisely, if we attempt to use
Euclidean space and time coordinates, and neglect the effect of a
gravitational field on the coordinate system, we will find that
the definition of these coordinates is not precise.  In other
words, the definition of these Euclidean coordinates is
ill-defined when a gravitational field is present. To demonstrate
the magnitude this effect, consider the Schwarzschild metric
\begin{equation} \label{SchwarzschildMetric2}
-ds^2=g_{ij}dx^idx^j = (1-\frac{r_g}{r})c^2dt^2 -
\frac{dr^2}{1-\frac{r_g}{r}} -r^2 (\sin^2 \theta d\phi^2 +
d\theta^2)
\end{equation}
where the gravitational radius is given by $r_g= 2 G M/c^2$.  For
Earth, $M=$5.98 $\times$10$^{24}$ kg, so $r_g=0.88$ cm.   The
physical length $dl$ in a static space-time is given
by~\cite{LLClassicalFields}
\begin{equation} \label{spaceMetric}
dl^2 = \gamma_{\alpha \beta} \, dx^\alpha dx^\beta
=\frac{dr^2}{1-\frac{r_g}{r}} + r^2 (\sin^2 \theta d\phi^2 +
d\theta^2)
\end{equation}
where the 3-dimensional spatial metric $\gamma_{\alpha \beta}$ is
related to the 4-dimensional  metric $g_{ij}$ by
\begin{equation}\label{spaceMetric2}
\gamma_{\alpha \beta} = g_{\alpha \beta}  - \frac{g_{0\alpha}
g_{0\beta} }{g_{00}}
\end{equation}
The physical distance between two points at radial coordinates
$r_1$ and $r_2$, and at the same coordinate angle $\theta$ and
$\phi$, is given by
\begin{widetext}
\begin{equation}\label{3DLength}
l=\int_{r_1}^{r_2} dl = \int_{r_1}^{r_2}
\frac{dr}{\sqrt{1-\frac{r_g}{r}}} = \left[
\frac{(r-r_g)}{\sqrt{1-\frac{r_g}{r}}} +
 r_g  \log(\sqrt{r} +\sqrt{r-r_g})
\right]_{r_1}^{r_2}
\end{equation}
\end{widetext} Now imagine the first point is on the Earth's
equator, so that $r_1=6.378\times10^6$ m is the Earth's equatorial
radius, and the second point is at the altitude of a GPS satellite
where $r_2=2.6561 \times 10^7$ m.  Since the Earth's gravitational
radius is $r_g=2 G M/c^2 = 0.88$ cm,  the length $l$ in
Eq.~(\ref{3DLength}) can be expanded in the small parameter
$x=r_g/r_1 << 1$, giving
\begin{equation}\label{length}
l= r_2-r_1 + \frac{1}{2} \, r_g \log{\frac{r_2}{r_1}}
\end{equation}
The first term on the right, $r_2-r_1$, is the Euclidean length
between the coordinate points at radial coordinate $r_1$ and at
$r_2$. The second term  is the  correction to the physical length
$l$ due to the presence of a gravitational field. The magnitude of
this correction, for the $r_1$ and $r_2$ values cited above, is
$\frac{1}{2} r_g \log{\frac{r_2}{r_1}}=0.63$ cm. This shows that
the physical length $l$ is longer by $0.63$ cm than the difference
between radial coordinate values in a Euclidean space that has
zero gravitational field.  The gravitational field has the effect
of stretching the physical space between coordinate values.

The implication of the effect is that a 4-dimensional relativistic
frame of reference (coordinate system) must be implemented (rather
than a system of 3-dimensional coordinates plus a time scale) if
we want to have an unambiguous definition of a space-time
coordinate system. In other words, a Euclidean system of
3-dimensional coordinates, plus a time scale, will have inherent
error  of $\frac{1}{2} r_g \log{\frac{r_2}{r_1}}$ over a distance
$r_2-r_1$, due to the warping of space-time due to the presence of
the gravitational field.

In the above calculation, I have neglected the Earth's quadrupole
moment $J_2 =  1.0826800\times10^{-3}$  of the  mass distribution.
The inclusion of this quadrupole moment can be expected to modify
the gravitational correction (second term) in Eq.~(\ref{length})
by additional terms of order $J_2 r_g$, leading to an additional
length correction on the order of $ J_2 r_g
\log{\frac{r_2}{r_1}}$, which is of the order of $10^{-2}$ mm over
distances $r_2 - r_1=2.0\times10^7$ m.

Another way of stating the effect of the gravitational warping of
coordinates is to say that, if we used Euclidean geometry, the
accuracy of clock synchronization is limited to  $(0.63 {\rm
cm})/c = 2.1\times 10^{-11} \, {\rm s} =21$ ps, even if perfect
clocks and equipment are used.  This value of 21 ps assumes the
previous values of $r_1$ and $r_2$. Reference to Table I shows
that, for some applications, we must have time synchronization to
better than 21 ps.

The gravitational warping of coordinates is closely associated
with (but distinct from) the slowing down of light in a
gravitational field, also known as the Shapiro time delay effect.
The average speed of light can be computed along a radial path
from $r_1$ to $r_2$.  The path is a null geodesic given by
$ds^2=0$, which gives $c^2 dt^2 = dr^2/(1-r_g/r)^2$.  The
coordinate time for light to traverse this path is
\begin{equation}\label{nullGeodesic2}
\int_{t_1}^{t_2} dt = \frac{1}{c}\int_{r_1}^{r_2}
\frac{dr}{1-\frac{r_g}{r}}
\end{equation}
The average speed of light over this path is then given by
\begin{equation}\label{aveSpeed}
\frac{l}{t_2 -t_1} = c\left[ 1 - \frac{1}{2} \frac{r_g}{r_2-r_1}
\, \log \frac{r_2}{r_1} \right]
\end{equation}
The second term in the above equation is the correction to the
speed of light due to the presence of the Earth's mass, and for
the previous values of $r_1$ and $r_2$ has the small magnitude
\begin{equation}\label{valueSlowed}
\frac{\Delta c}{c} = - \frac{1}{2} \frac{r_g}{r_2-r_1} \, \log
\frac{r_2}{r_1} = -3.13\times 10^{-10}
\end{equation}
The negative sign in Eq.~(\ref{aveSpeed}) means that light is has
travelled slower than in vacuum in the absence of a gravitational
field.  This correction to the speed of light depends on the
direction of light travel.

\section{Clock Synchronization}

As discussed in the introduction, accurate clock synchronization
is the backbone of applications such as high-accuracy navigation,
communication, geolocation, and space-based interferometer
systems. Also, as previously discussed, clock synchronization
cannot be divorced from the general problem of navigation in
space-time (determining the position and time of an observer). The
key point here, is that clock synchronization (divorced from the
problem of determining spatial coordinates) is not a covariant
concept, whereas navigation in space-time {\it is} a covariant
concept, and consequently it can be formulated in covariant
equations, such as was done in Subsection V-A, using the world
function.  See Eq.~(\ref{CovariantNavigation}). However, if there
exists a system of coordinates in which the  clocks are all
stationary at constant known spatial positions, then we can
discuss their synchronization.  However, this is generally not the
case because satellites that carry clocks orbit the Earth.

In the next Subsection B, we discuss synchronization of clocks
that are stationary, at known spatial positions of an arbitrary
system of coordinates, in the presence of a gravitational field.
The system of coordinates need not be rigid and can be rotating.
Section C discusses the peculiar way that GPS clocks are
``synchronized".  Since the GPS satellites are moving, the clocks
are (approximately) locally synchronized to a coordinate time in a
certain metric, and hence the meaning of synchronization is
different for GPS satellites than for stationary clocks.

In pre-relativistic physics, clock synchronization was a
straight-forward concept wherein two clocks were simply set to the
same time.  With the advent of highly accurate atomic clocks on
board satellites,  more accurate schemes for synchronizing remote
clocks have been developed. Furthermore, the clocks to be
synchronized are in relative motion as well as in different
gravitational potentials. The accuracy of clocks has improved to
the point that previously small unobservable effects must be
accounted for by a comprehensive theory.  A metric theory of
gravity, such as general relativity, is the vehicle of choice for
dealing with clock synchronization. Within the theory of
relativity there are a number of clock synchronization schemes.
However, one central concept--that of simultaneous event--is at
the heart of the issue of clock synchronization.  Simultaneity of
two events is a definition, and not an absolute covarian concept,
unless the two events are co-located at the same event in
space-time.

\subsection{Eddington Slow Clock Transport}

Clocks on maritime vessels of the 1700 and 1800's were used to
navigate the oceans.   An accurate clock  was kept on the ship,
and it was used to determine longitude of the vessel's current
position~\cite{Sobel}. The basic idea was to carry an accurate
clock, and be in possession of the ``correct time". Using this
time, and a view of the sky, the ship's longitude could be
computed based on a theory of Earth rotation.  This scheme of
having the correct time by slowly transporting a clock is often
called ``Eddington slow clock transport"~\cite{Eddington1924}.

From the point of view of special relativity, where $g_{00}=-1$
and $g_{\alpha \beta}=\delta _{\alpha \beta}$, $\alpha,
\beta=1,2,3$.  (constant and uniform gravitational field), a clock
can be transported with arbitrarily small velocity and still
maintain the ``correct time". From Eq.~(\ref{propertime}), when
the velocity of a clock is arbitrarily small, $d x^\alpha/d x^0
\rightarrow 0$, $\alpha=1,2,3$, the relation between proper time
$\tau$ and coordinate time $t=x^0/c$, reduces to $\Delta \tau =
\Delta x^0 /c$. So a  slow moving clock, which keeps proper time
time $\tau$, can by definition be made to keep coordinate time
$x^0$. Such a clock can be slowly transported over large distances
and can be used to synchronize other remote clocks. Unfortunately,
Eddington slow clock transport is too restrictive for applications
to clocks on satellites, because the speed of the satellites is
not small, typically on the order of $v/c=d x^\alpha/d
x^0\sim10^{-5}$, and because the gravitational potential  between
different satellites varies significantly.  See Section IX for a
discussion of the effect of satellite motion and gravitational
potential on proper time.

\subsection{Einstein Synchronization}

Another method of synchronizing {\it stationary} clocks is based
on exchanging electromagnetic signals between clocks at {\it known
spatial locations}. This type of synchronization is often called
Einstein synchronization, and is based on a particular definition
of simultaneity of (spatially separated) events. Perhaps the
clearest discussion of the definition of simultaneity, within the
context of general relativity, is given by L. D. Landau and E. M.
Lifshitz~\cite{LLClassicalFields}. We use their definition of
simultaneity in the clock synchronization argument presented
below.

\begin{figure}
\includegraphics{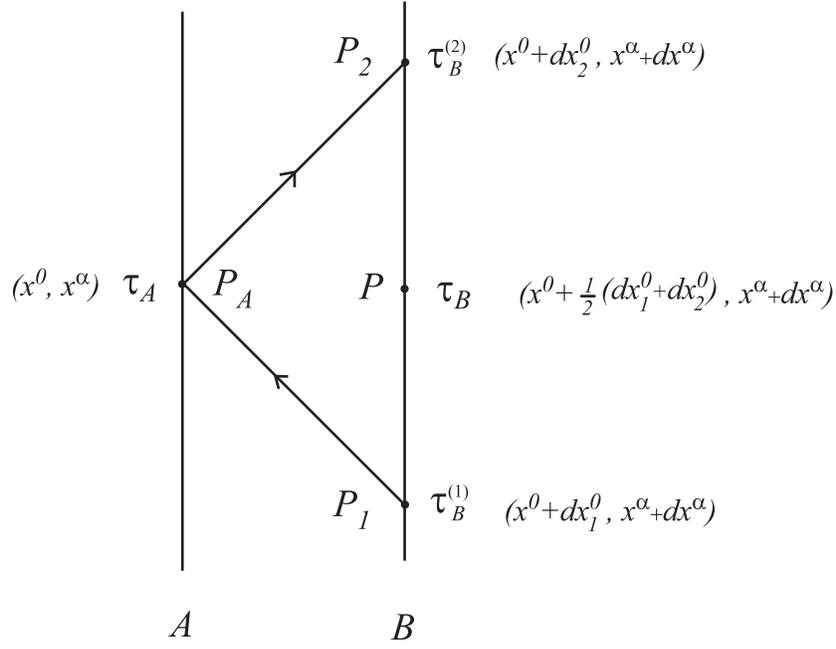}
\caption{\label{fig:EinsteinSynch} The world line of clock $A$ and
clock $B$ are shown.  Clock $B$ is synchronized to clock $A$ by
electromagnetic signals. A signal from clock $B$ is sent out at
event $P_1$, the signal reflects off the face of clock $A$ at
event $P_A$, and returns to clock $B$ at event $P_2$.}
\end{figure}

Consider two clocks at rest in some frame of reference
(4-dimensional coordinates).  Clock $A$ has world line defined by
 constant spatial coordinates $x^\alpha$,
and clock $B$ has world line defined by spatial coordinates
$x^\alpha + dx^\alpha$, $\alpha=1,2,3$, see
Figure~\ref{fig:EinsteinSynch}. Clock $B$ is to be synchronized to
clock $A$ by exchange of electromagnetic signals. At event $P_1$,
with coordinate time $x^0+dx^0_1$, an electromagnetic signal is
sent from clock $B$ to clock $A$.  Clock $A$ receives the signal
at event $P_A$, which has coordinates $(x^0,x^\alpha)$, and
immediately reflects the signal back to clock $B$, where it is
received at event $P_2$ with coordinates $(x^0 + dx^0_2,x^\alpha +
dx^\alpha)$.  This reflected signal (which is received by $B$ at
$P_2$) contains the proper time reading, $\tau_A$, of clock $A$ at
event $P_A$.

The question then arises: what point $P$ (coordinate time) on
world line of clock $B$ is simultaneous with event $P_A$ on world
line of clock $A$? Detailed consideration of this problem leads to
the conclusion that there is no preferred way to define a point on
world line $B$ that is simultaneous with event $P_A$.  Therefore,
we {\it arbitrarily} take the midpoint between $P_1$ and $P_2$ as
defining the point $P$ that is simultaneous with $P_A$.  For
electromagnetic propagation, the space-time interval must vanish:
\begin{eqnarray}\label{zeroInterval}
-ds^2  & = &  g_{ij}dx^i dx^j  \\
   & =  &  g_{\alpha \beta}dx^\alpha dx^\beta
+ 2 g_{0 \alpha} dx^0 dx^\alpha + g_{00}\left( dx^0 \right)^2 = 0
\nonumber
\end{eqnarray}
Solving for the coordinate time $dx^0$, we get  two solutions,
corresponding to the two directions of signal travel between
$x^\alpha$ and $x^\alpha +dx^\alpha$:
\begin{widetext}
\begin{eqnarray}
dx^{0}_1 & = &  -\frac{1}{g_{00}} \left(  g_{0\alpha} dx^\alpha -\sqrt{(g_{0\alpha} g_{0\beta} - g_{00} g_{\alpha \beta})dx^\alpha dx^\beta} \right)  \\
dx^{0}_2 & =  &  -\frac{1}{g_{00}} \left( g_{0\alpha} dx^\alpha
+\sqrt{(g_{0\alpha} g_{0\beta} - g_{00} g_{\alpha \beta})dx^\alpha
dx^\beta} \right) \label{dxSoln1}
\end{eqnarray}
\end{widetext}
Note that if $dx^{0}_2>0$ then $dx^{0}_1 <0$. Therefore, the
coordinate time of event $P$, which is simultaneous with event
$P_A$, is given by (using the midpoint definiton of simultaneity):
\begin{equation}\label{eventPTime}
  x^0 - \frac{g_{0\alpha}dx^\alpha}{g_{00}}
\end{equation}
Since events $P$ and $P_A$ are simultaneous (by definition), we
consider clock $A$ and $B$ to be synchronized when their proper
times are equal at these events:
\begin{equation}\label{synchronized}
  \tau_B = \tau_A
\end{equation}
This corresponds to shifting the origin (epoch) of proper time.

The reflected signal from clock $A$ arrives at $B$ at the time
$\tau_B^{(2)}$.The proper time interval between event $P$ and
$P_2$ is given in terms of coordinate time by the integral:
\begin{equation}\label{timeSynchDiff}
 \tau_B^{(2)} - \tau_B  =  \frac{1}{c} \int_P^{P_2} ds
\end{equation}
Now, consider clocks $A$ and $B$ to be stationary, so that their
spatial coordinates, $x^\alpha$ and $x^\alpha + dx^\alpha$, are
constant in time in the chosen coordinate system.  Then the proper
time elapsed on clock $A$ between point $P$ and $P_A$ is given by
\begin{widetext}
\begin{eqnarray}
 \tau_B^{(2)} - \tau_B & =  & \frac{1}{c} \int_{x^0 +\frac{1}{2}\left(dx^{0}_1+ dx^{0}_2 \right)}^{x^0 +  dx^{0}_2}  \,\, \sqrt{-g_{00}} \,\,\,  dx^0 \label{firstLine1} \\
            & = & \frac{1}{c} \left[
            \frac{1}{\sqrt{-g_{00}}}\sqrt{(g_{0\alpha} g_{0\beta} - g_{00} g_{\alpha \beta})dx^\alpha dx^\beta} \right]_P \label{firstLine2}
\end{eqnarray}
\end{widetext}
where in the last line the quantities are evaluated at $P$, at
coordinate time given by Eq.~(\ref{eventPTime}). Using the
condition in Eq.~(\ref{synchronized}), Eq.~(\ref{firstLine2}) can
be written as
\begin{equation}\label{synchCondition}
 \tau_B^{(2)} = \tau_A + \frac{1}{c} \left[  \frac{1}{\sqrt{-g_{00}}}\sqrt{(g_{0\alpha} g_{0\beta} - g_{00} g_{\alpha \beta})dx^\alpha
 dx^\beta}\right]_P
\end{equation}
Equation~(\ref{synchCondition}) gives the condition for clock B to
be synchronized to clock A, in terms of quantities that are
observed by clock B.  Note that $\tau_A$ is the proper time on
clock $A$, as observed by clock $B$.
Equation~(\ref{synchCondition}) gives the proper time that must be
set on clock $B$ at event $P_2$ so that clock $B$ is
``synchronized" with clock $A$ at the earlier event $P$.  Note
that the synchronization was done between clocks $A$ and $B$ that
are separated by an infinitesimal spatial interval $dx^\alpha$. In
practice, for small gravitational fields, this infinitesimal
interval can correspond to large distances. For the case of a flat
space-time with a Minkowski metric, $g_{00}=-1, g_{\alpha
\beta}=\delta_{\alpha \beta}$, Eq.~(\ref{synchCondition}) gives
$\tau_B^{(2)} = \tau_A + l/c$ where the known spatial distance
between clocks is  $l= g_{\alpha \beta} dx^\alpha dx^\beta$.

Some comments are in order on the practical application of the
synchronization condition in Eq.~(\ref{synchCondition}).  From
Eq.~(\ref{eventPTime}), we know the space-time coordinates of
point $P=(x^0- \frac{g_{0\alpha} dx^\alpha}{g_{00}}, x^\alpha +
dx^\alpha)$, and the synchronization condition in
Eq.~(\ref{synchCondition}) depends on the metric components
$g_{ij}$. For high-accuracy clock synchronization, we must know
the metric components $g_{ij}$ in the coordinate system in which
the clocks are at rest. For the accuracy needed in some
applications (see Table~\ref{tab:TimeingPrecision}), the metric is
not known to sufficient accuracy.  An even more serious problem is
that the spatial positions of clock A and B must be known a
priori, i.e., the clock positions are not determined by the
synchronization protocol, and the synchronization protocol depends
on these positions. Finally, in the real world, satellites are in
motion with respect to one another and there is no single (simple)
coordinate system in which more than two satellites are at rest.

In conclusion, it becomes evident that for most applications, the
critical problem is not clock synchronization (correlating clock
times), but that of navigation (correlation of clock positions and
times).  See Section V subsection~A.

Consequently, in real applications such as in the satellite system
known as GPS, a different synchronization scheme is used. In the
GPS scheme, satellite clocks keep (approximately) the coordinate
time in the underlying ECI coordinate frame.  We discuss GPS clock
synchronization in the next section.

In practice, Einstein clock synchronization is the basis of the
applied technique known as two-way satellite time transfer (TWSTT)
which in practice gives very accurate time synchronization between
two points that are in common view of the same communication
satellite~\cite{Kirchner1991,Francis2002}.

\subsection{GPS Clock Synchronization}

The Global Positioning System (GPS) is a U.S. Department of
Defense Satellite System consisting of approximately 24 satellites
orbiting the Earth. The system consists of four satellites in each
of six inclined orbital planes of 55$^\circ$. Each satellite of
the GPS carries atomic clocks, and sends pseudorandom code signals
that are the basis of providing accurate time and navigation
signals~\cite{Kaplan96,Hofmann-Wellenhof93,ParkinsonGPSReview,Bahder2003}.
The GPS uses a unique time synchronization scheme, wherein the
clocks send time, encoded on chips of the pseudorandom code, and
these chips (with time stamps on them) are received at Earth
monitoring stations.  The stations determine any clock corrections
needed based on the broadcasts from the satellites.  The GPS
satellites move at a fast speed (approximetely $v/c\sim10^{-5}$)
and the clocks suffer relativistic time dilation of approximately
7 $\mu$s per day. On the other hand, the satellites are at a high
altitude, and the clocks run fast (as compared with coordinate
time in the ECI frame metric), by about 45 $\mu$s per day, due to
the high gravitational potential in which they
operate~\cite{Bahder2003}. The net effect is that, with respect to
a clock on the Earth's surface, the clocks appear to run fast by
approximately 38 $\mu$s per day.

Each GPS satellite has an orbit that is approximately circular.
Consequently, all the clocks on the satellites behave in roughly
the same way--they run fast by 38 $\mu$s per day, or in terms of
frequency, as observed from the surface of the Earth the
oscillators run fast by 1 part in 4.4$\times$10$^{-10}$, see
Section IX, subsection E. If something were not done to counter
this effect, all GPS satellite clocks would appear to run fast.
Consequently, a ``factory offset" is applied to the frequency of
each clock oscillator (in software) in the amount
-4.4$\times$10$^{-10}$. The clocks on board the GPS satellites
then keep (approximately) coordinate time in the ECI frame, and
the clocks appear to (approximately) keep correct time as seen
from the Earth's surface. This ``factory offset" does not
compensate for the different orbital eccentricities of the
satellites, since for each satellite.  The factory also does not
compensate for the relative velocity between satellites and
different moving GPS receivers, see Ref.~\cite{Bahder2003} for a
detailed discussion of the role of the Doppler effect in GPS
measurements.

\begin{figure}
\includegraphics{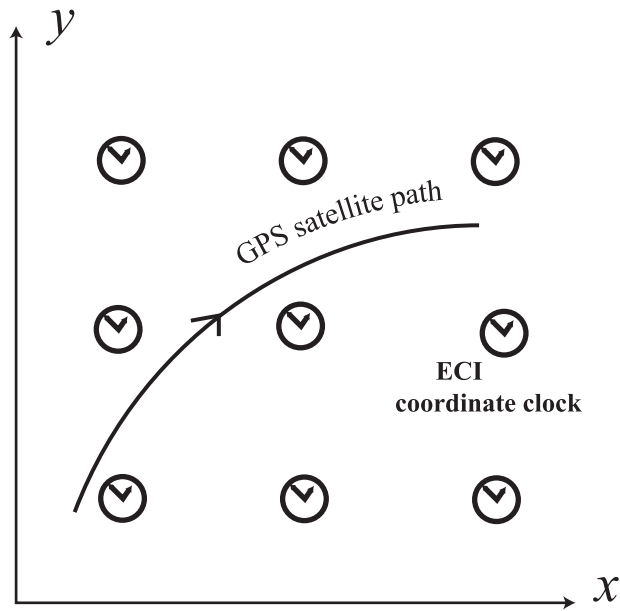}
\caption{\label{fig:GPStimeSynch}The motion of a GPS satellite in
3-dimensional space is shown as it passes (stationary) coordinate
clocks in the ECI frame.  At any instant, the GPS satellite clock
agrees with time on the underlying ECI frame coordinate clocks.}
\end{figure}

The ``factory offset" applied to the GPS satellite oscillators can
be understood as another scheme for clock synchronization.  There
exists a coordinate time $t=x^0/c$, in some metric around the
Earth, see Eq.~(\ref{EarthMetric}).  The coordinate time is a
global quantity, as compared to the proper time, which depends on
the given world line.  By definition, clocks on the Earth's geoid
keep coordinate time in the ECEF frame.  In order to synchronize
the GPS satellite clocks to coordinate time in the ECI frame, the
``factory offset" is applied. With this offset, the GPS clocks
keep approximate coordinate time in the ECI frame of reference. As
each satellite moves through space, we can imagine that it passes
``hypothetical" coordinate clocks (that are stationary in the ECI
frame and keep ECI coordinate time). The ``factory offset" has the
effect that a GPS satellite clock instantaneously  agrees
(closely) with each ECI frame coordinate clock that it passes, see
Figure~\ref{fig:GPStimeSynch}.  So the GPS satellite clocks are
(approximately ) synchronized to coordinate clocks in the
underlying ECI frame.  Actually, this is an unusual idea--for
(GPS) clocks in one frame (comoving frame of a GPS satellite) to
keep coordinate time in another (ECI) frame. As mentioned above,
in the case of GPS, this synchronization is approximate, because
only a constant rate offset is applied before satellite launch,
which cannot compensate for effects, such as the eccentricity of
each satellite orbit (called $e \sin E$ effect), the Earth's
quadrupole moment (or $J_2$ effect), the effects of other forces
(such as solar pressure and atmospheric drag) on each satellite,
and imperfect atomic clocks. The satellite ground tracking system
for the GPS monitors satellite clocks and attempts to compensate
for these effects by calculating satellite clock corrections and
uploading these corrections to the satellites so that  users of
the GPS may apply them to get accurate space-time navigation.

\subsection{Quantum Synchronization}

During the past several years, alternative schemes for clock
synchronization~\cite{MandelOu1987,Jozsa2000,Chuang2000,Yurtsever2000,burt2001,Jozsa2000a,preskill2000,Giovannetti2001,Bahder2004,Valencia2004}
have been proposed based on quantum information theory and
entanglement of quantum states~\cite{Ekert2000,Nielson2000}.
Quantum effects may be exploited for clock synchronization since
entangled (or correlated) photon pairs are found to interfere
destructively at a beam splitter~\cite{MandelOu1987}. Such photon
pairs are believed to be created at a single space-time event,
within 100 fs of each other~\cite{MandelOu1987}. Exploiting this
effect, two relatively simple synchronization schemes have been
very recently proposed based on the interference of entangled
photon pairs states, which can be created by parametric
downconversion in a crystal that lacks a center of inversion
symmetry~\cite{MandelOu1987}.  One of these schemes is capable of
clock synchronization in free space~\cite{Bahder2004}, while the
other relies on there being a difference of group velocities for
each of the photons in the entangled pair in an optical medium,
such as exists in an optical fiber~\cite{Valencia2004}. Very
recently, an experimental demonstration of quantum clock
synchronization has been carried out in the
laboratory~\cite{Valencia2004}.

The quantum mechanical clock synchronization proposals do not
included the basic relativistic effects present when clocks on
board satellites are to be synchronized: the fast relative motion
of clocks and the variation in gravitational potential between
satellites in various orbital regimes, such as low Earth orbit
(LEO), geosynchronous orbit (GEO), and highly elliptical orbit
(HEO).  In the next section, we describe the resulting
syntonization problem that arises for satellites at differing
gravitational potentials.

\section{The Syntonization Problem}

Satellites in multiple orbital regimes may have their clocks
synchronized by means of exchange of optical signals, as described
in the previous section.  This means that at one coordinate time,
all clocks can be made to read the same value (for example, the
same coordinate time). Satellite applications require that the
synchronization between members of the satellite ensemble  be
maintained for some time period, or alternatively, the clock
differences must be known at a given elapsed time from the
synchronization epoch.  It is well known that the proper time, and
consequently the hardware time (see Section III), will run at a
different rate on each satellite with respect to coordinate time
of some given metric. This difference in rate of proper time is
due to the motion of the clock (time dilation) and gravitational
potential effects (sometimes called the gravitational red shift
effect).

Since the hardware time on all real clocks is affected by their
motion and position in the space (local gravitational potential),
the only reasonable measure of time is coordinate time, as used in
a metric theory of gravity, such as general relativity. Coordinate
time is a mathematical construct that is global, which means that
it is the same everywhere in the space. In distinction, proper
time depends on the history (world line) of the clock.  In this
Section, we compute the difference between proper time on board a
satellite, and coordinate time, for satellites in various orbits
about the Earth.

Since the clock on board a satellite is located at a different
spatial position than a reference clock on the Earth's surface,
these clocks can be compared by exchange of electromagnetic
signals. The relation between these two clocks must be
established, such that their relative motion and their respective
gravitational potentials are taken into account.   At any instant
in time, the relative velocity of the satellite with respect to
the ground leads to the satellite clock running slower than the
ground clock, due to relativistic time dilation.  The higher
gravitational potential at the satellite leads to the satellite
clock running faster than the ground clock. Generally speaking,
clocks in low-orbiting satellites run slow due to the predominant
time dilation effect, due to high orbital velocity and small
altitude above the Earth's surface.  On the other hand, clocks in
high-orbit satellites generally run faster than ground clocks
because the gravitational potential effect is predominant.

While a satellite clock can be compared to an Earth-bound
reference clock, there is nothing special about the Earth-bound
clock as a reference clock. In fact, from the point of view of an
ECI coordinate system, the Earth bound clock moves in a circle,
just like the satellite clock. Furthermore, the actual reading on
an ideal clock depends on its world line, i.e., its past history
of velocity and gravitational potential. All ideal (and hardware)
clocks suffer from this complication. Therefore, the comparison of
all clocks must be made to a standard time or ``clock" that is
global and does not depend on its history.  Such a quantity is the
coordinate time associated with some metric that describes the
space-time in the vicinity of the Earth. Coordinate time is a
mathematical construct that is the same in all of space-time. The
complication is that coordinate time (and generally each spatial
coordinate) is arbitrary in a metric theory such as general
relativity, and depends on an arbitrary choice of a space-time
coordinates. In the next section, we will choose a space-time
metric that has the desirable property that coordinate time
corresponds (approximately) to the proper time kept by an ideal
clock on the Earth's geoid (surface of equal geopotential).
\bigskip

\subsection{Choice of Metric in Vicinity of the Earth}

The definition of coordinate time comes from choosing a specific
metric for the space-time.  In general relativity, the coordinates
$x^i$, $i=0,1,2,3$, are mathematical entities that are never
observed. The metric of space-time depends on these coordinates,
and takes into account the warping of space-time due to the
presence of the gravitational field.  For the same physical
space-time, we can choose different coordinates. Of course, all
observations are independent of the coordinates, and so the choice
of coordinates is arbitrary.  However, it is convenient to choose
space-time coordinates so that coordinate time has some physical
meaning. Perhaps the most reasonable choice is to take the metric
for space-time of the form~\cite{AshbyInParkinsonGPSReview}
\begin{widetext}
\begin{equation}\label{EarthMetric}
-ds^2= -(1+\frac{2}{c^2} V) (d{\bar x}^0)^2 + (1-\frac{2}{c^2} V)
\left[ (dx^1)^2 + (dx^2)^2 + (dx^3)^2 \right]
\end{equation}
\end{widetext} where ${\bar x}^0 ,x^1,x^2,x^3$ are geocentric
equatorial coordinates, where $x^3$ coincides with the Earth's
axis of rotation and increasing positive values point to North.
The Earth is taken to be an oblate spheroid with potential given
by~\cite{Caputo1967}
\begin{equation}
V(r,\theta)= -\frac{G M}{r} \left[ 1 - J_2 \left( \frac{R_e}{r}
\right)^2 P_2(\cos(\theta))  \right] \label{EarthPotential}
\end{equation}
where, $r^2= (x^1)^2 + (x^2)^2 + (x^3)^2$ and $\theta$ is the
polar angle measured from the $x^3$ axis.   In
Eq.~(\ref{EarthPotential}), $G$ is Newton's gravitational
constant, $M$ is the mass of the earth,  $P_2(x) =(3 x^2 -1)/2$ is
the second Legendre polynomial, $R_e$ is the Earth's equatorial
radius, and $J_2$ is the Earth's quadrupole moment, whose value is
approximately $J_2= 1.0826800\times 10^{-3}$.  The metric in
Eq.~(\ref{EarthMetric}) is the solution~\cite{LLClassicalFields}
of the linearized version of Einstein
Eqs.~(\ref{EinsteinFieldEquations}).

The coordinate time can be given a simple interpretation.
Transform the metric in Eq.~(\ref{EarthMetric}) to rotating ECEF
coordinates $y^i$, using the transformation
\begin{eqnarray}
{\bar x}^0 & = & {\bar y}^0  \nonumber \\
x^1 & = & \cos(\frac{\omega}{c} {\bar y}^0) \, y^1  - \sin(\frac{\omega}{c} {\bar y}^0) \, y^2 \nonumber \\
x^2 & = & \sin(\frac{\omega}{c} {\bar y}^0) \, y^1 + \cos(\frac{\omega}{c} {\bar y}^0) \, y^2 \nonumber \\
x^3 & = & y^3 \label{CoordinateTransformation1}
\end{eqnarray}
Note that the coordinate time in the rotating frame, ${\bar
y}^0={\bar x}^0$.  In these ECEF rotating coordinates, the metric
is given by
\begin{widetext}
\begin{eqnarray}\label{RotatingCoordMetric}
-ds^2 & = &  -\left[ 1 + \frac{2 V}{c^2} -
\frac{\Omega^2}{c^2}\left[ (y^1)^2 + (y^2)^2 \right] + \frac{2
 V}{c^2} \frac{\Omega^2}{c^2} \left[ (y^1)^2 + (y^2)^2
\right] \right] (d{\bar y}^0)^2 \nonumber \\
   &   & + (1-\frac{2 V}{c^2} ) \left[
2\frac{\Omega}{c}(y^1 dy^2 - y^2 dy^1)d{\bar y}^0 + (dy^1)^2 +
(dy^2)^2 + (dx^3)^2 \right]
\end{eqnarray}
\end{widetext}
From Eq.~(\ref{RotatingCoordMetric}), we see that stationary
clocks in the ECEF coordinates (clocks which satisfy
$dy^\alpha=0$) that are at the same value of geopotential $\phi$,
where
\begin{equation}\label{geoPotDef}
\phi = V - \frac{1}{2}\Omega^2 \left( (y^1)^2 + (y^2)^2 \right)
\end{equation}
have the same rate of proper time with respect to coordinate time
$y^0$. The term geopotential is used because $\phi$  takes into
account the effect of angular velocity $\Omega$ of Earth rotation.
We have neglected the small cross-term $\frac{2 V }{c^2}
\frac{\Omega^2}{c^2} R^2 \sim 10^{-21}$.

Using the observation that clocks at a constant value of
geopotential run at the same rate, it is advantageous to define
the new coordinate time $t$
\begin{equation}\label{NewCoordTime}
c t = x^0 = y^0 = \left( 1 + \frac{\phi_o}{c^2} \right) {\bar y}^0
= \left( 1 + \frac{\phi_o}{c^2} \right) {\bar x}^0
\end{equation}
where $\phi_o$ is the geopotential on the Earth's equator:
\begin{equation}\label{geopotential}
\phi_o = -\frac{GM}{R_e}(1+\frac{1}{2}J_2) -\frac{1}{2}\Omega^2
R^2_e
\end{equation}
 The
dimensionless magnitude of this term is $ \phi_o/c^2 = -6.96928
\times 10 ^{-10}$.  Using the transformation in
Eq.~(\ref{NewCoordTime}), the metric becomes
\begin{widetext}
\begin{eqnarray}
-ds^2  & =  &   -\left[ 1 + \frac{2}{c^2} \left(  \phi - \phi_o
\right) \right] (dy^0)^2  + \left( 1 - \frac{2 V + \phi_o}{c^2}
\right)
2\frac{\Omega}{c} (y^1 dy^2 - y^2 dy^1) dy^0   \nonumber  \\
    &   &  + (1-\frac{2 V}{c^2} ) \left[  (dy^1)^2 + (dy^2)^2 +
(dx^3)^2 \right]   \label{RotatingCoordMetric2}
\end{eqnarray}
\end{widetext}
Eq.~(\ref{RotatingCoordMetric2}) gives the space-time metric in
ECEF rotating coordinates $y^i$.  Note that an ideal clock that is
stationary in EFEC coordinates (with $dy^\alpha=0$), has proper
time
\begin{equation}
d\tau=ds/c= \frac{1}{c} \left[ 1 + \frac{2}{c^2} \left(  \phi -
\phi_o \right) \right]^{1/2} \, dy^0
\end{equation}
When this clock is located on the geoid, then $ \phi = \phi_o$,
and $d\tau =dy^0/c$, so this ideal clock keeps coordinate time.
Hence a good hardware clock on the geoid can be used as a standard
to keep coordinate time, $x^0$, in the rotating ECEF coordinates.
Note that by Eq.(\ref{CoordinateTransformation1}) the coordinate
time in rotating ECEF coordinates is the same as coordinate time
in ECI coordinates, so this same clock keeps coordinate time,
$x^0$, in the ECI frame coordinates $x^i$.

Using the coordinate time transformation in
Eq.~(\ref{NewCoordTime}), the metric in the ECI coordinates, given
in Eq.~(\ref{EarthMetric}), becomes
\begin{widetext}
\begin{equation}\label{EarthMetricECI}
-ds^2=g_{ij}dx^i dx^j = -\left[ 1 + \frac{2}{c^2} \left( V -
\phi_o  \right) \right] (dx^0)^2 + \left( 1-\frac{2}{c^2} V
\right) \left[ (dx^1)^2 + (dx^2)^2 + (dx^3)^2 \right]
\end{equation}
\end{widetext}
Equation~(\ref{EarthMetricECI}) gives the metric in ECI
coordinates.  The coordinate time that enters into the metric,
$x^0$, is the time kept by ideal clocks on the geoid. This result
was the singular goal of the time transformation given in
Eq.~(\ref{NewCoordTime}).

Note however,  that in the ECI metric in
Eq.~(\ref{EarthMetricECI}), the proper time interval $ds$ on a
stationary clock in ECI coordinates (with $dx^\alpha=0$), is not
equal to coordinate time interval $dx^0$ because in general $V\ne
\phi_o$.

The ECI coordinate metric, given in Eq.~(\ref{EarthMetricECI}),
will be used below as the basis for comparing satellite clocks in
different orbits.  The time kept by ideal satellite clocks (proper
time) will be compared to the global coordinate time $x^0$, which
is the time kept by ideal clocks on the geoid.

\subsection{Integration of Geodesic Equations} In order to
compare the proper time on a satellite clock to the coordinate
time in ECI metric given in Eq.~(\ref{EarthMetricECI}), the
satellite world line must be known.   The satellites mass is
negligible compared to that of the Earth, so the satellite is
essentially a freely falling test particle that follows a geodesic
in 4-dimensional space-time.  The geodesic equations are given by
\begin{equation}\label{geodesicEq1}
  \frac{d^2x^i}{ds^2} + \Gamma^i_{jk} \frac{dx^j}{ds}
  \frac{dx^k}{ds}=0
\end{equation}
where the coordinates are given by  $x^i=(x^0,x^\alpha)$,
$\alpha=1,2,3$.  The four velocity can be written as
\begin{equation}\label{4-vel}
  {\rm v}^i(s)=
  \frac{dx^i(s)}{ds}=\frac{dx^0}{ds}(1,\frac{dx^\alpha}{dx^0})\equiv
  \frac{dx^0}{ds} v^i
\end{equation}
where $v^i = (1, dx^\alpha/dx^0)= (1,dx^\alpha /d(c t))$.  The
4-velocity is normalized,
\begin{equation}\label{4VelNorm}
-1 = \left( \frac{dx^0}{ds} \right)^2  \, \left[ -\alpha + \beta
\delta_{\kappa \gamma} v^\kappa v^\gamma \right]
\end{equation}
where $\alpha$ and $\beta$ are functions of $s$.

The four geodesic differential Eq.~(\ref{geodesicEq1}) must be
supplemented by initial conditions: initial 4-velocity ${\rm
v}^i(s=0)$ and an initial 4-dimensional position $x^i(s=0)$. These
initial conditions are taken from the classical Newtonian orbital
mechanics in 3-dimensional Euclidean space.

Now introduce Cartesian perifocal coordinates $(z^0,z^1,z^2,z^3)$,
defined so the origin coincides with one focus and the $+z^1$-axis
contains the periapsis, and the $z^1$--$z^2$ plane contains the
elliptical orbit~\cite{BateEtAl1971}.  In these perifocal
coordinates $z^i$, the Cartesian coordinate position ${\bf r}$ and
velocity ${\bf v}$ are given by
\begin{eqnarray}
{\bf r} & = & r \cos \nu {\bf P} + r \sin \nu {\bf Q}   \label{PosVel1} \\
{\bf v} & = &  \left(\frac{GM}{a (1-e^2)}\right)^{1/2} \left[
-\sin \nu {\bf P} + (e +\cos \nu) {\bf Q} \right] \label{PosVel2}
\end{eqnarray}
where  $\nu$ is the true anomaly, $a$ is the semimajor axis of the
orbit, and $e$ is the orbital eccentricity. In
Eq.~(\ref{PosVel1})--(\ref{PosVel2}), ${\bf P}$ and ${\bf Q}$ are
three-dimensional orthogonal unit basis vectors along the $z^1$-
and $z^2$-axis. In perifocal coordinates, the radial coordinate of
the orbit is given by
\begin{equation}\label{radialCoord}
  r=\frac{a(1-e^2)}{1+e \cos \nu}
\end{equation}
For the initial conditions, take $\nu=0$, so the Cartesian
position and velocity components are given by
\begin{eqnarray}
{\bf r}(0) & = & (z^1(0),z^2(0),z^3(0)) = (a(1-e),0,0)    \label{PosVelInit1} \\
{\bf v}(0) &  = & \left(\frac{dz^1}{dz^0}
,\frac{dz^2}{dz^0},\frac{dz^3}{dz^0} \right)_{s=z^0 =0}  \nonumber  \\
        &  =   &  \left(0,
\left( \frac{GM}{a c^2} \frac{1+e}{1-e}\right)^{1/2}, 0 \right)
\equiv \frac{dz^\beta(0)}{dz^0} \label{PosVelInit2}
\end{eqnarray}
Note that in Eq.~(\ref{PosVelInit2}) I have taken the origin of
coordinate time $z^0 = s =0$, when the satellite is at the
periapsis. This is an arbitrary choice made for convenience, and
does not affect the results computed over one orbital period.

The transformation between the perifocal coordinates, $z^i$, and
the geocentric equatorial coordinates, $x^\alpha$, is given by
\begin{eqnarray}
x^0 & = & z^0   \label{PeriToGeoTransformation0} \\
x^\alpha & = & d^\alpha_{~\beta}  \, z^\beta,    \,\,\,\,
\alpha,\beta=1,2,3  \label{PeriToGeoTransformation1}
\end{eqnarray}
where the orthogonal transformation matrix is given by
\begin{widetext}
\begin{equation}\label{TransMatrix}
d^\alpha_{~\beta} = \left(
\begin{array}{ccc}
\cos (\Omega )\,\cos (\omega ) - \sin \Omega \, \sin \omega \,
\cos i & -\cos\Omega \, \sin\omega -\sin\Omega\,\cos\omega\,\cos i
&
\sin\Omega \, \sin i \\
\sin\Omega \, \cos\omega +\cos\Omega \sin\omega \,\cos i &
-\sin\Omega \, \sin\omega +\cos\Omega \cos\omega \,\cos i &
-\cos\Omega\,\sin i \\
\sin\omega \, \sin i & \cos \omega \, \sin i & \cos i \
\end{array}
\right)
\end{equation}
\end{widetext}

The 4-dimensional initial conditions are taken as
\begin{eqnarray}
x^0(0)      & = & 0             \label{4-dInitialCond1}   \\
x^\alpha(0) & = & \, d^\alpha_{~\beta}  \,\, z^\beta(0)  \label{4-dInitialCond2} \\
v^0(0)      & = & 1  \label{4-dInitialCond3} \\
v^\alpha(0)    & = &     d^\alpha_{~\beta} \,\,
\frac{dz^\beta(0)}{dz^0}  \label{4-dInitialCond4}
\end{eqnarray}
Using the metric in Eq.~(\ref{EarthMetricECI}), and the
normalization of the 4-velocity
\begin{equation}\label{4-VNorm}
-1 = g_{ij} {\rm v}^i {\rm v}^j
\end{equation}
I find that
\begin{equation}\label{ddx0Rel}
\frac{dx^0}{ds}= \left[ \alpha - \beta \delta_{\kappa
\gamma}v^\kappa v^\gamma \right]^{-1/2}
\end{equation}
where
\begin{eqnarray}
\alpha  & = & 1+\frac{2}{c^2}(V-\phi_o) \equiv  1 + {\tilde \delta} \label{alpDef} \\
\beta  & = & 1 - \frac{2}{c^2} V  \equiv 1-\delta \label{betaDef}
\end{eqnarray}
and both $\alpha$ and $\beta$ are functions of $s$ through the
coordinates $x^\alpha(s)$.

The 4-velocity can then be expressed as
\begin{equation}   \label{4-VelDefFinal}
{\rm v}^i(s)= \left[ \alpha - \beta \delta_{\kappa \gamma}v^\kappa
v^\gamma \right]^{-1/2} (1,v^\alpha)
\end{equation}
So the initial condition on the 4-velocity become
\begin{equation}\label{4-VelInitialCondition}
{\rm v}^i(0)= \{  \left[ \alpha - \beta \delta_{\kappa
\gamma}v^\kappa v^\gamma \right]^{-1/2} \}_{s=0}
\left(1,v^\alpha(0) \right)
\end{equation}
where $v^\alpha(0)$ is given by Eq.~(\ref{4-dInitialCond4}). In
all formulas, Greek indices take values $\alpha=1,2,3$, and Latin
indices $i=0,1,2,3$.

Using Eq.~(\ref{4VelNorm}) and (\ref{4-vel}), the rate of change
of satellite proper time with coordinate time can be written in
terms of the 4-velocity vector components as
\begin{equation}\label{ProperCoordinateTimeRate}
  \frac{ds}{dx^0} = \left[ \frac{\alpha}{1+\beta \delta_{\kappa \gamma } {\rm v}^\kappa {\rm v}^\gamma}
  \right]^{1/2}, \,\,  \,\,\,\, \kappa,\gamma=1,2,3
\end{equation}

Equations~(\ref{geodesicEq1}) can be integrated using the initial
conditions given in
Eq.~(\ref{4-dInitialCond1})--(\ref{4-dInitialCond4}). The
satellite orbital elements enter the differential equations
through Eq.~(\ref{PosVelInit1})--(\ref{PosVelInit2}) and
Eq.~(\ref{TransMatrix}).

\begin{figure}
\includegraphics{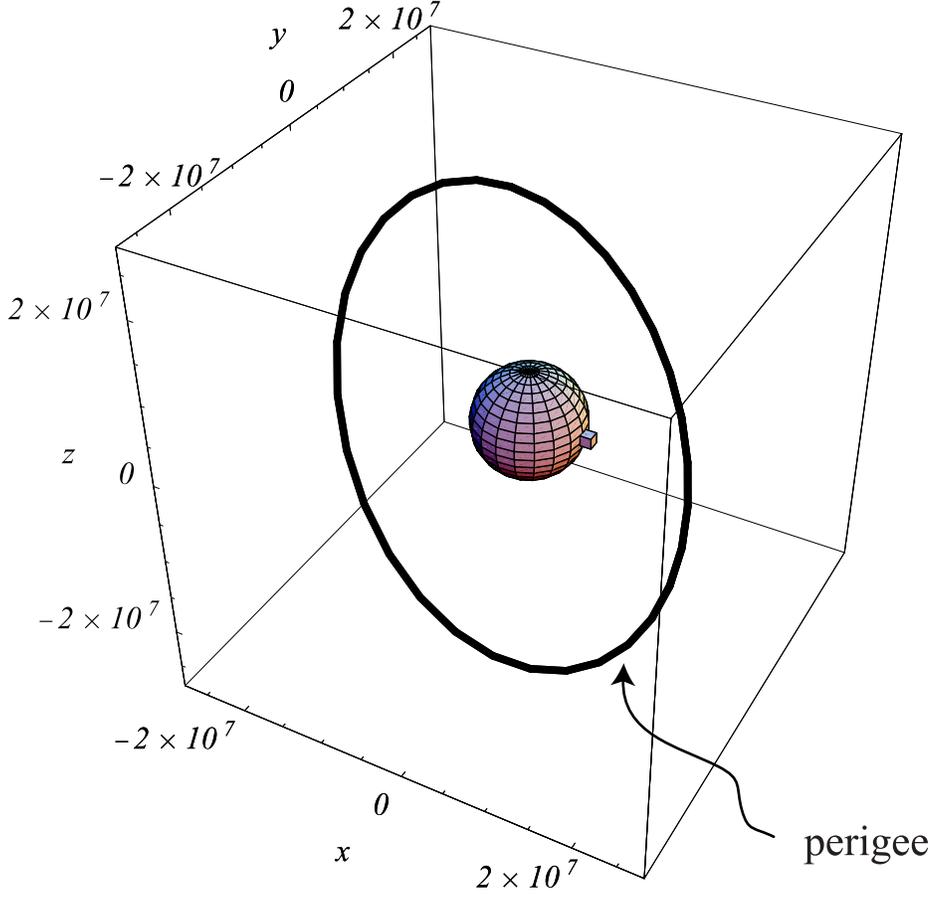}
\caption{\label{fig:Doppler3DGraphic} The orbit orientation is
shown in a real-space 3-dimensional view at time $x^0 = t =0$,
with perigee placed at $z<0$ and at a maximum value of the x-axis
of the orbit. (The perigee is in the lower right side of the
figure.) At time $x^0 = t =0$, the receiver (shown as cube) is on
the equator on the surface of the Earth at $(x,y,z)=(R,0,0)$,
where $R$ is Earth's equatorial radius.}
\end{figure}

The goal is to compute the difference between satellite proper
time and coordinate time.  We also compute the frequency shift
(gravitational plus Doppler shift) as observed on Earth, from a
satellite transmitting at a known frequency.   The absolute
orientation of the satellite orbit is not of interest here.
Therefore, we set the longitude of the ascending node
$\Omega=\pi/2$ and the argument of periapsis $\omega=3 \pi/2$.
Setting these two orbital parameters in this way puts the apogee
on the negative $x-$axis, on the $z>0$ half-space. The perigee is
on the negative $x-$axis, at $z<0$.  See
Figure~\ref{fig:Doppler3DGraphic}.

\subsection{Proper Time Minus Coordinate Time for
Various Orbits}

The difference between proper time $\tau$  kept by the satellite
clock, and coordinate time $\Delta t= \Delta x^0/c$, kept by
clocks on the geoid (in metric given by
Eq.~(\ref{EarthMetricECI})) is given by
\begin{equation}\label{TimeDiff}
\Delta \tau(s) - \Delta t(s)  = \frac{1}{c} \int_{s_1}^{s} (1 -
\frac{dx^0}{ds}) ds
\end{equation}
where  $dx^0 / ds$ is given by
Eq.~(\ref{ProperCoordinateTimeRate}). Equation~(\ref{TimeDiff})
gives the time difference, $\Delta \tau(s) - \Delta t(s)$, at a
given value of proper time measured from some epoch, or starting
time.  We choose the arbitrary starting time $s_1 =c \tau = c t =
x^0 =0$ to be when the satellite is at periapsis.

The gravitational field of the Earth is such that the satellite
orbit does not close on itself. In the integration of the
differential equations for this model, we have arbitrarily chosen
$s=s_1=0$ as the point at which a satellite is at periapsis. We
take the orbital period to be the value of $s=s_p$ at which the
satellite is at its closest approach, as defined by the minimum of
the Euclidean distance squared $d^2(s)$:
\begin{equation}\label{EuclidDist}
d^2(s) = \sum_{\alpha=1}^{3}  \left( x^\alpha(s) - x^\alpha(s_1)
\right)^2
\end{equation}

For satellites orbiting Earth, we have the small the quantities
$\nu<<1$, $\delta<<1$ and ${\tilde \delta}<<1$, and the integrand
in Eq.~(\ref{TimeDiff}) can be approximated by
\begin{equation}\label{integrand1}
1 - \frac{dx^0}{ds} = \frac{1}{2} \left( {\tilde \delta} - \nu^2
\right) - \frac{1}{8} \left[ 3 {\tilde \delta}^2 - \nu^4 - 4 \nu^2
\delta - 2 \nu^2 {\tilde \delta} \right] +O(5)
\end{equation}
where  we have dropped terms that we call $O(5)$.  Terms of
$O(1)\sim10^{-5}$, and so  $\nu\sim 10^{-5} \sim O(1)$, $\delta
\sim 10^{-10} \sim O(2)$, ${\tilde \delta} \sim 10^{-10} \sim
O(2)$, for typical Earth orbiting satellites. Terms in the
integrand in Eq.~(\ref{TimeDiff})  have been dropped that are of
order 10$^{-25}$. For a satellite with an orbital period that is
as large as one day (86400 sec), the error in dropping these terms
is on the order of 10$^{-20}$ sec. So our results, within this
model, are accurate to 10$^{-20}$ sec per day.

\subsection{Proper Time Minus Coordinate Time: Numerical Results for Various Orbits}

A computer program was written in the Mathematica programming
language to solve the general relativistic geodesic equations of
motion given in Eq.~(\ref{geodesicEq1}), subject to the initial
conditions on the 4-velocity in Eq.~(\ref{4-VelInitialCondition}).
The purpose of this program is to illustrate the relativistic
effects on satellites in various orbital regimes, and not to
predict accurate orbits that include all satellite perturbations.
Consequently, the effects of atmospheric drag and solar pressure
have been neglected, but in the future these effects could be
included. The computer program takes as input the satellite orbit
semimajor axis $a$, eccentricity $e$, and inclination $i$, as well
as the gravitational constant times the Earth mass, $GM$, and the
value of the earth's quadrupole moment, $J_2$. The values of these
parameters are shown in Table~\ref{tab:NumericalValues}.
\begin{table*}
\caption{\label{tab:NumericalValues}Numerical value of the
parameters used in the computer program.}
\begin{ruledtabular}
\begin{tabular}{ccccc}
$c$  (m/s)                 &  $GM$ (m$^3$/sec$^2$)      &   $J_2$                     &  $R$ (m) &  Earth Rotation Rate $\Omega$ (radian/s)   \\
\hline
2.99792458$\times$10$^8$   &  3.986005$\times$10$^{14}$  &  1.0826800$\times$10$^{-3}$ & 6.378137$\times$10${^6}$   &  7.2921151467$\times$10$^{-5}$ \\
\end{tabular}
\end{ruledtabular}
\end{table*}

As mentioned above, the orbit is arbitrarily taken to have the
longitude of the ascending node $\Omega=\pi/2$ and the argument of
periapsis $\omega=3 \pi/2$. Choosing these parameters in this way
puts the apogee on the negative $x-$axis, on the $z>0$ half-plane,
and the perigee on the negative $x-$axis, at $z<0$. The computer
program solves the geodesic equations of motion numerically for
the (equatorial geocentric) coordinates $x^i(s)$, $i=0,1,2,3$, as
a function of the proper time $s$ along the orbit. The equations
of motion are integrated from $s=0$ at perigee, to $s=s_p$ at the
closest approach to the initial position (approximately one
orbital period) which defines the orbital period.  At perigee,
both the coordinate time and proper time are arbitrarily taken to
be equal, and are set to zero.  As the satellite  progresses in
its orbit beyond perigee, proper time deviates from coordinate
time, due to relativistic time dilation and gravitational
potential effects.

In order to illustrate the numerical values of the effects
discussed above, we use a representative satellite from each of
four orbital regimes: low-Earth-orbit (LEO), geostationary orbit
(GEO), Global Positioning System (GPS) satellite orbits, and
highly-elliptical orbit (HEO). The numerical values used in the
computer program are shown in Table~\ref{tab:SatelliteParameters}.
These values were obtained from CelesTrak WWW on the the World
Wide Web from web site: {\rm http://celestrak.com}.
\begin{table*}
\caption{\label{tab:SatelliteParameters}For satellites with given
orbital parameters, semimajor axis $a$, eccentricity $e$,
inclination $i$, and value of Earth's quadrupole moment $J_2$, the
computed values of proper time minus coordinate time, $\Delta \tau
- \Delta t$, are shown per orbital period, and per day. The value
per day is the average of this difference over one Earth solar
day.}
\begin{ruledtabular}
\begin{tabular}{cccccccc}
satellite & semimajor axis & eccentricity &  inclination & $J_2$                 & $\Delta t$ period & $\Delta \tau - \Delta t$   &  $\Delta \tau - \Delta t$    \\
          &  (meters)      &              &    (degrees) &                         &     (minutes)      &  $\mu$s per period         &   $\mu$s per day             \\ \hline
LEO\footnote{COSMOS 2366 Russian military spacecraft}   &    7.3635$\times 10^6$  &  0.00292   &   82.9    &1.0826800$\times 10^{-3}$ &  105.12         &     -1.290509              &      -17.678433 \\
LEO\footnote{Same as above LEO but taking $J_2=0$.}    &     7.3635$\times 10^6$ &   0.00292    &  82.9        & 0.0                   & 104.81          &    -1.301039                &    -17.875853    \\
GEO\footnote{Nominal values for a GEO satellite.}                      & 4.2164174 $\times 10^7$ & 0.0 & 0.0  & 1.0826800$\times 10^{-3}$ &  1435.96    & 46.4512489   &  46.5818860   \\
GEO\footnote{Same as GEO above, but with $J_2=0$.} & 4.2164174 $\times 10^7$ & 0.0 & 0.0  & 0.0                      &   1436.0     & 46.4230537   &  46.5501514  \\
HEO\footnote{Highly Elliptical Orbit:1998-054 Molniya 1-91} & 2.70365$\times10^7$   &  0.747194      &62.8&   1.0826800$\times 10^{-3}$&   743.08      &  20.1582623  &    39.0644760  \\
HEO\footnote{Same as above HEO but taking $J_2=0$.}         & 2.70365$\times10^7$   &  0.747194      &62.8&   0.0                      &   737.37       & 19.9308525  &    38.9226991 \\
GPS\footnote{International Designator 1996-041}    & 2.66965$\times 10^7$&  0.0017418 &   55.03     &1.0826800$\times 10^{-3}$&   723.573310      &     19.438916       &        38.6858366      \\
GPS\footnote{Same as GPS above, but with $J_2$=0}  &  2.66965$\times 10^7$ & 0.0017418    &   55.03     &   0.0               &   723.504421        &   19.420036       &        38.6519441       \\
\end{tabular}
\end{ruledtabular}
\end{table*}
We compute the difference between elapsed proper time and
coordinate time, $\Delta \tau - \Delta t$, where $\Delta \tau$ is
the elapsed proper time, and $\Delta t$ is the elapsed coordinate
time, at which the satellite is at perigee.  We arbitrarily choose
the coordinate time and proper time to be zero when the satellite
is at perigee. Then, up to an additive constant, the coordinate
time, $t= x^0/c$, where $x^0$ appears in the metric in
Eq.~(\ref{EarthMetricECI}), is equal to the proper time kept by a
reference clock on the geoid, such as the Master Clock at the U.S.
Naval Observatory in Washington, D.C.

For each satellite, Table~\ref{tab:SatelliteParameters} shows the
the computed values of the proper time minus coordinate time
difference, $\Delta \tau - \Delta t$, from perigee to
(approximate) perigee, accumulated over one orbital period, as
determined by solving the geodesic equations of motion, and
evaluating Eq.~(\ref{TimeDiff}). The periods of the satellites
vary widely, so to compare the magnitude of the effects, the last
column in Table~\ref{tab:SatelliteParameters} shows the difference
$\Delta \tau - \Delta t$  integrated over one solar day. This
value is simply taken to be the average of $\Delta \tau - \Delta
t$ over one satellite period, multiplied by the ratio (24
hour)/(satellite period). The calculations in the
Table~\ref{tab:SatelliteParameters} are done for a finite value of
Earth quadrupole $J_2$, and for $J_2=0$. Comparing the time
difference $\Delta \tau - \Delta t$ for $J_2 = 0$ and for finite
$J_2$ gives an estimate of the size of the effect of the Earth's
quadrupole on proper time kept by a satellite clock. The coupling
to the quadrupole terms is larger for lower altitudes (since the
terms proportional to $J_2$ vary as $1/r^3$) and depends on the
inclination of the orbit.

For the low-Earth-Orbit (LEO) satellite, the apogee is at altitude
of 1007 km and perigee at 964 km. The difference between proper
time and coordinate time, $\Delta \tau - \Delta t = -17.678433$
$\mu$s per day, which means that less proper time $\tau$ (aboard
the satellite) elapses over one solar day than coordinate time $t$
on the geoid . This is due to the predominant effect of time
dilation resulting from the high orbital speed of a LEO satellite.
Also shown in Table~\ref{tab:SatelliteParameters} is the effect of
the non-zero value of the Earth's quadrupole moment, $J_2$.  For
the LEO satellite, the finite value of $J_2$ has the effect of
changing the difference $\Delta \tau - \Delta t$ by 197.42 ns.

For high altitude GEO satellite, the difference between proper
time and coordinate time, $\Delta \tau - \Delta t$, is of opposite
sign to that of a LEO satellite.  For a GEO satellite, the proper
time elapses faster than coordinate time by 46.5818860 $\mu$s per
day. The gravitational shift  (causing proper time to run fast)
dominates the time dilation (which causes proper time to run
slow), due to the satellites high altitude and relative slower
speed. Also,  the GEO satellite is less influenced by the finite
size of Earth's quadrupole moment $J_2$, which is only 31.7346 ns
per day, as compared to 197.42 ns for the LEO satellite.
Furthermore, for a GEO satellite, the contribution to $\Delta \tau
- \Delta t$ from the $J_2$ quadrupole effect is opposite to that
for a LEO satellite.

\begin{figure*}
\includegraphics{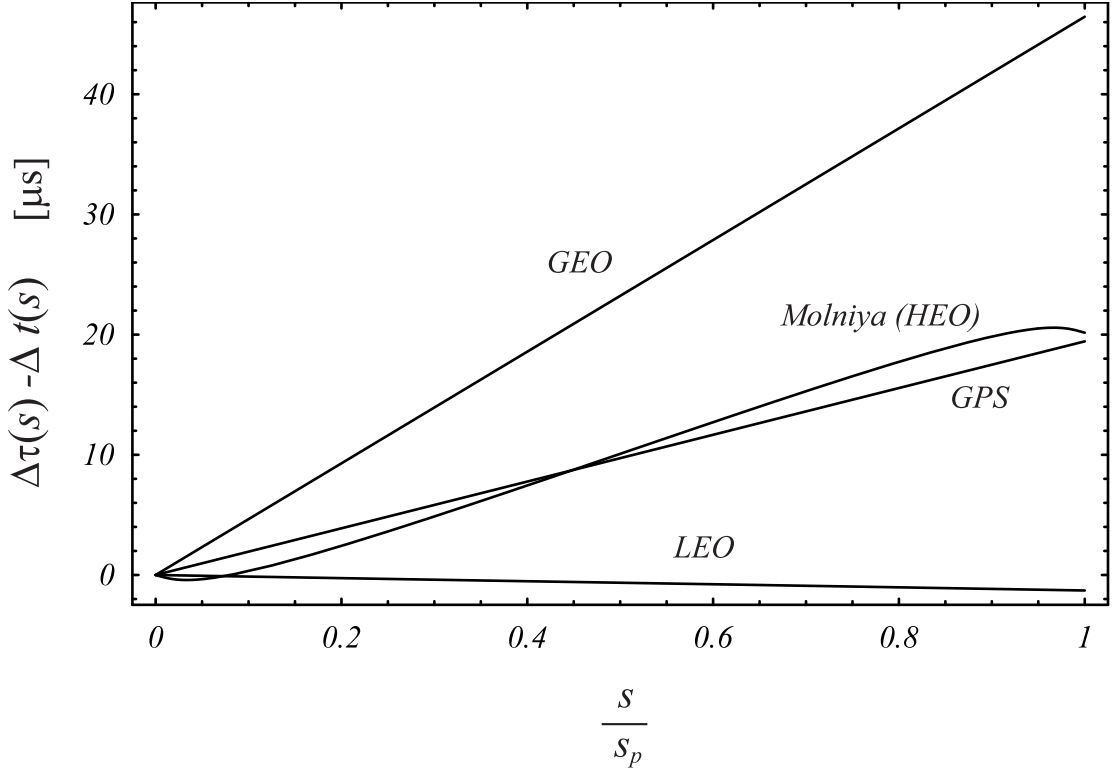}
\caption{\label{fig:TimeDiffPlot} The proper time minus the
coordinate time, $\Delta \tau(s) - \Delta t(s)$  is shown for the
LEO, GEO, HEO (Molniya) and GPS satellites vs.
$s/s_p=\tau/\tau_p$, the fraction of proper time period along the
orbit.  On the abscissa, to the accuracy of the plot, proper time
difference is approximately equal to coordinate time difference,
$\Delta \tau \sim  \Delta t$, where t is coordinate time measured
in seconds.}
\end{figure*}

For the (high-altitude) GPS satellite, apogee is at 20365 km
altitude and perigee is at 20272 km altitude. The predominant
effect is due to change of gravitational potential, which leads to
more elapsed proper time than coordinate time, by 38.6858366
$\mu$s per day. Since the altitude of GPS satellites is lower than
GEO satellites, the increase of proper time  is smaller. For GPS,
the effect of nonzero value of $J_2$ is also similar to the GEO
satellite, but has the larger value  33.8925 ns, due to the lower
altitude which leads to a greater influence of $J_2$

The LEO, GEO, and GPS satellites have nearly circular orbits. In
contrast, the HEO satellite has an eccentricity of 0.747194, and
consequently a perigee of 6835 km (457 km altitude) and apogee
47238 km (40,860 km altitude).  This places the HEO satellite both
lower and higher than the nearly circular orbit LEO satellite
during different portions of its orbit.  The altitude of the HEO
satellite at perigee and apogee is also lower and higher than the
GPS satellite altitude. Consequently, the HEO satellite has proper
time running more slowly than coordinate time when it is at low
altitude (like the LEO satellite) and it has proper time running
faster than coordinate time when its is at high altitude (like GEO
and GPS satellites). The net effect for the HEO satellite is that
39.0644760 $\mu$s per day more proper time has elapsed than
coordinate time.  This size of this net effect is similar to the
GPS satellite because the HEO satellite spends little time at the
low altitude portion of its orbit.  The effect of $J_2$ for the
HEO satellite is 141.777 ns, which is 4 times the same effect for
the GEO satellite, and 0.7 times this effect for the LEO
satellite.

The  results described above are the integrated value of proper
time minus coordinate time, over one complete orbital period. It
is useful to plot the value of proper time minus coordinate time
as a function of the fraction of the orbital period covered,
$s/s_p$, see Figure~\ref{fig:TimeDiffPlot}. This figure shows that
for each satellite, the proper time $\Delta \tau=s/c$ diverges
from the coordinate time $\Delta t=\Delta x^0/c$, at a different
rate, along its orbit. Note that time is scaled to the orbital
period along each orbit, so that $s/s_p=1$ occurs at the
respective period of each orbit. The plot in
Figure~\ref{fig:TimeDiffPlot} shows a geometrical comparison,
however, it does not permit a comparison of the effect in time.

\begin{figure*}
\includegraphics{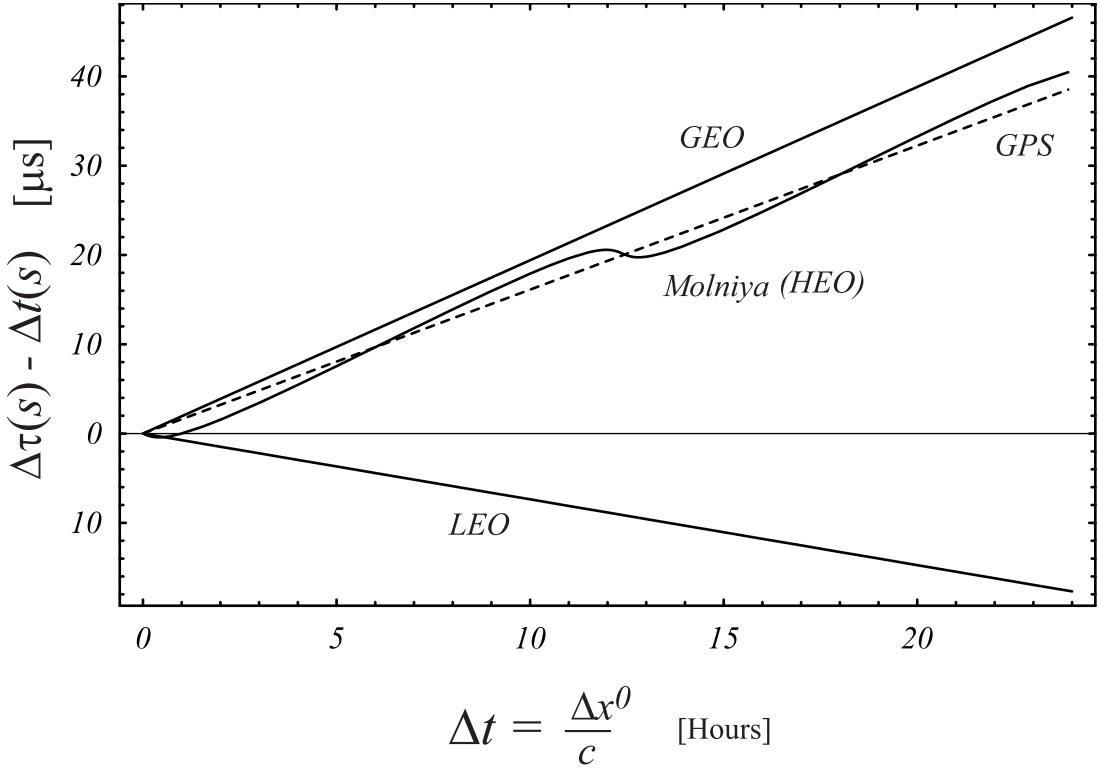}
\caption{\label{fig:TimeDiffPlot2} The proper time minus the
coordinate time difference, $\Delta  \tau(s) - \Delta t(s)$,  is
shown for the LEO, GEO, HEO (Molniya) and GPS satellites vs.
coordinate time $\Delta t$ in units of hours. Coordinate time
$\Delta t$ is equivalent to time kept by an ideal reference clock
on the Earth's geoid.}
\end{figure*}

For some practical purposes, it is preferable to look at the
proper time minus coordinate time difference, $\Delta \tau -
\Delta t$ as a function of coordinate time $\Delta t = \Delta x^0
/ c$. In this way, the effects are shown as they would appear in
real time, say with respect to the USNO Master Clock.
Figure~\ref{fig:TimeDiffPlot2} shows $  \Delta \tau - \Delta t$
vs. $\Delta t$ for the LEO, GEO, HEO (Molniya) and GPS satellites.
For each satellite, on the scale of this graph, the plots are
almost linear functions. The obvious exception is the HEO
(Molniya) orbit satellite, where, due to its high eccentricity,
the time difference is highly nonlinear. Actually, even for the
LEO, GEO, and GPS satellites, which have nearly circular orbits,
the rate of change of  $\Delta \tau(s) - \Delta t(s)$ with time
$t$ is non-constant, due to the interaction of the orbital
eccentricity, inclination, finite value of $J_2$, and general
relativistic precession effects. These effects are not visible on
this graph, but can be significant, depending on the desired level
of time synchronization that is required (see Table
\ref{tab:TimeingPrecision}) and the time interval over which the
synchronization must be maintained (stability).  The nonlinear
variation of proper time with satellite time can best be seen on a
plot of the frequency shift, see the next Subsection.

Since the rate at which proper time diverges from coordinate time
is nearly constant for the LEO, GEO, and GPS satellites, these
values could be used to compensate for the drift of proper time
away from coordinate time, by applying a ``factory offset" to the
satellite clock frequency, as has been done in the GPS.  This
could even be done after the orbit is established and the
compensation could even be changed periodically. However,
inspection of the time synchronization that is required for
various applications, see for example Table
\ref{tab:TimeingPrecision}, shows that for many applications this
compensation is not feasible, due to the non-constant rates (for
reasons mentioned above), as well as due to other forces (such as
solar pressure and atmospheric drag) that perturb the satellite
orbit and lead to proper time changes that cannot be modelled
sufficiently accurately. Instead, it is likely that for the most
accurate applications of time synchronization, the synchronization
will have to be redone periodically, according to the required
stability (accuracy required over given time interval).

\subsection{Observed Doppler and Gravitational Frequency Shift}

When a satellite clock is manufactured, it has a certain
prescribed hardware clock rate compared to coordinate time, in
some system of coordinates.  Alternatively, we can say that the
clock, or oscillator in the clock, is calibrated to a given
frequency $\omega_s$ when it is compared to a local frequency
standard.  The word local here means that during the calibration
procedure, the satellite clock is siting at rest with respect to
the frequency standard and that these two devices are at the same
gravitational potential (they are co-located and at rest relative
to each other). We then refer to the frequency $\omega_s$ as the
proper frequency of the satellite oscillator or clock.

\begin{figure}
\includegraphics{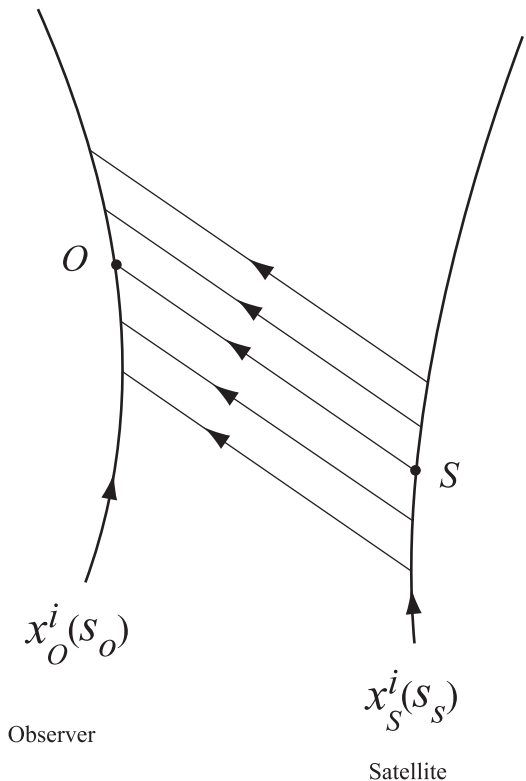}
\caption{\label{fig:ClockWorldLine} The world line of the
satellite (transmitter) and observer (receiver) are shown
schematically. During satellite proper time $ds_s$, the satellite
clock sends $N$ uniformly-spaced pulses, which are received during
observer proper time $ds_o$. A typical reception event $O$ is
connected to an emission event $S$ by a null geodesic.}
\end{figure}

When the satellite is launched, its clock is assumed to operate
normally.  In fact, if the frequency standard were placed aboard
the satellite, and a comparison was made between the satellite
clock and the frequency standard, we would expect to obtain
$\omega_s$ for the satellite oscillator frequency,  to within the
calibration accuracy. However, if the frequency standard is
located on the Earth's geoid, and the satellite clock sends out
electromagnetic pulses, then the frequency of the pulses observed
on the geoid (with respect to the frequency standard on Earth)
will be $\omega_o$, which is different than the proper frequency
of the satellite clock, $\omega_s$. This frequency shift is due to
the relative motion of satellite and receiver (Doppler effect) and
also due to the difference in gravitational potential between
satellite and receiver.  In what follows, we compute this effect.

Consider a satellite orbiting Earth on world line $x^i_s(s_s)$,
$i=0,1,2,3$, where the functions $x^i_s()$ specify the world line
of satellite $s$, and $s_s$ is the parameter related to proper
time along the world line, $\tau = s_s / c$. See
Figure~\ref{fig:ClockWorldLine}. Now consider an observer on a
different world line, $x^i_o(s_o)$.  The satellite broadcasts a
periodic electromagnetic signal, and within the geometric optics
approximation, we say that these signals travel along null
geodesics, so the reception event $O$ and transmission event $S$
are connected by a null geodesic. In terms of the world function
we have
\begin{equation}\label{null1}
\Omega(S,O)=0
\end{equation}
where the coordinates of the events are given by
\begin{eqnarray}
S & = & (x_s^0,x_s^\alpha) \label{Sevent} \\
O & = & (x_o^0,x_o^\alpha), \,\,\,\, \alpha=1,2,3  \label{Oevent}
\end{eqnarray}

We compute the frequency measured by the observer, in terms of the
frequency transmitted by the satellite. Prior to the measurement,
the satellite and observer are assumed to have calibrated their
equipment, when they were on a common world line (before launch),
as discussed above. We define
\begin{eqnarray}
\omega_o & = &  \mbox{frequency measured by observer at reception event}~O  \label{om1} \\
\omega_s & = &  \mbox{satellite proper frequency transmitted at event $S$,} \\
          &  &   \mbox{as measured with respect to  frequency standard on board satellite} \label{om2}
\end{eqnarray}
During a proper time $d \tau = ds_o/c$ measured by the observer,
there are $N$ cycles of the radiation received. These cycles were
sent by the satellite during a proper time $d s_s/c$ , as measured
by the satellite clock.  Assuming that the signal does not pile up
anywhere, e.g., a static space-time, we can say that
\begin{equation} \label{dopplerN}
N= \omega_o \frac{ds_o}{c}=\omega_s \frac{ds_s}{c}
\end{equation}
Note that $d s_o$ is measured at event $O$ and that $ ds_s$ is
measured at event $S$ and that these two events are arbitrarily
separated (but connect by a null geodesic).  We can then relate
the world lines of the observer and satellite by
\begin{eqnarray}
d s_o^2 & = & -g_{ij}(O) \,  dx^i_o dx^j_o \label{dxo} \\
d s_s^2 & = & -g_{ij}(S) \, dx^i_s dx^j_s \label{dxs}
\end{eqnarray}
where $dx^i_o$ and $dx^i_s$ are increments along the observer and
satellite world lines, and $g_{ij}(O)$ and $g_{ij}(S)$ is the
metric evaluated at events $O$ and $S$, respectively.  The ratio
of frequencies can then be written as
\begin{equation}\label{freqRatio2}
\frac{\omega_o}{\omega_s} = \frac{d s_s}{d s_o} = \left[
\frac{g_{ij}(S) \, \frac{dx^i_s}{d \lambda} \frac{dx^j_s}{d
\lambda}}{g_{ij}(O) \, \frac{dx^i_o}{d \lambda} \frac{dx^j_o}{d
\lambda}} \right]^{1/2}
\end{equation}
where $\lambda$ is a common parameter for the two world lines.
Note that the frequency ratio in Eq.~(\ref{freqRatio2}) depends on
two space-time events, $S$ and $O$, so this ratio is a two-point
tensor, see Section V and Section X-C.

We compute the observer's world line by taking the observer to be
stationary on the surface of the rotating Earth. Using ECEF
(Earth-centered Earth fixed) coordinates, $y^i$, $i=0,1,2,3$, the
spatial coordinates of the observer, $y^\alpha$, are constant. The
transformation from ECI Cartesian-like space-time coordinates
$x^i$, to ECEF Cartesian space-time coordinates $y^i$, is of the
form given in Eq.~(\ref{CoordinateTransformation1}), with $\omega
\rightarrow \Omega$, where $\Omega$ is the angular frequency of
rotation of the Earth, with respect to an ECI coordinate system
with a common $z$-axis. In the ECI coordinates $x^i$, taking the
observer to be located on the $y^1$-axis on the equator, the
observer's world line is then given by choosing $y^1=R$, $y^2=0$,
$y^3=0$, where $R$ is Earth's equatorial radius.

As a metric in the vicinity of the Earth, we choose
Eq.~(\ref{EarthMetric}), with $\bar{x}^0 \rightarrow x^0$ and we
neglect the Earth's quadrupole moment, taking $J_2=0$, so that
$V(r,\theta) \rightarrow \phi = -G M/r$. The world function for
this metric is given by:
\begin{equation}
\Omega(x^i_1,x^j_2) =  \frac{1}{2}\eta_{ij} \Delta x^i \Delta x^j
+ \frac{1}{2}\delta_{ij} \Delta x^i \Delta x^j \frac{2GM}{c^2} \,
\frac{1}{|{\bf x}_2 - {\bf x}_1|}
 \log \left( \frac{\tan(\frac{\theta_1}{2})}{\tan(\frac{\theta_2}{2})} \right)
\label{WorldFunction2}
\end{equation}
where $c \Delta t \equiv x^0_2 - x^0_1$, and  $\theta_1$ and
$\theta_2$ are defined by
\begin{equation}
\cos \theta_a = \frac{{\bf x}_a \cdot ( {\bf x}_2 -  {\bf x}_1 )
}{|{\bf x}_a| |{\bf x}_2 - {\bf x}_1|}, \;\;\;\; a=1,2
\label{cosineDef2}
\end{equation}

We choose the parameter $\lambda=x^0$ in Eq.~(\ref{freqRatio2}),
so the ratio of frequencies becomes
\begin{equation}\label{freqRatio3}
\frac{\omega_o}{\omega_s} = \left[ \frac{1+\delta_s -(1-\delta_s)
\delta_{\alpha \beta} v^\alpha_s v^\beta_s} {1+\delta_o
-(1-\delta_o) \delta_{\alpha \beta} v^\alpha_o v^\beta_o}
 \right]^{1/2}
\end{equation}
where
\begin{eqnarray}
\delta_o & = & \left[ \frac{2}{c^2} \phi \right]_o =  \left[ -\frac{2GM}{c^2 r}\right]_o  \label{opoint} \\
\delta_s & = & \left[ \frac{2}{c^2} \phi \right]_s = \left[ -\frac{2GM}{c^2 r}\right]_s \label{spoint} \\
v^2_o    &  = & \delta_{\alpha \beta} v^\alpha_o v^\beta_o  \label{ov}  \\
v^2_s    &  =  & \delta_{\alpha \beta} v^\alpha_s v^\beta_s  \label{sv} \\
v^\alpha_o & = & = \frac{d x^\alpha_o}{d x^0}  \label{voalpha}  \\
v^\alpha_s & = & = \frac{d x^\alpha_s}{d x^0}  \label{vsalpha}
\end{eqnarray}
and the subscripts $o$ and $s$ on the square brackets indicate
evaluation at events $O$ and $S$, respectively.

For typical Earth orbiting satellite applications, the
$\delta$-terms and velocity terms in Eq.~(\ref{freqRatio3})  are
all small:  $\delta_o \sim \delta_s \sim v^2_o \sim v^2_s \sim
O(2)$ where $O(2)\sim 10^{-10}$.

It is then convenient to look at the fractional frequency shift,
making the expansion
\begin{eqnarray}
\frac{\omega_o}{\omega_s} - 1 = \frac{\Delta \omega}{\omega_s} & =
& \frac{1}{2}(\delta_s - \delta_o + v^2_o -
v^2_s) -\frac{1}{8}(\delta_s - \delta_o + v^2_o - v^2_s)^{2}  \nonumber \\
   &   &  + \frac{1}{2} \left[ (\delta_o -v^2_o)^2 -\delta_o v^2_o +\delta_s v_s^2 - (\delta_s -v^2_s)(\delta_o-v_o^2)
\right]  + O(5)   \label{SatDopplerFinal}
\end{eqnarray}
where $O(5) \sim 10^{-25}$ for typical Earth-orbiting satellites.

As in Eq.~(\ref{geodesicEq1})--(\ref{ProperCoordinateTimeRate}),
we use the classical mechanical conditions to integrate the
geodesic equations, for the current metric. So the satellite world
line is known for the given initial conditions.   The satellite
world line is parametrized by coordinate time: $x_s^\alpha(x^0)$.
The observer's world line, $x^\beta_o(x^0)$, is also parametrized
by coordinate time.  We know the observer's world line because we
assume that the observer is located on the surface of the Earth on
the $y^1$ axis. The calculation is then done as follows. For a
given (emission event) satellite coordinate time, $x^0_s$, and
spatial position, $x_s^\alpha$, we compute the observer's
reception event coordinate time $x^0_o$ by solving:
\begin{equation}\label{null2}
\Omega(x^0_s,x^\alpha(x^0_s), x^0_o,x^\beta_o(x^0_o))=0
\end{equation}
The emission and reception event coordinates, in
Eq.~(\ref{Sevent}) and (\ref{Oevent}), are then known. These
coordinates are used in Eq.~(\ref{SatDopplerFinal}) to plot the
observed frequency shift, $\frac{\omega_o}{\omega_s} - 1$  as a
function of the reception event time $x^0_o$.

A computer program in the Mathematica programming language was
written to carry out the required computations.  The program does
not include the (relatively large) atmospheric signal propagation
delays, although these effects could be included in the future
work. Furthermore, no allowance is made for the Earth obscuring
the signal from the satellite to observer.

\subsection{Doppler plus Gravitational Frequency Shift: Numerical Results for Various Orbits}

\begin{figure}
\includegraphics{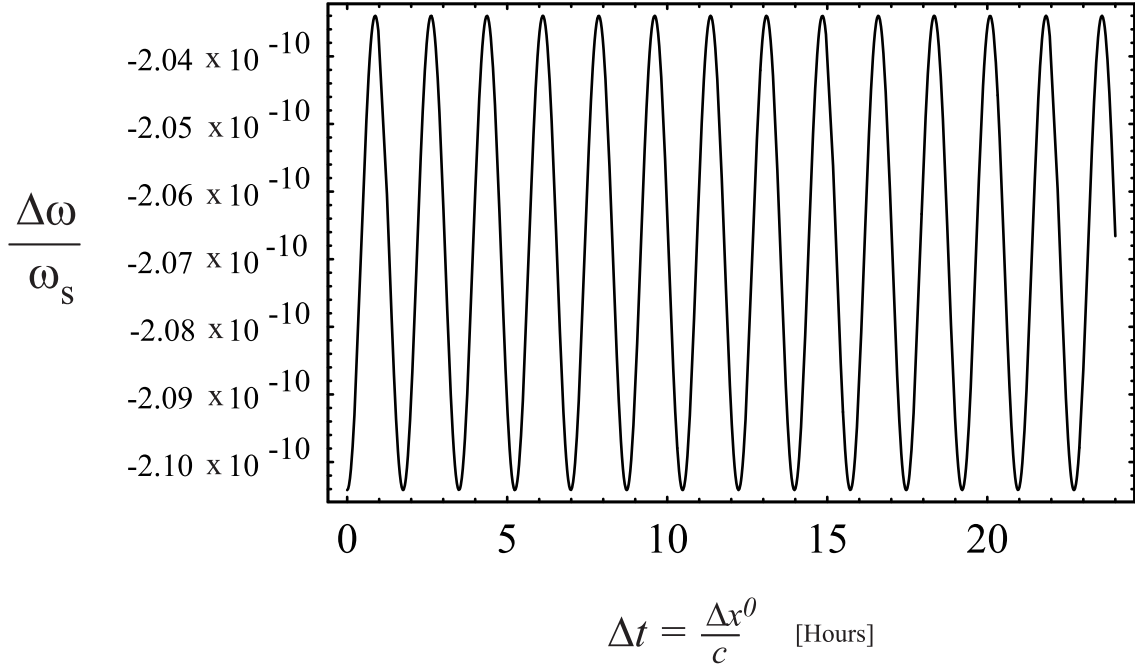}
\caption{\label{fig:FreqShift-LEO} The frequency shift (Doppler
plus gravitational) as observed from the geoid for the LEO
satellite is shown vs. coordinate time $\Delta t=\Delta x^0/c$ in
units of hours.}
\end{figure}

\begin{figure}
\includegraphics{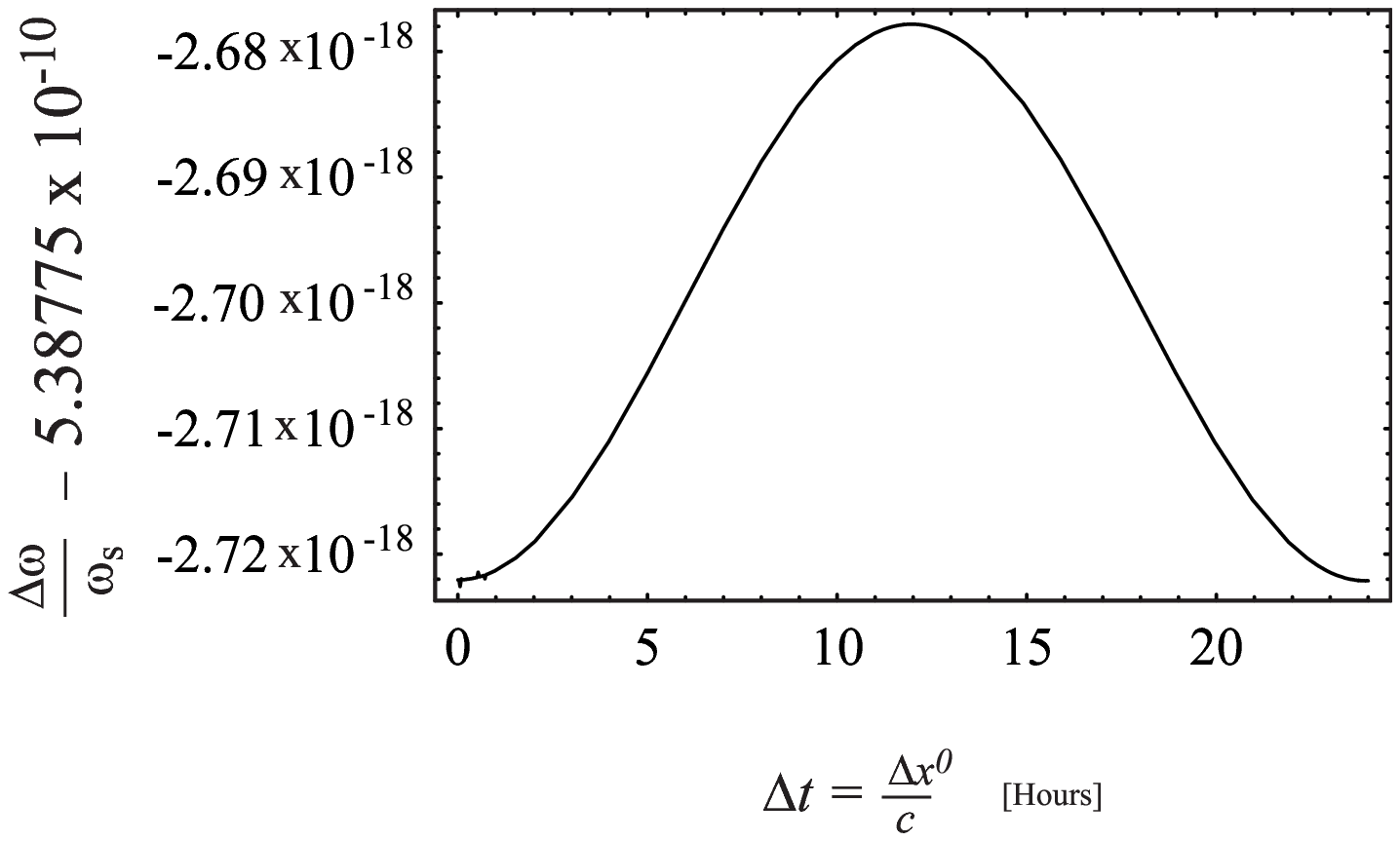}
\caption{\label{fig:FreqShift-GEO} The frequency shift (Doppler
plus gravitational) as observed from the geoid for the GEO
satellite is shown vs. coordinate time $\Delta t=\Delta x^0/c$ in
units of hours.}
\end{figure}

\begin{figure}
\includegraphics{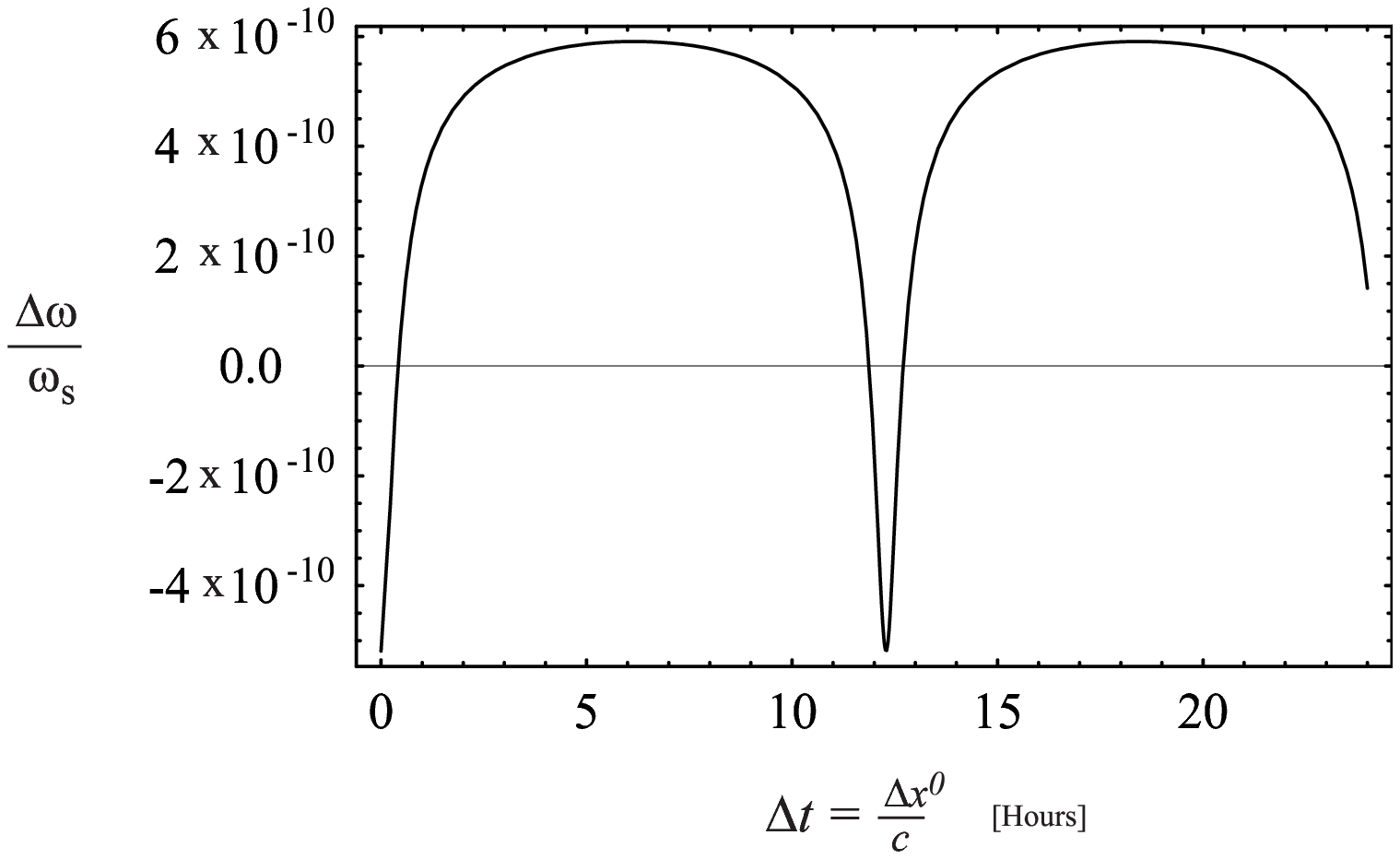}
\caption{\label{fig:FreqShift-HEO} The frequency shift (Doppler
plus gravitational) as observed from the geoid for the HEO
satellite is shown vs. coordinate time $\Delta t=\Delta x^0/c$ in
units of hours.}
\end{figure}

\begin{figure}
\includegraphics{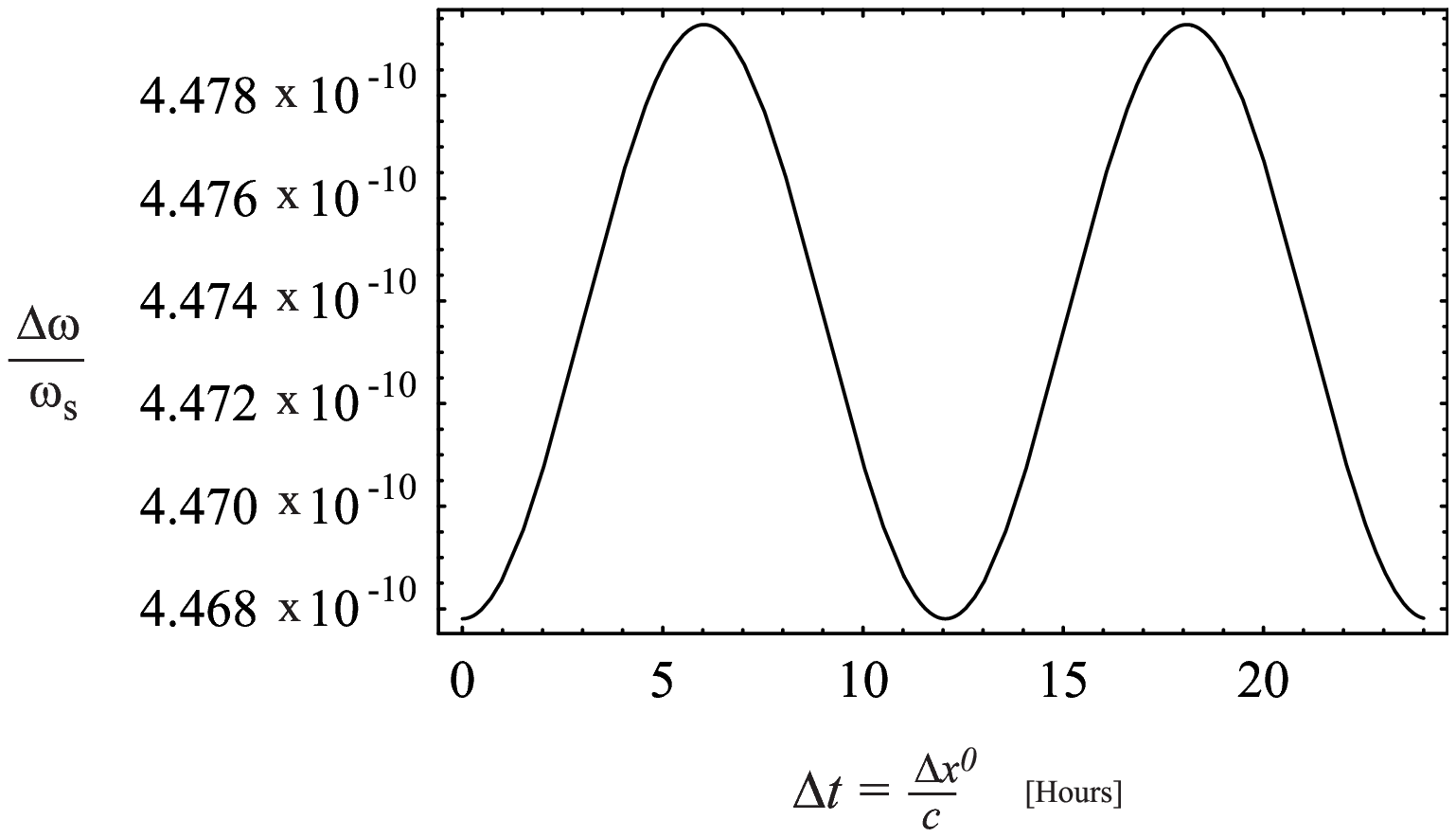}
\caption{\label{fig:FreqShift-GPS} The frequency shift (Doppler
plus gravitational) as observed from the geoid for the GPS
satellite is shown vs. coordinate time $\Delta t=\Delta x^0/c$ in
units of hours.}
\end{figure}

\begin{figure}
\includegraphics{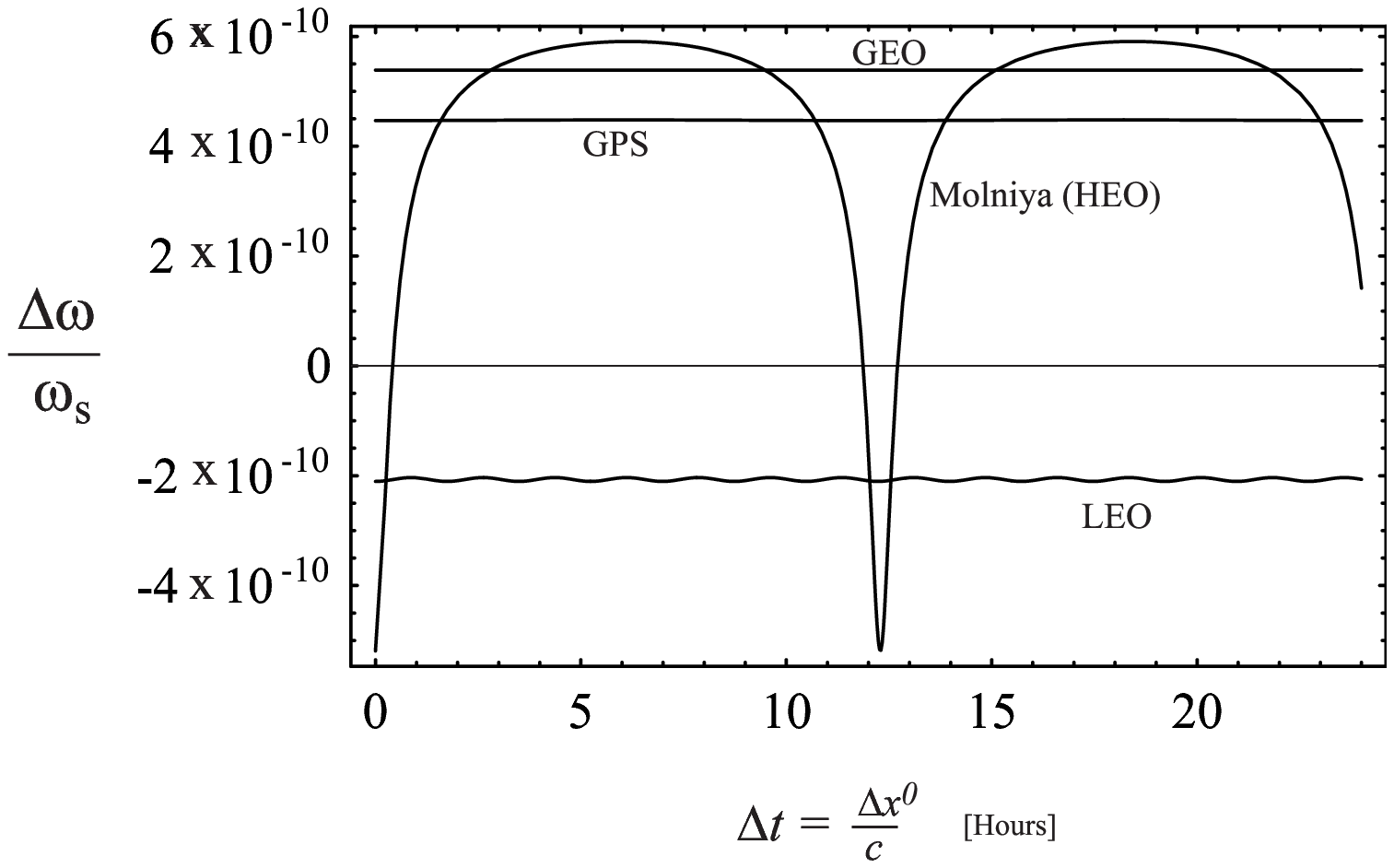}
\caption{\label{fig:FreqShift-all} The frequency shifts (Doppler
plus gravitational) as observed from the geoid is shown vs.
coordinate time $\Delta t=\Delta x^0/c$ in units of hours, for the
LEO, GEO, HEO (Molniya), and GPS satellites.}
\end{figure}

Proper time on board a satellite is not a linear function of
coordinate time. This nonlinear functional dependence can be seen
on a plot of the fractional frequency shift, $\Delta \omega
 / \omega_s$, where $\Delta \omega = \omega_o - \omega_s$, and $\omega_o$
is the satellite frequency as observed with respect to a reference
oscillator on the geoid, and $\omega_s$ is the proper frequency
emitted by the satellite--as determined by a frequency standard on
board the satellite. See Eq.~(\ref{SatDopplerFinal}) for the
definition of $\Delta \omega /\omega$ in terms of orbital
parameters.  (Here we assume that the satellite frequency standard
has been calibrated and that it has not been altered, as has been
done in GPS satellites by the ``factory offset".)   The plots in
Figures~\ref{fig:FreqShift-LEO}--\ref{fig:FreqShift-all} show the
fractional frequency shift, $\Delta \omega /\omega_s$ vs. $\Delta
t$, where $\Delta t$ is the elapsed coordinate time measured from
time of perigee, for the LEO, GEO, HEO, and GPS satellites. (At
perigee we have arbitrarily taken $\Delta t=0$.)

The frequency shifts, similar to those shown in
Figures~\ref{fig:FreqShift-LEO}--\ref{fig:FreqShift-all}, would be
observed at other locations than the Earth's geoid.  For example,
whenever the proper time is nearly equal to coordinate time,
$x^0$, approximately the same plot would be observed.  This is the
case for observers in laboratories located at altitude above the
geoid. Similar, but more complicated frequency shifts than plotted
in Figures~\ref{fig:FreqShift-LEO}--\ref{fig:FreqShift-all}  would
be observed by satellites. Note that the observed frequency shift
depends on position and velocity of both observer and satellite.
The dependence on spatial position of observer and satellite is
due to the fact that space is not homogeneous in a curved
space-time, due to the presence of the gravitational field.

The LEO satellite in Figure~\ref{fig:FreqShift-LEO} has an overall
shift to lower frequency, due predominantly to its high orbital
speed. The effect of time dilation dominates the effect due to the
gravitational potential difference. Superimposed on the overall
negative frequency shift, is a small-amplitude rapid variation in
the frequency shift due to the satellite's short orbital period.
This rapid variation is due to the slight orbital eccentricity and
the satellite's orbital inclination interacting with the Earth's
gravitational field.

The frequency shift of the GEO satellite has an overall positive
value of 5.38775$\times 10^{-10}$, and is dominated by the
gravitational potential frequency shift, causing a shift to higher
frequency when observed from the ground, see
Figure~\ref{fig:FreqShift-GEO}. The plot shows the residual
frequency shift, after subtraction of 5.38775$\times 10^{-10}$,
resulting from a slightly non-circular orbit.  The satellite
parameters have been chosen so that the orbital inclination is
zero.  The small variation in frequency shift is the result of
choosing an initial position and velocity that do not produce a
perfectly circular orbit (as may happen when real a satellite is
inserted into orbit). In the initial data used here, the result is
a small variation in the frequency shift on the order of $2\times
10^{-18}$.

In contrast, the HEO satellite in the Molniya orbit shows a
markedly different behavior, see Figure~\ref{fig:FreqShift-HEO}.
The frequency shift is negative for short periods of time when the
satellite is near perigee (near $\Delta t=0$) but is mostly
positive for a long time (for approximately 12 hours, roughly
between $\Delta t=0.5$ and $\Delta t=11.5$ hours) when the
satellite spends a long time at high altitude. The HEO satellite
makes a rapid transition from the extremes of frequency shift of
$\Delta \omega / \omega = +4\times 10^{-10}$ to $-4\times
10^{-10}$, in a time on the order one hour, whereas the satellite
period is approximately 12.4 hours.

\begin{figure}
\includegraphics{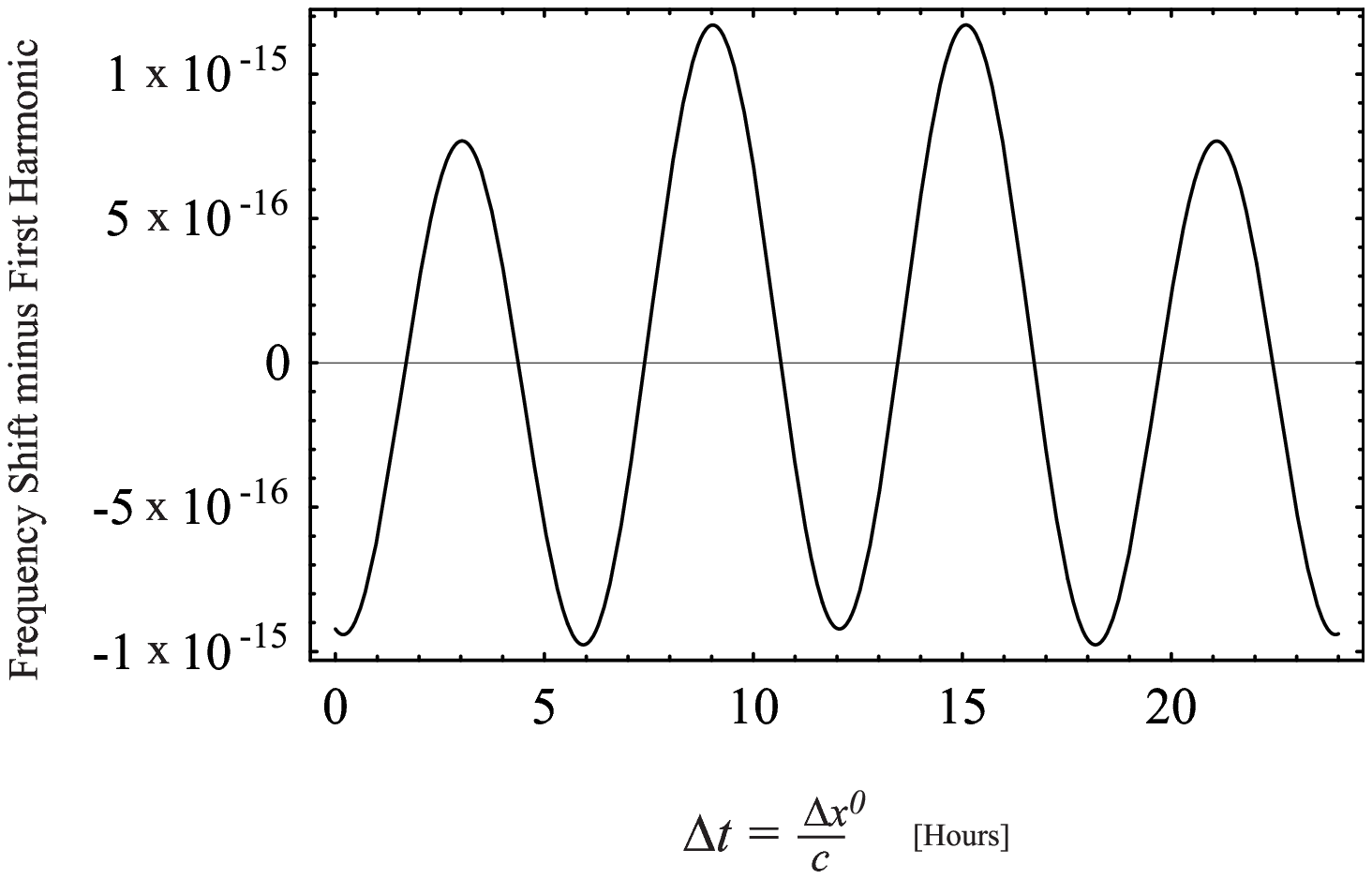}
\caption{\label{fig:FreqShift-GPS-HigherHarmonic} A plot of
Eq.~(\ref{fitFunctionDiff}) shows that the next higher harmonic,
is present in the fractional frequency shift observed on the geoid
for the GPS satellite.}
\end{figure}

The frequency shift for the GPS satellite is shown in
Figure~\ref{fig:FreqShift-GPS}.  The GPS satellite is in a high,
almost circular orbit, with an inclination of 55.3$^\circ$. The
net frequency shift is observed to be positive (higher frequencies
are observed on the geoid) with a slight variation due to orbital
eccentricity. The net shift is approximately
4.4$\times$10$^{-10}$, and in the actual GPS system, it is
compensated by the factory offset of -4.4$\times$10$^{-10}$. In
addition to this constant frequency shift, the frequency shift
also varies at the period of the GPS satellite. Furthermore, plots
of frequency shift for all satellites typically have non-zero
amplitudes at higher Fourier components than the satellite period.
As an example, for the GPS case, we can fit the function
\begin{equation} \label{fitFunction}
A + B  \sin(2 \pi f \, t + \phi)
\end{equation}
where $A$, $B$, $\omega^\prime$ and $\phi$ are constants, to be
fit to the computed function $\Delta \omega / \omega_s $. The
values of the best fit parameters are given by $A=4.4736 \times
10^{-10}$, $B=-5.78749\times 10^{-13}$, $f = 1.99009$ day$^{-1}$,
and $\phi/2 \pi=0.25011$ day. To demonstrate the presence of
higher harmonics, Figure~\ref{fig:FreqShift-GPS-HigherHarmonic}
shows a plot of the difference
\begin{equation} \label{fitFunctionDiff}
\frac{\Delta \omega} {\omega_s}  - \left(A + B  \sin( 2 \pi f \, t
+ \phi) \right)
\end{equation}
The plot shows that the next harmonic present is at twice the
basic GPS period.  From
Figure~\ref{fig:FreqShift-GPS-HigherHarmonic}, we see that the
amplitude for this next harmonic, which is on the order of
1$\times$10$^{-15}$, is significantly smaller than for the
fundamental frequency. Even though we have chosen a spherical
gravity model for the Earth for the frequency shift calculation,
higher harmonics are present here because the GPS orbit is
slightly eccentric.  In general, higher harmonics are present in
all satellites because the orbit is eccentric and because the
Earth has a non-spherical gravitational field. In the frequency
shift calculation presented here, we have neglected the effect of
the Earth quadrupole moment, however, it can easily be
incorporated into the calculation.

\section{Geolocation in Curved Space-Time}
Location of a source of electromagnetic radiation by using
multiple receivers is termed
geolocation~\cite{Ho1993,Fang1995,Niezgoda1994,Ho1997,Pattison2000}.
The source (called the emitter) can radiate a continuous wave (cw)
signal or a pulse. In either case, there are two methods commonly
employed to locate the emitter: time difference of arrival (TDOA)
and frequency difference of arrival (FDOA).  The TDOA technique is
based on differences of time of flight of the signal from the
emitter to each receiver.  FDOA is based on the Doppler effect and
the difference of frequencies observed by each receiver.

Often the emitter of interest is located on the surface of the
Earth.   This is a helpful mathematical constraint that reduces
the number or receivers needed to locate the emitter. Since
satellites are often used to receive signals, and they are costly
resources, this constraint helps reduce the number of satellites
required for geolocation.

Electromagnetic waves propagate at an (almost) constant velocity
in an ECI system of coordinates. However, in this system of
coordinates, an emitter that is stationary on the Earth's surface,
is rotating with respect to the ECI coordinates, and hence the
constraint that the emitter is on the Earth surface is actually a
time-dependent constraint (for an Earth model that is not a
surface of revolution).
\bigskip

\subsection{Time Difference of Arrival (TDOA) Geolocation}

In this section, the world function formalism is used to write the
general equations for TDOA geolocation, in an arbitrary system of
coordinates, either  ECEF or ECI coordinates. Since the world
function of space-time is an invariant, the general equations for
geolocation are covariant: they are valid in any system of
coordinates.  I assume that the emitter is located at a space-time
event whose coordinates are $x_o^i$, $i=0,1,2,3$. Similarly, I
assume that three satellites, $s=1,2,3$, receive the emitter
signal at space-time events with coordinates $x^i_s$. The
satellites' ephemerides are assumed known, and each satellite
carries a clock so the three reception times $x^0_s$, $s=1,2,3$,
are known as well as the spatial coordinates of the reception
events. So all satellite space-time coordinates $x^i_s$ are known.
The time of emission at the emitter, $x_o^0$, and the spatial
position of the emitter, $x_o^\alpha$, $\alpha=1,2,3$, constitute
four unknowns.

The electromagnetic waves from emitter to each satellite travel on
null geodesics that connect the emission event  to the three
reception events, at each of the three satellites. In an arbitrary
system of coordinates, the emitter  coordinates are related to the
satellite coordinates by the three equations
\begin{equation}\label{nullGeodesics}
\Omega(x^i_o,x^j_s)=0 \,\,\,\,\, s=1,2,3
\end{equation}
where $\Omega$ is the world function of the space-time.  The
causality conditions requiring that the reception occur after
emission, $x^0_o < x^0_s$, must also be added for a unique
solution.

The constraint that the emitter is on the Earth surface means that
the emitter coordinates lie on a 4-dimensional hypersurface, i.e.,
the two dimensional surface of the Earth sweeps out a
3-dimensional hypersurface (over time) given by:
\begin{equation}\label{EarthConstraint}
\chi(x^i_o)=0
\end{equation}
Equations~(\ref{nullGeodesics}) and (\ref{EarthConstraint}) form a
system of four nonlinear equations for the four space-time
coordinates $x^i_o$ of the emitter.  Since this system of
equations is nonlinear, it may have multiple solutions, or, no
solutions at all.

If the emitter is not on the surface of the Earth, then four
satellites (not three) are needed to provide  four equations
(s=1,2,3,4) of the form of Eqs.~(\ref{nullGeodesics}).

As an example of the application of TDOA to locate an emitter,
assume the emitter is on the Earth surface and that the space-time
surrounding Earth is well-modelled by the Schwarzschild metric.
The world function for a Schwarzschild space-time is
known~\cite{Synge1960,Bahder2001}.   Using Cartesian-like
coordinates~\cite{Bahder2001} and solving
Eq.~(\ref{nullGeodesics}) by iteration leads to
\begin{widetext}
\begin{equation}
\label{timeTransferScwarzschild} x^0_s = x^0_o +   |{\bf x}_s -
{\bf x}_o|
 + \frac{G M}{ c^2 }  \left[
 2 \log \left( \frac{\tan(\frac{\theta_o}{2})}{\tan(\frac{\theta_s}{2})} \right)
 +  \cos \theta_o - \cos \theta_s \right] \equiv x^0_o + \xi({\bf x}_s, {\bf x}_o)
\end{equation}
\end{widetext} where ${\bf x}_s$ and ${\bf x}_o$ are the spatial
coordinates of satellites and emitter, and $|{\bf x}_s - {\bf
x}_o|$ is the three-dimensional Euclidean distance between
satellite $s$ at reception time and emitter at emission time. When
three satellites, $s=1,2,3$, make time of arrival measurements, we
can form two independent TDOA equations:
\begin{eqnarray}
x^0_1 - x^0_2 & = & \xi({\bf x}_1, {\bf x}_o) - \xi({\bf x}_2, {\bf x}_o)\label{TDOA2Eq1}\\
x^0_2 - x^0_3 & = & \xi({\bf x}_2, {\bf x}_o) - \xi({\bf x}_3,
{\bf x}_o) \label{TDOA2Eq2}
\end{eqnarray}
Equations~(\ref{TDOA2Eq1})--(\ref{TDOA2Eq2}) are the TDOA
equations which contain the effects of the gravitational field on
delay of signal propagation (Shapiro time delay, which is on the
order of 40 ps). To these two equations, I must add the constraint
that the emitter is on the Earth surface.
Equations~(\ref{TDOA2Eq1})--(\ref{TDOA2Eq2}) are in Schwarschild
coordinates, which are essentially ECI coordinates. The
transformation from these ECI coordinates, ${\bf x}=
(x^1,x^2,x^3)$, to rotating ECEF coordinates, ${\bf y} =
(y^1,y^2,y^3)$, is given by
\begin{eqnarray}
y^0 & = & x^0 \nonumber \\
y^1 & = & \cos(\frac{\omega}{c} x^0) \, x^1 + \sin(\frac{\omega}{c} x^0) \, x^2 \nonumber \\
y^2 & = & -\sin(\frac{\omega}{c} x^0) \, x^1 + \cos(\frac{\omega}{c} x^0) \, x^2 \nonumber \\
y^3 & = & x^3 \label{CoordinateTransformationxx}
\end{eqnarray}
In three-dimensional notation, there exists a rotation matrix $R$,
given by Eq.~(\ref{CoordinateTransformationxx}), that relates the
spatial coordinates at a given coordinate time $x^0$
\begin{equation}\label{Transf}
{\bf y} = R(x^0) \cdot {\bf x}
\end{equation}
The coordinate time $x^0$ is taken to be the same in ECI and ECEF
coordinates.

For an emission event time $x^0_o$, the rotating ECEF coordinates
are related to the Schwarzschild coordinates by
\begin{equation}\label{emitterCoord2}
  {\bf y}_o = R(x^0_o) \cdot {\bf x}_o
\end{equation}

In rotating ECEF coordinates, the Earth's surface is given by
\begin{equation}\label{EarthSurface}
f({\bf y})=0
\end{equation}
The constraint in Eq.~(\ref{EarthConstraint}) that the emitter is
located on the Earth surface is then time-independent and can be
written in ECEF rotating coordinates as
\begin{equation}\label{constraintECEF}
f({\bf y}_o)= f(R(x^0_o) \cdot {\bf x}_o)=0
\end{equation}
Note that the unknown time of emission $x^0_o$ enters explicitly
in Eq.~(\ref{constraintECEF}).  This time can be eliminated by
using Eq.~(\ref{timeTransferScwarzschild}) (with $s=1$) in
Eq.(\ref{constraintECEF}), leading to the constraint equation
\begin{equation}\label{FinalConstraintEq}
f(R(x^0_1 - \xi({\bf x}_1, {\bf x}_o)) \cdot {\bf x}_o) = 0
\end{equation}
The three equations to be solved for the three spatial ECI
coordinates ${\bf x}_o$ of the emitter are
Eq.~(\ref{TDOA2Eq1})--(\ref{TDOA2Eq2}) and
(\ref{FinalConstraintEq}).  When the coordinates ${\bf x}_o$ are
found, they are to be substituted into
Eq.~(\ref{timeTransferScwarzschild}) to compute the emission time
$x^0_o$, in terms of the known satellite reception times $x^0_s$.
This time $x^0_o$ is then used in Eq.~(\ref{emitterCoord2}) to
compute the emitter ECEF rotating coodinates ${\bf y}_o$.

\subsection{Doppler Effect in a Gravitational Field}

The previous section described the equations needed to carry out
gelocation by the technique of TDOA. An alternative and
complimentary method is based on measurements of frequency
difference of arrival (FDOA).

The frequency that an observer sees emitted by a distant source is
not the same as the frequency transmitted by that source.  More
specifically, the observed frequency differs from the proper
(transmitted) frequency because of relative motion between source
and observer (Doppler effect) and because of a gravitational
potential difference between the position of the source and
position of the observer. Most often, the Doppler effect and the
gravitational potential effect are treated as separate effects. In
a static space-time, there is no difference.  However, in a
general space-time (not static or stationary), these two effects
are inseparable.  One expression describes both effects. Below, we
follow closely the derivation by Synge~\cite{Synge1960}.
\begin{figure*}
\includegraphics{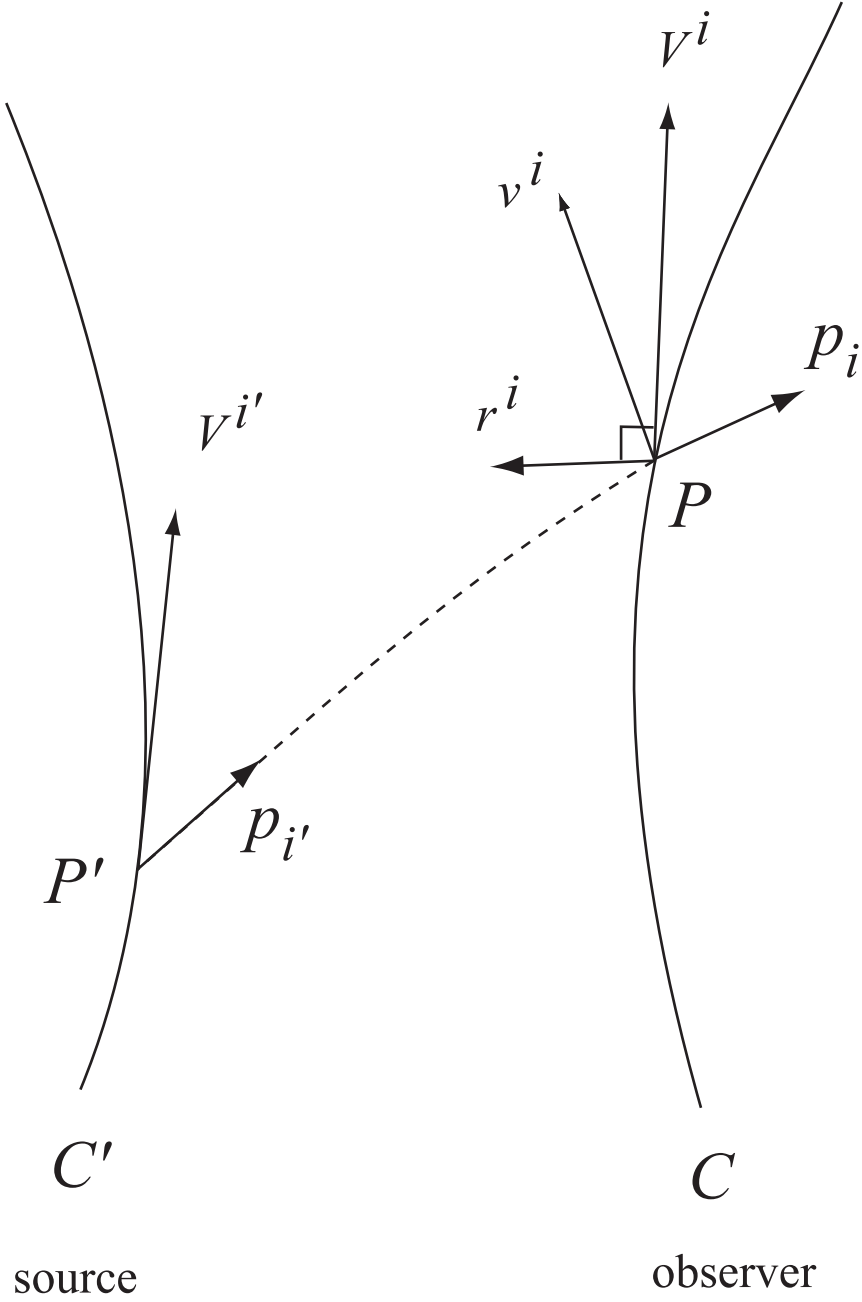}
\caption{\label{fig:doppler} The world lines of the emitter
$C^\prime$ and receiver $C$ are shown, together with their four
velocity,  $V^{i^\prime}$ at emission event $P^\prime$, and $V^i$
at reception event $P$. The photon at emission time has proper
momentum $p^{i^\prime}$ and at reception time, has momentum
$p^i$.}
\end{figure*}
Consider a space-time with an arbitrary gravitational field (not
static or stationary).  The source of electromagnetic radiation
(emitter)  travels on world line $C^\prime$ and the observer
(satellite with receiver) has world line $C$, see
Figure~\ref{fig:doppler}.  The emitted photon has momentum
$p^{i^\prime}$ in the comoving coordinate system of the emitter.
In the comoving system of coordinates of the observer (satellite)
the photon has momentum $p^i$.  For the comoving system of
coordinates of the observer, we define a tetrad $\lambda^i_{(a)}$,
$a=0,1,2,3$.  The three spatial components of the tetrad play the
role of orthonormal laboratory unit vectors. This tetrad frame
does not have to be a Fermi frame (it does not need to be a
non-rotating frame). We choose the time-like basis of the tetrad
to be the observer 4-velocity $\lambda^i_{(0)} =V^i$. We also use
the fact that the three space-like basis vectors of the tetrad are
orthogonal to this timelike basis vector $\lambda^i_{(\alpha)} \,
\lambda_{i(0)} = 0$, for $\alpha=1,2,3$, and orthonormal to each
other $\lambda^i_{(\alpha)} \, \lambda_{i(\beta)} = 0$, for
$\alpha,\beta=1,2,3$ and $\alpha\ne \beta$.  The inner product
between the tetrad basis is the Minkowski matrix:
\begin{equation}\label{teradDot}
\lambda^i_{(a)} \, \lambda_{i(b)} = \eta_{(ab)}
\end{equation}
where $\eta_{(ab)} = {\rm diag}(-1,+1,+1,+1)$.  As usual, tetrad
indices are raised and lowered with $\eta_{(ab)}$.

Define the 3-velocity of the source on $C^\prime$, relative to
observer on  $C$, by the three invariant projections on the tetrad
at $P$:
\begin{equation}\label{velocityProject}
  v_{(\alpha)}=v_i \, \lambda ^i_{(\alpha)}
\end{equation}
However, the velocity $V^{i^\prime}$ of the source and the
velocity of the observer $V^i$ cannot be compared directly because
they are at different space-time points, $P^\prime$ and $P$,
respectively. Therefore, the 4-velocity vector at $P^\prime$ must
be parallel translated to $P$, defining the vector $v_i$ at $P$ by
\begin{equation}\label{Vparalell}
v_i=g_{ij^\prime} \, V^{i^\prime}
\end{equation}
where $g_{ij^\prime}$ is the parallel propagator from $P^\prime$
to $P$.  Taking  $v^i v_i=-1$, so that $v^i$ is the velocity in
units of $c$, the invariant components on the tetrad basis are
related by
\begin{equation}\label{tetraddVel}
  v^{(0)} = \left(  1 +\delta_{\alpha \beta} \,  v^{(\alpha)}  v^{(\beta)}
  \right)^{1/2} =- v_{(0)}
\end{equation}
Then if $v^{(0)}=1$, the other components $v^{(\alpha)}=0$,
$\alpha=1,2,3$, so that, as compared at point $P$, the velocity of
source and observer are equal.

Now define at point $P$ the unit vector $r^i$ orthogonal to the
observer 4-velocity
\begin{equation}\label{r-Def}
  r_i V^i=0
\end{equation}
such that the vector $r_i$ lies in the 2-element which contains
the tangent vector $V^i$ to $C$ and the tangent vector $p_i$ to
the null geodesic $P^\prime \, P$ that connects the source and
observer.  Then by Eq.~(\ref{r-Def}), $r_i \lambda^i_{(0)} =
r_{(0)} =  0$.   Define the speed of recession of the source with
respect to the observer by
\begin{equation}\label{recessionSpeed}
  v_R = v_i \, r^i = v_{(\alpha)} \, r^{(\alpha)}
\end{equation}
since $r^{(0)}=0$.

Since the curve $P^\prime \, P$ is a null geodesic, the emitted
photon proper momentum $p_{i^\prime}$ is related to the photon
momentum $p_i$ at $P$ by parallel transport. Furthermore, the
scalar product of two 4-vectors is invariant under parallel
transport:
\begin{equation}\label{4VectorDotInvariant}
p_{i^\prime} V^{i^\prime} = p_i v^i
\end{equation}

The energy of the emitted photon in the comoving frame of
reference of the emitter is
\begin{equation}\label{emittedPhotonE}
 -E^\prime=  p_i^\prime V^{i^\prime} = p_{(0)} v^{(0)} +
 p_{(\alpha)} v^{(\alpha)}
\end{equation}

The energy of a photon, with momentum $p^i$, seen in the comoving
frame of the observer with 4-velocity $V^i$, is given by
\begin{equation}\label{photonEnergyObserved}
  E= - p_i V^i = - p_i \, \lambda^i_{(0)} = -p_{(0)} = p^{(0)}
\end{equation}

From the definition of unit vector $r^i$, i.e., that it exists in
the 2-element that contains the tangent $V^i$ at $P$ to $C$, and
the null geodesic tangent vector $p_i$, and the fact that $p_i$ is
a null vector, the  vector $p_i$ can be written as a decomposition
\begin{equation}\label{pVectorDefinition}
  p^i = \alpha (V^i - r^i)
\end{equation}
where $r_i r^i = 1$ and the magnitude $\alpha$ is to be
determined.  Multiplying Eq.~(\ref{pVectorDefinition}) by $V_i$, I
find that $p^i V_i = -\alpha = -E$.  Therefore,
Eq.~(\ref{pVectorDefinition}) can be written as
\begin{equation}\label{pVectorDefinition2}
  p^i = E (V^i - r^i)
\end{equation}

From Eq.~(\ref{emittedPhotonE}), and using $p_{(0)}=-E$,
\begin{equation}\label{Eexpression}
E^\prime = E v^{(0)} - p_{(\alpha)} v^{(\alpha)}
\end{equation}
The term $p_{(\alpha)} v^{(\alpha)} = p_i \lambda^i_{(\alpha)}
v^{(\alpha)}$ and using Eq.~(\ref{pVectorDefinition2}) for $p_i$
leads to

\begin{eqnarray}\label{P-expression}
p_{(\alpha)} v^{(\alpha)} & = & E(V_i- r_i) \lambda^i_{(\alpha)}
v^{(\alpha)}  \label{P-expression1} \\
    &  =  & E\left[ V_i  \lambda^i_{(\alpha)} - r_i \lambda^i_{(\alpha)}
    \right]  \label{P-expression2} \\
      &  = & -E r_{(\alpha)} v^{(\alpha)} = -E v_R \label{P-expression3}
\end{eqnarray}
since $V_i \lambda^i_{(\alpha)}=0$ because the frame is
orthogonal, and we used Eq.~(\ref{recessionSpeed}) for the
definition of the recessional velocity $v_R$.  Now use
Eq.~(\ref{P-expression3}) in Eq.~(\ref{Eexpression}) for
$p_{(\alpha)} v^{(\alpha)}$ and Eq.~(\ref{tetraddVel}) for
$v^{(0)}$ to obtain
\begin{equation}\label{Doppler1}
  E = \frac{E^\prime}{\left(  1+ v^2 \right)^{1/2} + v_R}
\end{equation}
where

\begin{equation}\label{speedSqared}
v^2 =\delta_{\alpha \beta} v^{(\alpha)} v^{(\beta)}
\end{equation}
and $\alpha,\beta=1,2,3$.  In Eq.~(\ref{Doppler1}), we have the
following definitions: \newline

\noindent $E$=photon energy observed by satellite at $P$ \\

\noindent $E^\prime$=photon energy emitted in (proper) comoving frame of emitter at $P^\prime$ \\

\noindent $v^2$=square of speed in terms of components on tetrad, defined by Eq.~(\ref{speedSqared}) \\

\noindent $v_R$ = recessional velocity defined by
Eq.~(\ref{recessionSpeed})\newline

Equation~(\ref{Doppler1}) shows that the frequency of an
electromagnetic signal observed by a satellite, $E$, is related to
the frequency emitted, $E^\prime$, by motional (velocity) effects.
Of course, the role of the gravitational field in producing a
frequency shift (gravitational red shift effect) is contained in
Eq.~(\ref{Doppler1}) in the geometry of the curved space-time. The
significance of Eq.~(\ref{Doppler1}) is that it is a general
expression, that is valid in an arbitrary space-time and does not
assume a static or stationary space-time. However,
Eq.~(\ref{Doppler1})  does assume that geometrical optics is valid
because the geodesic law for light propagation was used.

\subsection{Observed Doppler Shifts}

The observed energy of a photon, $E$,  is related to the observed
angular frequency $\omega$, by $E= \hbar \omega$.  In a system of
coordinates where the photon 4-momentum is $p_i$, and the observer
4-velocity is $V^i$, the observed photon frequency is given by
$E=\hbar  \omega = - p_i V^i$.  Similarly, the energy of the
photon emitted by the emitter, in the comoving frame of the
emitter, is related to the frequency by $E^\prime= \hbar
\omega^\prime = - p_{i^\prime} V^{i^\prime}$, where $p_{i^\prime}$
is the 4-momentum at the emission event and $V^{i^\prime}$ is the
4-velocity of the emitter.

The fractional shift in frequency, between the emitted and
received frequencies, due to motional doppler effects and
gravitation is given by
\begin{equation}\label{FreqShift2}
  \frac{E^\prime - E}{E^\prime} = \frac{p_{i^\prime} V^{i^\prime} -p_i V^i }{p_{i^\prime} V^{i^\prime}}
\end{equation}
The right side of Eq.~(\ref{FreqShift2}) can be written in terms
of the derivative of world function of the space-time. The world
function connects two points, $P^\prime$ and $P$, by a geodesic.
The geodesic connecting $P^\prime$ and $P$ may be expressed
parametrically as $x^i(u)$, for $i=0,1,2,3$, for $u_0 \le u \le
u_1$, where $P: x^i(u_o)$ and $P:x^i(u_1)$.

\begin{figure}
\includegraphics{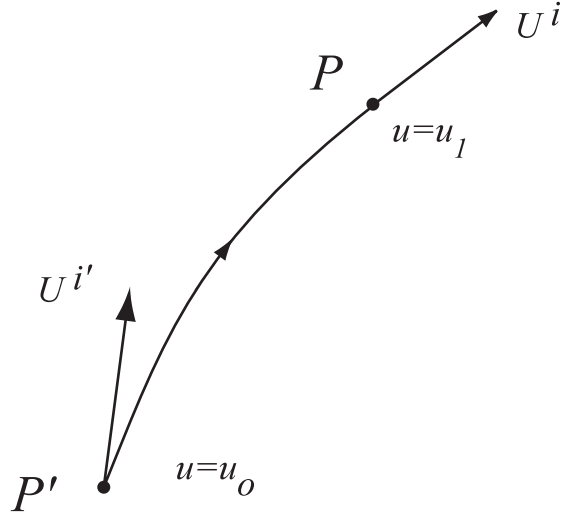}
\caption{\label{fig:UnitTangents} A geodesic path is shown between
points $P^\prime$ and $P$, with unit vectors $U^{i^\prime}$ and
$U^i$ at the end points.}
\end{figure}

The world function is a two-point scalar, depending on points
$P^\prime$ and $P$.  The  covariant derivatives of the world
function with respect to point $P^\prime$ and $P$ are given by
\begin{eqnarray}
\Omega_{i^\prime}  & = &
 \Omega_{i^\prime}(P^\prime,P)
= \frac{\partial \Omega}{\partial x^{i^\prime}} = -(u_1 - u_o)U_{i^\prime} \label{WFderiv1} \\
 \Omega_i  &  = &  \Omega_i(P^\prime,P)  =
 \frac{\partial \Omega}{\partial x^i}= (u_1 -  u_o)U_{i}  \label{WFderiv2}
\end{eqnarray}

The photon momentum, $p_{i^\prime}$ at point $P^\prime$, is
related to the photon momentum, $p_{i}$ at point $P$, by parallel
transport.  Also, the photon momentum is in the direction of the
tangent to the geodesic. Also, the magnitude of a vector (photon
momentum) is constant under parallel transport. Therefore, the
momentum is proportional to the tangents at point $P^\prime$ and
at $P$
\begin{eqnarray}
U_{i^\prime} & = & \alpha p_{i^\prime} \label{Up1} \\
U_{i} & = &  \alpha p_{i} \label{Up2}
\end{eqnarray}
where $\alpha$ is the same proportionality constant. Using
Eq.(\ref{WFderiv1})--(\ref{Up2}), the frequency difference in
Eq.~(\ref{FreqShift2}) can be written as
\begin{equation}\label{FreqShiftFinalExp}
\frac{\omega^\prime - \omega}{ \omega^\prime} =
  \frac{\Omega_{i^\prime} V^{i^\prime} + \Omega_{i} V^{i}}{\Omega_{i^\prime} V^{i^\prime}}
\end{equation}
Eq.~(\ref{FreqShiftFinalExp}) gives the relation between the
frequency $\omega$ observed by a satellite at the reception event
$P$ and the frequency of the emitted electromagnetic signal at the
emission event $P^\prime$.  The satellite has 4-velocity $V^i$ and
the emitter has 4-velocity $V^{i^\prime}$. The  frequency
$\omega^\prime$ is the proper frequency of the electromagnetic
signal broadcast by the emitter in its comoving frame, i.e., the
frame in which the emitter is at rest.
Eq.~(\ref{FreqShiftFinalExp}) contains the effect of the
gravitational potential (gravitational red shift) as well as the
effect associated with the relative motion of the source and
observer, which is usually called the Doppler effect.

The significance of the expression in
Eq.~(\ref{FreqShiftFinalExp})  is that it can be evaluated in any
system of coordinates (any reference frame).  For example,
Eq.~(\ref{FreqShiftFinalExp}) can be evaluated in inertial or
rotating coordinates.  The key ingredient to carrying out the
calculation is that the world function must be computed for the
given space-time.  This has already been done for the
Schwarzschild space-time, which models the Earth as a
sphere~\cite{Bahder2001,Synge1960}.  A similar calculation of the
world function should be done to include the effects of the
Earth's quadrupole potential $J_2$.  However, the gravitational
effects due to $J_2$ are three orders of magnitude smaller than
the monopole terms, and for some applications, may be negligible
when dealing with computations of signal propagation time or
frequency shift. For example, if the time delay due to the
monopole contribution is of the order of 40 ps, then the effect of
$J_2$ is expected to be three orders of magnitude smaller.
Furthermore, the doppler effect is not a cumulative effect such as
time dilation, so these terms do not increase in size with time.
Therefore, for many purposes the world function for the
Schwarzschild space-time is sufficient.

\subsection{Frequency Difference of Arrival (FDOA) Geolocation}

In this section, we derive the basic relations for geolocation by
frequency difference of arrival (FDOA) in a curved space-time.
Geolocation means that measurement of the frequency of an emitter
by several satellites (in relative motion, and at known positions
in space-time) can be used to locate the emitter in space-time.
Here we will take into account the frequency shifts due to the
relative motion of satellite and emitter, as well as due to the
gravitational potential effect.  When satellites in different
orbital regimes, e.g., LEO, GEO and HEO, are combined, the
gravitational potential differences can lead to significant errors
in emitter positions.  We neglect the (relatively large) effects
that can result from atmospheric propagation delays.

As derived in section X, subsection C,
Eq.~(\ref{FreqShiftFinalExp}) gives the basic expression that
connects the frequency emitted (proper frequency) with the
frequency observed, taking account the relative motion as well as
different gravitational potentials. Assume that the emitter sends
a signal at event $P_0=(x_o^0,x^\alpha_o)$ and that this signal is
received by a satellite at event $P_s=(x_s^0,x^\alpha_s)$. The
emitted frequency as measured at the emitter using a reference
oscillator is $\omega_o$. The frequency observed at the satellite,
$\omega_s$, is observed by using a reference oscillator at the
satellite identical to that used at the emitter to determine the
emitter's frequency. The fractional frequency difference can be
written in terms of the world function and the 4-velocities of the
emitter and satellite receiver:
\begin{equation}\label{FreqShiftVeryFinalExp}
\frac{\omega_o - \omega_s}{ \omega_o} =
  \frac{\Omega_{i_o} V^{i^o} + \Omega_{i_s} V^{i_s}}{\Omega_{i_o}
  V^{i_o}} \equiv R
\end{equation}
where for practical applications $R<<1$.  In
Eq.~(\ref{FreqShiftVeryFinalExp}), we have the following
definitions
\begin{eqnarray}
 \Omega_{i_o}   & = &  \frac{\partial \, \Omega(P_o,P_s))}{\partial x^i_o} \nonumber \\
 \Omega_{i_s}   & = &  \frac{\partial \, \Omega(P_o,P_s))}{\partial x^i_s}  \nonumber \\
V^{i_o} & = & \frac{d x^{i}_{o}}{d s} = \mbox{4-velocity of emitter} \nonumber \\
V^{i_s} & = & \frac{d x^{i}_{s}}{d s} = \mbox{4-velocity of satellite} \nonumber \\
\omega_o & = & \mbox{proper frequency of emitter, with 4-velocity~} V^{i_o} \nonumber \\
\omega_s & = & \mbox{frequency of emitter, as observed at } \nonumber \\
  &  &  \mbox{satellite with 4-velocity~} V^{i_s} \nonumber
\end{eqnarray}

The frequency observed by a given satellite can be written as
\begin{equation}\label{observedFreqBySat}
  \omega_s = \omega_o (1-R)
\end{equation}
where $R$ depends on the 4-velocities of emitter and receiver, and
the world function of space-time.  For the Minkowski metric given
in Appendix B, explicit evaluation of
Eq.~(\ref{FreqShiftVeryFinalExp}) gives

\begin{equation}\label{SpecialRelativityShift}
\omega_s = \omega_o =
  \frac{\gamma_s (1 - {\bf n}\cdot {\bf v}_s )}{\gamma_o (1 - {\bf n}\cdot {\bf v}_o )}
\end{equation}
where ${\bf v}_o $ and ${\bf v}_s $ are the velocities of the
emitter and satellite (in units of $v/c$), and the unit ${\bf n}$
vector is defined as ${\bf n}= ( {\bf r}_s - {\bf r}_o )/|{\bf
r}_s - {\bf r}_o|$, where ${\bf r}_o $ and ${\bf r}_s$ are the
spatial position of emitter and satellite at points of emission
and reception.

The measured quantities are frequency differences, which can be
formed from difference of satellite frequency given in
Eqs.~(\ref{observedFreqBySat}).  We do not give the final
complicated expressions here for frequency differences.  However,
oscillator frequency errors clearly contribute to errors in
estimated emitter position.  For example,  preliminary analysis
shows that the order of magnitude error in emitter location is
given by
\begin{equation}\label{emitterError}
|\delta {\bf x}_o | = \frac{\delta \omega}{\omega} \, \frac{|{\bf
x}_s - {\bf x}_o |}{|{\bf v}_s - {\bf v}_o |/c }
\end{equation}
where ${\bf x}_o$, ${\bf v}_o$ and ${\bf x}_s$, ${\bf v}_s$ are
the position and speed of the emitter and receiver, respectively,
and $\frac{\delta \omega}{\omega} $ is the fractional frequency
error. For example, for the typical values of satellite speed $v_s
\sim10^{-5}$, and satellite oscillator frequency error
$\frac{\delta \omega}{\omega} \sim 10^{-10}$, the order of
magnitude of emitter position error is $| \delta {\bf x}_o | \sim
$100 m.  A frequency error on the order of $10^{-10}$ can also be
expected to occur when a LEO and a GEO satellite are used together
for geolocation without compensating for the gravitational
frequency shift.  In such as case,  the use of two satellites can
lead to a similar position error on the order of 100 m. Detailed
investigation of Eq.~(\ref{observedFreqBySat}) for high-accuracy
geolocation is left for future work.

\section{Summary}

In this article we have considered the elements of the general
problem of navigation in space-time, as well as the restricted
problem of clock synchronization, within the context of a metric
theory of gravity, such as general relativity. In most real
applications, such as the GPS, a user is interested in determining
his space and time coordinates, rather than just time. General
relativity deals with the effect of motion and gravitational
potential differences on clocks, and it highlights features of the
space-time navigation problem that must be included in any
(classical or quantum) theory of navigation (or clock
synchronization).  Several quantum mechanical schemes have been
proposed to synchronize
clocks~\cite{MandelOu1987,Jozsa2000,Chuang2000,Yurtsever2000,burt2001,Jozsa2000a,preskill2000,Giovannetti2001,Bahder2004,Valencia2004}.
At the present time, the effects of motion and gravitational
potential differences have not been explicitly incorporated into
these quantum approaches to clock synchronization. Clearly the
magnitude of the relativistic effects is such that it must be
considered in future quantum approaches to navigation.  At the
present, there is no quantum theory of navigation in space-time
(analogous to the classical theory in Ref.~\cite{Bahder2003})
which permits a user to determine their four space-time
coordinates.

\begin{acknowledgments}
The author is grateful to Pete Hendrickson for numerous helpful
discussions. This work was supported by the Advanced Research and
Development Activity (ARDA).
\end{acknowledgments}

\appendix

\section{Significance of Accurate Clock Synchronization for Geolocation}

\begin{figure}
\includegraphics{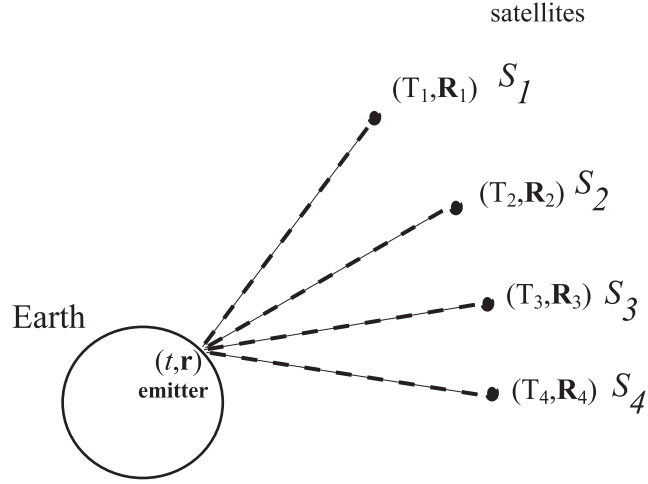}
\caption{\label{fig:geolocation} The time and position of an
electromagnetic emitter is shown at a space-time point $(t,{\bf
r})$, together with satellites, $S_a$, located at $(T_a,{\bf
R}_a)$, $a=1,2,3,4$.}
\end{figure}

Consider the problem of locating an emitter of electromagnetic
radiation located at a space-time point $(t,{\bf r})$ near the
Earth's surface.  The time of emission and position of this
emitter can be determined from signals received at fours
satellites, located at reception events $(T_a,{\bf R}_a)$,
$a=1,2,3,4$, see Fig.~\ref{fig:geolocation}.

In an inertial system of coordinates (such as the ECI frame), the
time and position of the emitter can be computed by solving the
four equations
\begin{equation} \label{LightTimeEq}
c (T_a-t) = |{\bf r} - {\bf R}_a|, \,\,\,\, a=1,2,3,4
\end{equation}
for $(t,{\bf r})$, where the satellite positions, ${\bf R}_a$, and
reception times, $T_a$, are known, and $c$ is the speed of light
in vacuum. In this Appendix, for simplicity, we neglect the
distinction between proper time and coordinate time.

By taking cyclic differences of Eqns.~(\ref{LightTimeEq}), we
cancel out the emission event time $t$ and obtain time difference
of arrival (TDOA) equations
\begin{eqnarray} \label{TDOAeqn}
c(T_1-T_2)&  = & |{\bf r} - {\bf R}_1| - |{\bf r} - {\bf R}_2| \\
c(T_2-T_3)&  = & |{\bf r} - {\bf R}_2| - |{\bf r} - {\bf R}_3| \\
c(T_2-T_4)&  = & |{\bf r} - {\bf R}_3| - |{\bf r} - {\bf R}_4|
\end{eqnarray}
where the measured quantities are the time differences
$T_a-T_{a+1}$ at the satellites.  The Eqns.~(\ref{TDOAeqn}) can be
solved for the coordinates of the emitter ${\bf r}$, in an
inertial coordinate system.  The value of ${\bf r}$ must then be
substituted into one Eqns.~(\ref{LightTimeEq}) to obtain the
emission time $t$.   The coordinates of the emitter, ${\bf r}$,
can then be transformed to geocentric (rotating with the Earth)
coordinates, ${\bf r}^\prime$, by a time-dependent rotation
\begin{equation} \label{CoordinateTransformation}
{\bf r}^\prime = \underline{{\bf D}}(t) \cdot  {\bf r}
\end{equation}
where $\underline{{\bf D}}(t)$ is the transformation matrix.

A similar problem occurs if we know that the emitter is located on
the surface of the Earth, so that we have the constraining
equation
\begin{equation} \label{EarthConstraint2}
|{\bf r}| = r_E
\end{equation}
where $r_E$=6378 km is the Earth's radius.   In this case, a
minimum of three satellites are sufficient to locate the emitter.
The first two of Eqns.~(\ref{TDOAeqn}) are solved together with
the constraining Eqns.~(\ref{EarthConstraint2}) for ${\bf r}$.
Then the geocentric (rotating with the Earth) coordinates, ${\bf
r}^\prime$, are found by the time-dependent rotation in
Eqns.~(\ref{CoordinateTransformation}).

The question then arises: How does the accuracy of the
synchronization of the clocks affect the accuracy of geolocation?
Assume that the clocks of two satellites are synchronized, and
that the satellite positions are known.
\begin{figure}
\includegraphics{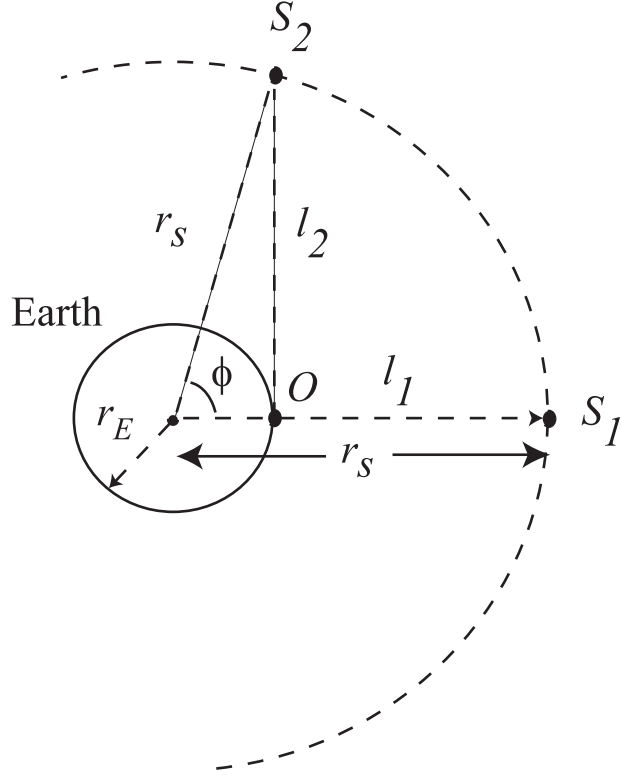}
\caption{\label{fig:GeoSats} The orientation of an electromagnetic
emitter $O$ is shown with respect to two satellites $S_1$ and
$S_2$.}
\end{figure}
The maximum TDOA occurs for satellites separated by an angle
$\phi$ given by (see Figure~\ref{fig:GeoSats})
\begin{equation}
\sin \phi=l_2/r_s
\end{equation}
The path length difference between these two satellites is
\begin{equation}
\Delta l = l_2 -l_1 = \sqrt{r_s^2-r_E^2}- (r_s-r_E)
\end{equation}
The maximum time difference of arrival (TDOA) for a signal is
\begin{equation}
\Delta t = \frac{\Delta l}{c} = \frac{l_2 -l_1}{c} =
\frac{\sqrt{r_s^2-r_E^2}- (r_s-r_E)}{c}
\end{equation}
For geosynchronous satellites, where $r_E=$6378 km and
$r_s=$42,164 km, the maximum (TDOA) is $\Delta t=$19.64 ms, and it
occurs when when one satellite has a longitude that is the same as
the emitter and the other satellite is at $\phi=$81.3$^{\circ}$,
see Figure~\ref{fig:GeoSats}. Due to geometric constraints, the
whole range difference is contained in the maximum time delay of
19.64 ms. If the clocks in the two satellites are synchronized
only to an accuracy of, say 10~ns, then the whole range difference
is compressed into a time delay of 19.64 ms.  The fractional range
error, and consequently the order of magnitude of the position
error (neglecting geometric dilution of precision (GDOP) factors)
is given by
\begin{equation}
\Delta x \sim \Delta l = (l_2 -l_1) \frac{10 {\rm ns}}{19.64 {\rm
ms}} \sim 3000 {\rm m} = 1.5 {\rm nm}
\end{equation}
An improvement in the clock synchronization, say by three orders
of magnitude, translates directly into three orders of magnitude
improvement in position accuracy, resulting in position accuracy
of the order of 3 m.

In the discussion in this appendix, we have neglected atmospheric
(tropospheric and ionospheric)  time-delay effects, accuracy of
computational correlation algorithms, and noise in the receiving
system.  An implicit assumption is also made that the system of
coordinates (inertial reference frame) is accurate, and that the
time-dependent transformation between inertial and geocentric
(rotating) coordinates is accurately known.

\section{Conventions and Notation}

Where not explicitly stated otherwise, we use the convention that
Roman indices, such as found on space-time coordinates $x^i$, take
the values $i=0,1,2,3$, and Greek indices take values
$\alpha=1,2,3$. Summation is implied over the range of any index
when the same index appears in a lower and upper position.  In
some cases, summation over Greek indices is implied when indices
both appear in upper position, such as in $x^\alpha dx^\alpha$.

If $x^i$ and $x^i + dx^i$ are two events along the world line of
an ideal clock, then the square of the proper time interval
between these events  is $d\tau=ds/c$, where the measure $ds$ is
given in terms of the space-time metric as $ ds^2 = -g_{ij}\, dx^i
\, dx^j $.  I choose $g_{ij}$ to have the signature $+$2. When
$g_{ij}$ is diagonalized at any given space-time point, the
elements can take the form of the Minkowski metric given by
$\eta_{00}=-1$, $\eta_{\alpha \beta}=\delta_{\alpha \beta}$. In
discussion of an observer carrying a tetrad, the Mnkowski matrix,
 $\eta_{(ab)} = \eta_{ab}$, is used.



\end{document}